\documentclass[prx,twocolumn,showpacs,amsmath,amssymb,superscriptaddress,  nofootinbib]{revtex4-1}

\usepackage{epsfig,epstopdf} 
\usepackage{graphicx}
\usepackage{dcolumn}
\usepackage{bm}
\usepackage{bbm}
\usepackage{ulem}
\usepackage{xcolor}
\usepackage{amsmath}
\usepackage{esint}
\usepackage{bbold}
\usepackage{slashed}
\usepackage{mathrsfs}
\usepackage{subfigure}
\usepackage{verbatim}
\usepackage{dsfont}
\usepackage{float}
\usepackage{wasysym}

\usepackage[breaklinks]{hyperref}
\hypersetup{colorlinks=true, linkcolor=blue, citecolor=blue, filecolor=blue, urlcolor=blue}

\AtBeginDocument{%
    \newwrite\bibnotes
    \def\bibnotesext{Notes.bib}
    \immediate\openout\bibnotes=\jobname\bibnotesext
    \immediate\write\bibnotes{@CONTROL{REVTEX41Control}}
    \immediate\write\bibnotes{@CONTROL{%
    apsrev41Control,author="08",editor="1",pages="1",title="0",year="1"}}
     \if@filesw
     \immediate\write\@auxout{\string\citation{apsrev41Control}}%
    \fi
}%

\newcommand{\nc}{\newcommand}
\nc{\be}{\begin{equation}} \nc{\ee}{\end{equation}}
\nc{\bea}{\begin{eqnarray}} \nc{\eea}{\end{eqnarray}}
\nc{\bean}{\begin{eqnarray*}} \nc{\eean}{\end{eqnarray*}}
\nc{\dg}{\dagger}
\nc{\ua}{\uparrow} \nc{\da}{\downarrow}
\nc{\lag}{\langle} \nc{\rag}{\rangle}

\newcommand\T{\rule{0pt}{2.6ex}}       
\newcommand\B{\rule[-1.2ex]{0pt}{0pt}} 

\begin{document}

\title{Theory of Anomalous Floquet Higher-Order Topology: Classification, Characterization, and Bulk-Boundary Correspondence}

\author{Rui-Xing Zhang}
\email{ruixing@umd.edu}
\affiliation{Condensed Matter Theory Center, Department of Physics, University of Maryland, College Park, Maryland 20742-4111, USA}
\affiliation{Joint Quantum Institute, University of Maryland, College Park, MD 20742, USA}

\author{Zhi-Cheng Yang}
\affiliation{Joint Quantum Institute, University of Maryland, College Park, MD 20742, USA}
\affiliation{Joint Center for Quantum Information and Computer Science, University of Maryland, College Park, MD 20742, USA}

\date{\today}

\begin{abstract}

Periodically-driven or Floquet systems can realize anomalous topological phenomena that do not exist in any equilibrium states of matter, whose classification and characterization require new theoretical ideas that are beyond the well-established paradigm of static topological phases. In this work, we provide a general framework to understand anomalous Floquet higher-order topological insulators (AFHOTIs), the classification of which has remained a challenging open question. In two dimensions (2D), such AFHOTIs are defined by their robust, symmetry-protected corner modes pinned at special quasienergies, even though all their Floquet bands feature trivial band topology. The corner-mode physics of an AFHOTI is found to be generically indicated by 3D Dirac/Weyl-like topological singularities living in the phase spectrum of the bulk time-evolution operator. Physically, such a phase-band singularity is essentially a ``footprint" of the topological quantum criticality, which separates an AFHOTI from a trivial phase adiabatically connected to a static limit. Strikingly, these singularities feature unconventional dispersion relations that cannot be achieved on any static lattice in 3D, which, nevertheless, resemble the surface physics of 4D topological crystalline insulators. We establish the above higher-order bulk-boundary correspondence through a dimensional reduction technique, which also allows for a systematic classification of 2D AFHOTIs protected by point group symmetries. We demonstrate applications of our theory to two concrete, experimentally feasible models of AFHOTIs protected by $C_2$ and $D_4$ symmetries, respectively. Our work paves the way for a unified theory for classifying and characterizing Floquet topological matters. 

\end{abstract}

\maketitle

{\hypersetup{linkcolor=black}
\tableofcontents}

\section{Introduction}
The recognition of the role of topology in band theory has spawned a systematic classification and identification of topological band insulators that are distinct from atomic ones \cite{thouless1982,haldane1988model,kane2005z2,bernevig2006QSH,hasan2010colloquium,qi2011topological,bradlyn2017topo,po2018fragile,elcoro2020magnetic}. 
The nontrivial bulk band topology manifests itself as robust gapless states residing on lower-dimensional boundaries. Such bulk-boundary correspondence constitutes a hallmark of topological band theory, and plays an essential role in both experimental probe and theoretical characterization. Apart from internal symmetries (time reversal, particle-hole, and chiral symmetries), the ubiquitous crystalline symmetries in solids furnish a much richer set of topological crystalline insulators (TCIs) \cite{fu2011topological,liu2014topological,zhang2015topological,ando2015topological,wang2016hourglass,chang2017mobius,wieder2018wallpaper} that were previously overlooked in the tenfold-way topological classification. A well-known example is the three-dimensional (3D) mirror-protected TCI phase in bulk SnTe \cite{hsieh2012topological}, whose topological Dirac surface states have been experimentally observed using angle-resolved photoemission spectroscopy (ARPES) \cite{tanaka2012experimental}. 

Remarkably, crystalline symmetries can sometimes support a novel {\it higher-order} bulk-boundary correspondence \cite{benalcazar2017quantized,benalcazar2017electric,Langbehn2017reflection,song2017d-2,schindler2018higher,khalaf2018higher,wang2018high,yan2018majorana,hsu2018majorana,zhang2019helical,miert2018higher,wang2018weak,wieder2018axion,zhang2019higher,wu2019high,yan2019higher,benalcazar2019quantization,schindler2019fractional,zhang2020kitaev,zhu2020identifying,zhang2020holonomic,hsu2020inversion,vu2020time}, which is generally not possible in any internal-symmetry-protected topological insulators. Namely, for some TCIs, the bulk topology enforces gapless modes to live on boundary manifolds with codimension $d>1$, for example, 1D hinge modes on the surface of a bulk-3D system, and 0D corner modes on the edge of a bulk-2D system. Such TCIs are generally termed ``higher-order topological insulators" (HOTIs). Researchers have introduced various symmetry-based indicators \cite{khalaf2018symmetry,po2017symmetry} (i.e. functions of crystalline symmetry eigenvalues at high symmetry momenta) to achieve a simple and feasible topological diagnosis for HOTIs. So far, materialization of HOTI physics has been predicted in bismuth \cite{schindler2018higherBi} (experimentally supported by scanning tunneling microscopy studies), EuIn$_2$As$_2$ \cite{xu2019higher}, MoTe$_2$ \cite{wang2019MoTe2}, Cd$_3$As$_2$ \cite{wieder2020strong}, MnBi$_{2n}$Te$_{3n+1}$ \cite{zhang2020mobius}, etc.   

More recently, motivated by the experimental progress in synthetic photonic and cold atomic systems, such topological ideas have been extended beyond static electronic materials and apply to out-of-equilibrium systems as well, in particular, Floquet systems that are subject to time-periodic drives \cite{oka2009photovoltaic,kitagawa2010topo,lindner2011floquet,jiang2011majorana,cayssol2013floquet,rechtsman2013photonic,wang2013observation,usaj2014irradiated,oka2019floquet,nakagawa2020wannier,harper2020topology,zhang2020tunable}. For instance, the quasienergy Floquet bands of a periodically driven system can carry their own topological indices, analogous to the energy bands in a static system. However, such topological theory of Floquet bands is incapable of capturing {\it anomalous} Floquet topological phenomena. Specifically, Ref. \cite{rudner2013anomalous} first pointed out the possibility of a 2D anomalous Floquet topological insulator (AFTI) in class A, which hosts robust chiral edge modes even when all of its bulk Floquet bands carry trivial Chern numbers. In fact, the topological nature of such anomalous chiral edge mode configuration is encoded in the full time-evolution operator $U({\bf k}, t)$ \textit{within} one period $T$, instead of its Floquet band Hamiltonian. Following this seminal work, a complete classification of AFTIs from the ten Altland-Zirnbauer symmetry classes in all dimensions has been achieved \cite{roy2017periodic,yao2017topo}.

\begin{figure*}[t]
	\includegraphics[width=0.96\textwidth]{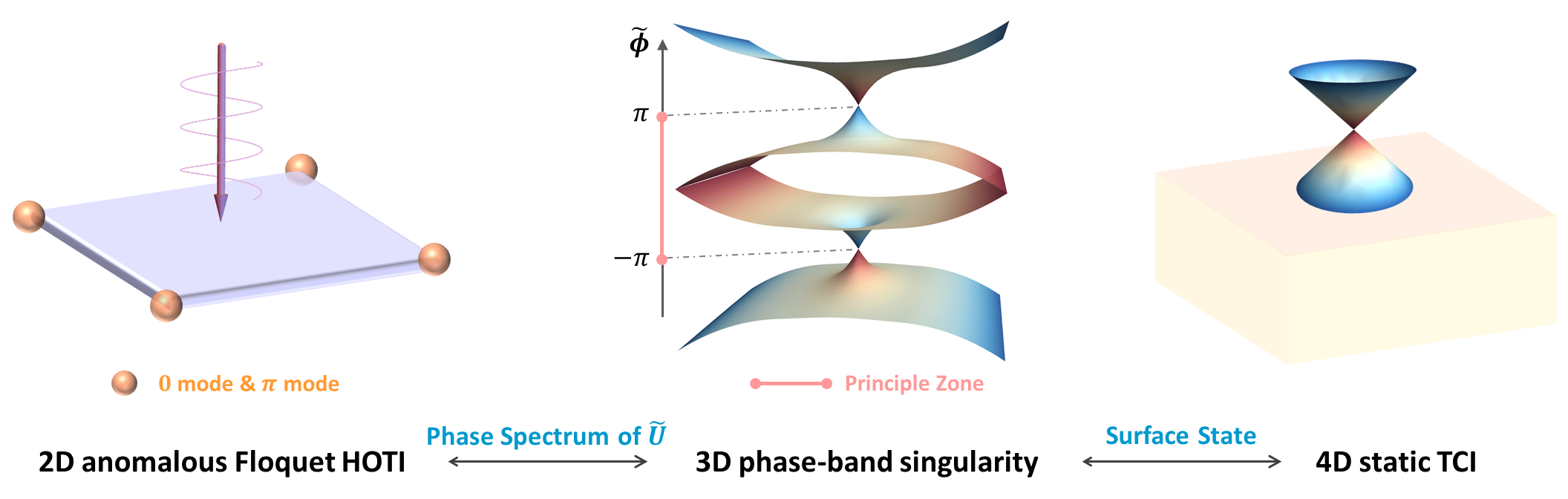}
	\caption{A 2D AFHOTI features Floquet corner modes at both quasienergies $0$ and $\pi$. It is characterized by the emergence of topological singularity on the principle zone boundary of the 3D phase band $\tilde{\phi}({\bf k}, t)$, i.e. the eigenspectrum of the periodized return map $\widetilde{U}({\bf k}, t)$. Such a phase-band singularity is further linked to the gapless surface state of a 4D static topological crystalline insulator.}
	\label{Fig:schematic}
\end{figure*} 

Following the breakthroughs of static HOTIs, it is natural to extend the concept of AFTIs to a higher-order version, i.e. anomalous Floquet higher-order topological insulators (AFHOTIs) \cite{peng2019floquet,rodriguez2019higher,chaudhary2019phonon,peng2020floquet,bomantara2020measurement,hu2020dynamical,huang2020floquet}. In 2D, such an AFHOTI, by definition, features robust 0D modes bound to the system's geometric corners, which are further energetically pinned at both quasienergies $0$ and $\pi/T$ \footnotemark[1]. However, theoretical proposals for AFHOTIs have so far been restricted to case-by-case studies, and the corresponding topological indices are also designed in a model-specific manner with limited generality. Moreover, the majority of AFHOTI proposals are in fact {\it not} bulk higher-order topological in the conventional sense. Namely, the corner modes of such a Floquet system can be symmetrically eliminated via closing the boundary quasienergy gap, instead of a bulk one. Following the convention in static HOTIs, one can call such Floquet systems {\it extrinsic} AFHOTIs, in sharp contrast to the {\it intrinsic} AFHOTIs which do feature a higher-order bulk-boundary correspondence. A typical example of an extrinsic AFHOTI is shown in Ref.~\cite{hu2020dynamical}, which introduced a 2D quantum walk model as a Floquet version of a quadrupole insulator \cite{benalcazar2017electric, benalcazar2017quantized}. Despite the phenomenological resemblance to an intrinsic AFHOTI, however, the crystalline symmetries (i.e. $D_2$ symmetry) in their model are insufficient to protect corner modes against generic adiabatic, $D_2$-preserving perturbations. Despite being inherently different, the crucial distinctions between extrinsic and intrinsic AFHOTIs have not been explored in depth in the literature. Specifically, lacking a general theory, the intrinsic AFHOTIs have so far remained far less well-understood than both their first-order cousins (i.e. AFTIs) and their static analogs (i.e. HOTIs). How to systematically characterize and classify higher-order anomalous Floquet topological phases remains an intriguing, crucial, and challenging open question in the fields of both topological phases and non-equilibrium dynamics.

\footnotetext[1]{The AFHOTIs should also be distinguished from conventional Floquet HOTIs \cite{bomantara2019coupled,seshadri2019generating,plekhanov2019floquet,zhu2020floquet}, where the corner modes only show up at {\it either} $0$ {\it or} $\pi/T$.  Such corner modes can be diagnosed from the topological indices of the bulk Floquet bands.}

In this work, we fill in this gap by providing a theoretical framework that allows for a systematic classification and characterization of general AFHOTIs. Unless otherwise stated, the AFHOTIs considered in this work always refer to the intrinsic ones. As a demonstration, in this work we focus on 2D AFHOTIs protected by both point group symmetries and chiral symmetry. The presence of chiral symmetry is necessary for all 2D AFHOTIs and is responsible for pinning their topological corner modes at quasienergies 0 and $\pi/T$. A key stepping stone in our theoretical framework is the deep connection between 2D anomalous Floquet higher-order topology and irremovable topological singularities in the 3D phase bands of the time-evolution operator. Such topological singularities live in the 3D $({\bf k}, t)$ parameter space and generally take the form of stable 3D Weyl or Dirac points across the principal zone boundary of the phase band. It is the very existence of such phase-band singularities that prohibits the AFHOTIs from being adiabatically deformed to any static system, making our AFHOTIs intrinsically dynamical phases of matter. Strikingly, all of the robust phase-band singularities exhibit unconventional dispersion relations that {\it cannot} be achieved in a 3D static lattice. Rather, such exotic 3D singularities resemble the boundary physics of static TCIs in 4D.

Specifically, our classification scheme for 2D AFHOTIs consists of two steps: (i) a classification of symmetry-protected topological phase-band singularities; (ii) a mapping from bulk phase-band singularities to boundary topological modes. For the first step, we start by exploring phase-band singularities protected by $n$-fold rotation groups $C_n$. We find that a stable $C_n$-protected singularity generically takes the form of a ``dynamical Weyl pair" (DWP), which is a pair of chiral-related Weyl points with a vanishing net monopole charge. We prove that such a single DWP is stable \textit{iff} $C_n$ anticommutes with chiral symmetry. This further allows us to define a set of {\it $C_n$ topological charges} that fully classify all stable DWPs for a given Floquet system with $C_n$ symmetry.

We then move on to classify the phase-band singularities protected by a dihedral group $D_n$, which is generated by a $C_n$ rotation and a mirror reflection $M_x$. The mirror symmetry in the $D_n$ group not only constrains the $C_n$ topological charges, but is also capable of protecting new type of singularities on their own. We identify a set of {\it mirror topological charges} to fully characterize mirror-protected singularities, which are well-defined only if mirror and chiral symmetries commute with one another. Combining the rotation and mirror topological charges, we obtain a complete classification of phase-band singularities for general class AIII $D_n$-symmetric Floquet systems in 2D. In particular, we emphasize that the $D_2$-symmetric Floquet quadrupole insulator model proposed in Ref.~\cite{hu2020dynamical} is {\it singularity-free} (see Appendix \ref{app:huang}) and carries {\it none} of the above crystalline topological charges, consistent with its extrinsic topological nature.

However, phase-band singularities \textit{do not} necessarily imply higher-order topology. 
Therefore, the second part of our results aims to establish the higher-order bulk boundary correspondence, i.e. the mapping between bulk singularities and Floquet corner modes. This is achieved by introducing a new theoretical technique dubbed ``phase-band dimensional reduction" (PBDR). Physically, PBDR maps a 2D Floquet system with phase-band singularities to a set of lower-dimensional building blocks, in this case, 1D class AIII AFTIs \cite{fruchart2016complex,liu2018chiral} with end modes. If the end modes of the resulting 1D AFTIs are both symmetry protected and topology enforced, they will correspond to robust corner modes in the original 2D Floquet system before the PBDR mapping. In this way, we succeed in establishing the higher-order bulk-boundary correspondence and defining the associated higher-order topological indices, which thus completes our classification for 2D AFHOTIs with point group symmetries.  In particular, we find that $C_n$-protected AFHOTIs generally admit a $\mathbb{Z}_2$ classification, while those protected by $D_n$ groups can admit a $\mathbb{Z}$ or $\mathbb{Z}\times 2\mathbb{Z}$ classification, depending on their specific symmetry properties.

The rest of the paper is organized as follows. Sec.~\ref{sec:preliminary} is a summary of our main results in this paper, which also contains a brief review of basic notions that are important to our theory. In Sec.~\ref{sec:ssh}, we revisit the theory of 1D AFTIs in class AIII and demonstrate how the anomalous Floquet topology in these 1D systems is encoded in their phase-band singularities. Sec.~\ref{sec:Cn} and~\ref{sec:Dn} classify $C_n$ and $D_n$-protected phase-band singularities in 2D class AIII Floquet systems, respectively. We introduce the PBDR technique in Sec.~\ref{sec:reduction}, where we also establish the higher-order bulk-boundary correspondence for each symmetry class, and conclude the classifications for 2D AFHOTIs with point group symmetries. As demonstrations, we apply our topological diagnostic to two concrete AFHOTI models in Sec.~\ref{sec:model}: one with $C_2$ symmetry and the other with $D_4$ symmetry. Finally, we close with discussions for future directions in Sec. \ref{sec:discussion}.

\section{Preliminaries and summary of results}
\label{sec:preliminary}

In this section, we start with some basic notions for general Floquet systems and further introduce the concept of phase-band topological singularities as an indicator of anomalous Floquet topological phenomena. We then define the concept of AFHOTIs in 2D and further specify the crucial role of both spectral and crystalline symmetries in protecting those exotic phases. Our key result in this work, the topological classification of 2D AFHOTIs with point group symmetries, is summarized in the last part.     

\subsection{Basic notions} 

The quantum dynamics of a time-periodic Bloch Hamiltonian $H({\bf k}, t)=H({\bf k}, t+T)$ is described by its time-evolution operator
\begin{equation}
U({\bf k}, t) = \mathcal{T} e^{-i\int_0^t H({\bf k}, t') dt'},
\end{equation}
where $\mathcal{T}$ denotes time ordering. At stroboscopic times, the Floquet operator $U({\bf k}, T)$ defines the Floquet Hamiltonian via
\begin{equation}
H_F^{(\varepsilon)}({\bf k}) \equiv \frac{i}{T}\log_{\varepsilon} U({\bf k}, T),
\end{equation}
whose eigenvalues $\{ \epsilon_n({\bf k}) \}$ define the quasienergy or Floquet bands. Here $\varepsilon$ labels the branch cut of the logarithmic function, with
\begin{equation}
\log_{\varepsilon}(e^{i\alpha}) = i \alpha,\ \text{for } \varepsilon-2\pi<\alpha<\varepsilon.
\end{equation}
For our purpose, we take $\varepsilon=\pi$ throughout our discussion and hence the superscript in $H_F^{(\varepsilon)}({\bf k})$ will be omitted for simplicity.
It is easy to see that $\{ \epsilon_n({\bf k}) \}$, the quasienergy spectrum of $H_F$, is periodic in units of $2\pi/T$, leading to a series of Floquet Brillouin zones. Therefore, we can restrict the quasienergies of $H_F$ within the principle zone $(-\frac{\pi}{T},\frac{\pi}{T}]$. Below we will drop the factor of $1/T$ in the quasienergies. Similarly to conventional electronic bands in static systems, a Floquet band can carry its own topological index. For example, a monolayer graphene can be turned into a Floquet Chern insulator in the presence of circularly polarized light \cite{usaj2014irradiated}, which features both a nonzero Chern number in its bulk Floquet bands and a chiral edge mode in the quasienergy spectrum of a finite system with open boundary. 

An anomalous Floquet topological system, however, features exotic boundary topological phenomena in its quasienergy spectrum, despite the triviality of all its Floquet bands. A known example is the 2D anomalous Floquet topological insulator (AFTI) in class A with chiral edge modes on the boundary, while each of its Floquet bands carries a vanishing Chern number \cite{rudner2013anomalous}. Such an anomalous Floquet topological phase, by definition, has no static analogs. Hence, its topological information is, in principle, encoded in the full time evolution process (e.g. $U({\bf k}, t)$ for $t\in(0, T]$), but not in a single Floquet operator $U({\bf k}, T)$.         

To understand the general anomalous Floquet topological physics, it is instructive to consider the eigenvalue spectrum of $U({\bf k}, t)$:
	\begin{equation}
	U({\bf k}, t)\ |\phi_n({\bf k}, t) \rangle = e^{-i \phi_n({\bf k}, t)} |\phi_n({\bf k}, t) \rangle,
	\end{equation}
	where $|\phi_n({\bf k}, t) \rangle$ denotes an eigenstate of the evolution operator, and $\{\phi_n({\bf k}, t)\}$ are known as the ``phase bands" in $({\bf k}, t)$ space \cite{nathan2015topo}.
When $t=T$, the phase band $\phi_n({\bf k}, T)$ is exactly the Floquet band $\epsilon_n({\bf k})$ (i.e. eigenvalues of $H_F$) up to a factor of $T$. Similarly to the Floquet bands, $\{\phi_n({\bf k}, t)\}$ are well-defined modulo $2\pi$, and can thus be restricted within the principal zone $(-\pi, \pi]$. Usually, it is convenient to define the return map (or the ``micromotion" operator)
\begin{equation}
\widetilde{U}({\bf k}, t) \equiv U({\bf k}, t) e^{i H_F ({\bf k}) t},
\end{equation}
which is periodic in time
\begin{equation}
\widetilde{U}({\bf k}, t) = \widetilde{U}({\bf k}, t+T),
\end{equation}
and contains the same information as $U({\bf k}, t)$~\cite{rudner2013anomalous}. The phase band $\{\widetilde{\phi}_n({\bf k}, t)\}$ for $\widetilde{U}$ can be defined similarly.

In particular, a {\it phase-band singularity} occurs when (i) the phase band manifold $\phi_n({\bf k,t})$ or $\widetilde{\phi}_n({\bf k}, t)$ is gapless across its principal zone boundary (e.g. $\widetilde{\phi}_n=\pm \pi$); (ii) the gaplessness has a topological origin and is irremovable without closing the bulk Floquet gap. We emphasize that the phase-band singularities generally serve as an indicator of general anomalous Floquet topological phenomena, including both AFTIs and AFHOTIs, which can manifest itself as point nodes, line nodes, etc. For example, the anomalous chiral edge modes of a 2D class A AFTI are determined by the net monopole charge of Weyl nodes in the phase band \cite{nathan2015topo}. As we will show later, the concept of phase-band singularity also plays a crucial role in our theory of AFHOTIs.

\subsection{Anomalous Floquet higher-order topology}

We define a 2D Floquet system as a symmetry-protected AFHOTI if the following conditions are satisfied:
\begin{enumerate}
	\item[(i)] its bulk Floquet spectrum is gapped around quasienergies 0 and $\pi$;
	\item[(ii)] the bulk Floquet bands are topologically trivial;
	\item[(iii)] it has no 1D anomalous (e.g. chiral or helical) edge modes penetrating the bulk Floquet gaps;
	\item[(iv)] it hosts robust 0D localized modes at quasienergies 0 and $\pi$;
	\item[(v)] both 0 and $\pi$ modes are protected symmetrically and topologically, which cannot be eliminated without breaking the symmetry or closing the bulk Floquet gap at 0 and $\pi$.
\end{enumerate} 

Let us make a few remarks on the above definition. First of all, a spectral symmetry (e.g. chiral symmetry ${\cal S}$) is required for pinning the 0D modes at the special quasienergies, without which the 0D modes can move freely in energy and merge with the bulk Floquet bands. We note that for some static obstructed atomic insulators with no chiral symmetry, one can still measure the fractional electron charge accumulated around each sample corner as a signature of the corner-mode physics \cite{benalcazar2019quantization,schindler2019fractional}, even when the corner localized eigenstates are shifted away from zero energy. However, such a fractional corner charge relies on a good definition of electron filling, which is somewhat tricky in periodically driven systems. In this case, only corner modes that are energetically pinned by chiral symmetry will be considered as higher-order topological in our theory.

Secondly, the higher-order bulk-boundary correspondence of an AFHOTI is {\it only} possible with the protection of certain spatial or space-time crystalline symmetry $G$. The symmetry protection guarantees the robustness of 0D modes against any adiabatic perturbations that respect $G$, i.e., smooth deformations to $U({\bf k}, t)$ that keep the bulk quasienergy gap open at 0 and $\pi$. In contrast, without $G$, one can always attach additional 1D Floquet topological states with 0 and $\pi$ end modes to the system's boundary, which can hybridize with the original 0D modes and shift them away from quasienergies 0 and $\pi$. The robustness against any symmetry-allowed perturbation is an important factor that distinguishes our definition of (intrinsic) AFHOTIs from the aforementioned extrinsic ones.

In addition, the triviality of bulk Floquet bands guarantees that the number of robust zero modes equals to that of the $\pi$ modes. In reality, both the zero and $\pi$ modes will simultaneously appear around certain geometric corners of a system with open boundary. Therefore, while we only mention explicitly one set of corner modes hereafter, the readers should always have in mind that there are two copies of such corner modes in the system at quasienergies 0 \textit{and} $\pi$, respectively.

Lastly, we note that an AFHOTI, by definition, cannot be diagnosed by any known topological index that is defined for static systems. This motivates us to establish a new theoretical framework for describing general AFHOTI phenomena, which is briefly summarized in the following subsections.

\subsection{Phase-band singularities for 2D class AIII Floquet systems}     
\label{sec:summary_singularity}  

In this work, we focus on the topological classification and characterization of 2D AFHOTIs protected by both chiral symmetry ${\cal S}$ and 2D point group symmetries. Namely, we will take the crystalline group $G$ as either a rotation group $C_n$ or a dihedral group $D_n$, for $n\in\{1,2,3,4,6\}$. Similar to 2D AFTIs, the topological nature of a 2D AFHOTI is generally encoded in certain types of singularities in the 3D phase-band manifold. This yields a simple and intuitive interpretation: the singularities are essentially a ``footprint" of of the topological quantum criticality that separates an AFHOTI from a trivial phase connected to a static limit.

\begin{figure}[t]
	\includegraphics[width=0.49\textwidth]{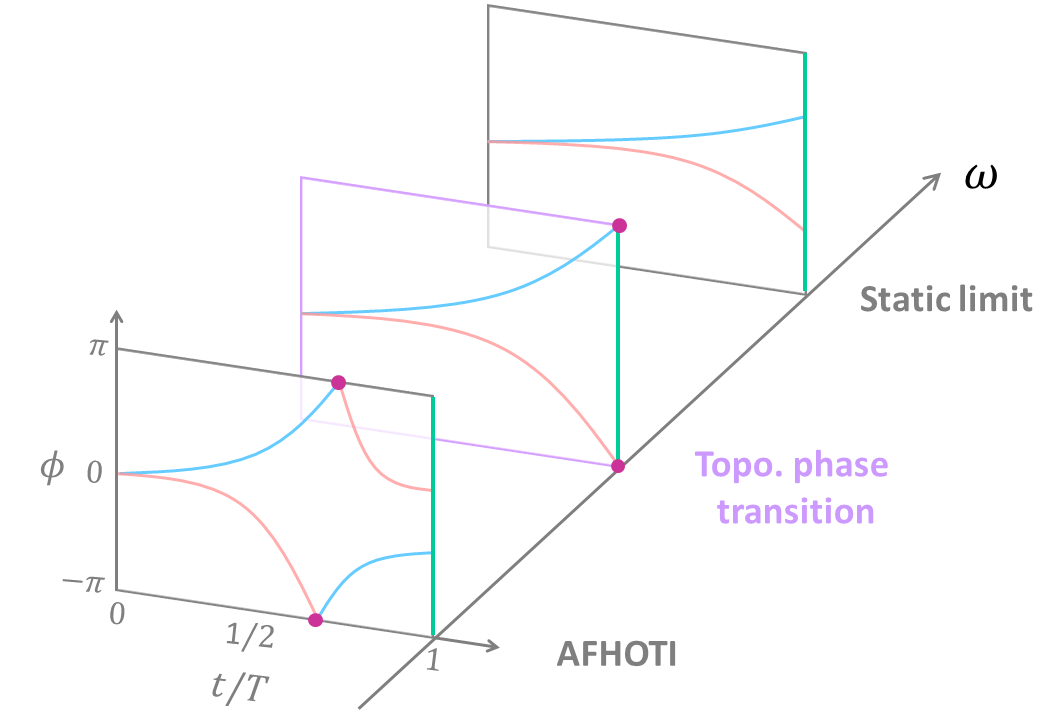}
	\caption{By gradually increasing the driving frequency $\omega$, an AFHOTI with phase-band singularity (the red dots) undergoes a topological phase transition with $\pi$-gap closing in the quasienergy spectrum, before turning into a static, singularity-free system. Note that the Floquet bands are the intersections between the phase band $\phi({\bf k}, t)$ and the green line at $t=T$. Therefore, at the topological transition point, the gapless Floquet bands around the $\pi$-gap is exactly the phase-band singularity in the original AFHOTI.}
	\label{Fig:phase_transition}
\end{figure}

To see this, let us first assume that an AFHOTI hosts a singularity in its phase bands $\phi({\bf k}, t)$, as shown in Fig.~\ref{Fig:phase_transition}. It is easy to see that by gradually increasing the driving frequency $\omega$ of this AFHOTI, we are effectively pushing the singularity away from $t=0$ along the time axis. In particular, at some critical frequency $\omega_c$ the singularity exactly sits at $t=T$ and closes the Floquet gap at quasienergy $\pi$. When $\omega>\omega_c$, the system remains singularity-free until it reaches the static limit with $\omega\rightarrow \infty$. In this case, all phases with $\omega>\omega_c$ (or $\omega>\omega_c$) are topologically and dynamically equivalent, since they are all connected to one another along the frequency path with no $\pi$-gap closing. As a result, the phase-band singularity at $\omega=\omega_c$ manifests itself as a topological phase transition between the AFHOTI and the static limit. Therefore, an AFHOTI, being statically impossible, must host phase-band singularities as an obstruction to the static limit. In other words, 
\begin{itemize}
	\item classifying the phase-band singularity is equivalent to classifying the quantum criticality that separates our target system from a static limit.
\end{itemize}

It should be emphasized that the presence of singularities is a {\it necessary yet insufficient} condition for AFHOTIs. In other words, not all $\pi$-gap-closing quantum criticalities indicate a transition from the static limit to an AFHOTI. Below, we will start from a systematic classification for phase-band singularities in general 2D class AIII Floquet systems with point-group symmetries, and leave discussions on their topological implications to Sec.~\ref{sec:preliminary_reduction}. \\

First of all, the phase-band singularities for 2D class AIII Floquet systems generically manifest themselves as collections of 3D Weyl nodes,\footnotemark[2]
whose locations and monopole charges are compatible with the crystalline symmetry group $G=C_n$ or $D_n$ as well as the chiral symmetry ${\cal S}$. In general, such a Weyl node can appear anywhere within the phase band Brillouin zone. However, for anomalous Floquet topology, we only need to focus on Weyl nodes across the \textit{principal zone boundary} at $\widetilde{\phi}=\pm \pi$, which completely determine the boundary modes at quasienergy $\pi$ and cannot be removed without closing the Floquet gap at $\pi$~\cite{nathan2015topo}. That all Floquet bands are topologically trivial requires the same number of boundary modes at quasienergy zero.
	Generically, the effective Hamiltonian of a 3D phase-band Weyl node can be written as:
	\begin{equation}
	h_{w} \equiv i\ {\rm log }\widetilde{U}({\bf k}, t)  \approx \pi \mathbb{1}_2 +  v_x k_x\sigma_x - v_y k_y\sigma_y + v_t t \sigma_z,
	\end{equation}
	where $\sigma_{x,y,z}$ are Pauli matrices, and the constant $\pi \mathbb{1}_2$ indicates that the Weyl node resides at the principal zone boundary.
\footnotetext[2]{In general, nodal loop singularities are also possible. However, such nodal loops are either contractible, in which case they can be shrunk to a point; or noncontractible (going around the entire Brillouin zone), which are generically unstable when translation symmetry is absent. Therefore, in this work, it suffices to consider point-like singularities (i.e. Weyl/Dirac nodes in 3D). An example of a contractible nodal loop is also considered in Sec.~\ref{sec:Dn}.} 

\begin{figure*}[t]
	\includegraphics[width=0.95\textwidth]{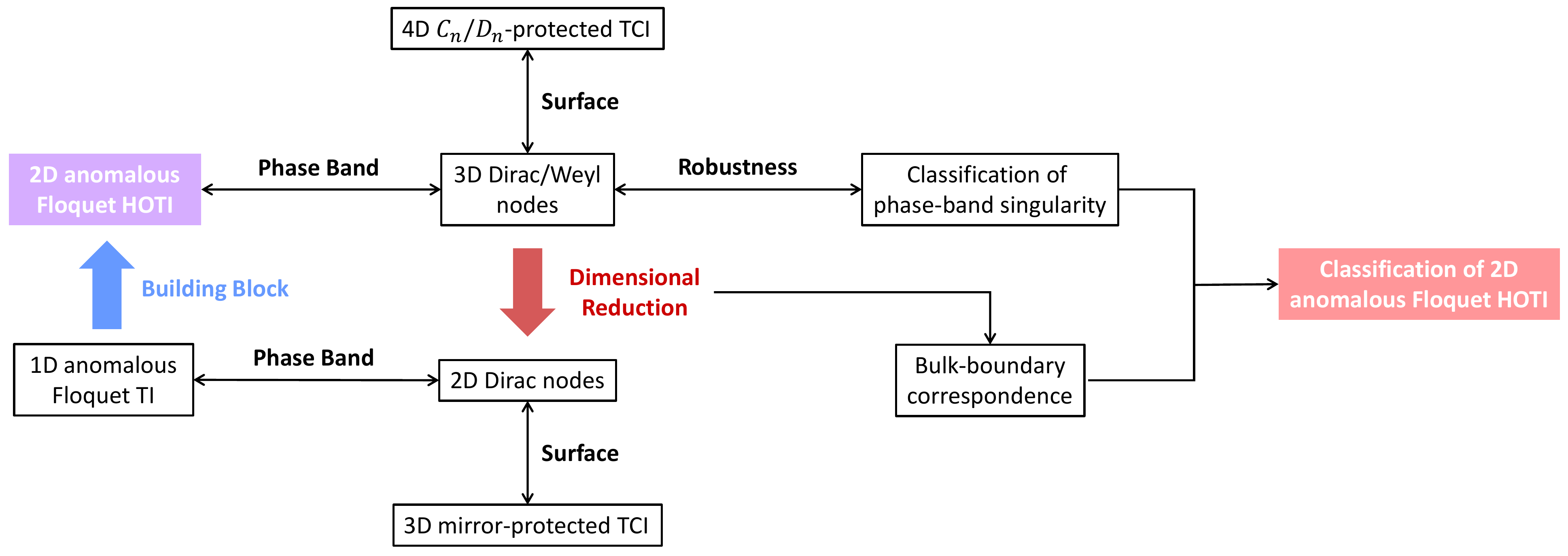}
	\caption{Logic flow of our classification scheme for 2D anomalous Floquet higher-order topology.}
	\label{Fig: Logic}
\end{figure*} 

To gain some intuitions of the symmetry constraints imposed on the phase-band Weyl node, we first note that chiral symmetry ${\cal S}$ acts as a {\it unitary time-reflection symmetry} on the return map:
\begin{equation}
\mathcal{S} \widetilde{U}({\bf k}, t) \mathcal{S}^{-1} = \widetilde{U}({\bf k}, T-t),
\label{eq:chiral}
\end{equation}
Thus, starting with a Weyl node carrying a Weyl charge $q$ at $({\bf k}, t)$, ${\cal S}$ will map it to another Weyl node with charge $-q$ at $({\bf k}, T-t)$. As a result, the phase-band Weyl nodes always come in chiral-related pairs, which guarantees a vanishing net Weyl charge and thus the absence of chiral edge modes. We dub such a chiral-related Weyl pair a {\it dynamical Weyl pair} (DWP). 

Naively, without any additional symmetry being imposed, such a pair of Weyl nodes can annihilate one another when brought together and hence is unstable. Nevertheless, when there is an additional $n$-fold rotation symmetry $C_n$ anticommuting with ${\cal S}$, a novel mechanism can be activated to prevent pair annihilation of certain DWPs on the rotation-invariant axes. Specifically, an on-axis Weyl node can be effectively viewed as a pair of counter-propagating 1D modes dispersing along $t$-direction. Then the right-moving and left-moving modes must transform as irreducible representations under $C_n$, which can thus be labeled by the $z$-component angular momentum $j_z=J$ and $j_z=J+1$, respectively. We note that bands with distinct $j_z$ values cannot hybridize to anticross on a rotation axis. This inspires us to define an {\it irrep-dependent Weyl charge} $q_J$ for an on-axis Weyl node. While the value of $q_J$ still characterizes the monopole charge, the additional subscript $J$ originates from the quantum number $j_z$ carried by the bands and thus provides the necessary symmetry information. In particular, two on-axis Weyl nodes carrying $q_{J_1}$ and $q_{J_2}$ can annihilate each other only if both $q_{J_1}=-q_{J_2}$ and $J_1=J_2$ are satisfied. 

As we show in Sec.~\ref{sec:Cn}, when $[C_n, {\cal S}]=0$, the chiral-related Weyl pair always carries opposite values of Weyl charges with the same $J$-label: $q_J$ and $-q_J$, thus they can always be trivialized as a whole. When $\{C_n, {\cal S}\}=0$, we find that such a DWP carries opposite Weyl charges with distinct $J$-labels: $q_{J}=-q_{J+\frac{n}{2}}$, and hence is stable by itself.
Notably, this implies that robust DWPs do not exist for $C_3$-symmetric systems, as $\frac{n}{2}$ must be an integer. The net topological singularities in the phase band is characterized by a set of  $C_n$ topological charges $\{Q_J^{(n)}\}$, which counts the net Weyl charge labeled by $J$ for all $C_n$-invariant axes. Further taking into account constraints from chiral symmetry, we find $\frac{n}{2}$ independent $C_n$ topological charges, leading to a $\mathbb{Z}^{\frac{n}{2}}$ classification for the $C_n$-protected topological singularities, as shown in Table~\ref{Table1}.

A dihedral group $D_n$ is generated by both a $C_n$ rotation and an additional mirror symmetry ${\cal M}$. Thus, the rotation charges $Q_J^{(n)}$ remain well-defined for dihedral groups. 
However, mirror symmetry imposes additional constraints on the rotation charges that further reduce the number of independent $Q_J^{(n)}$'s for each $n$.
Moreover, we demonstrate in Sec.~\ref{sec:Dn} that a mirror symmetry ${\cal M}$ can protect robust singularities by itself if $[{\cal M, S}]=0$, which defines a $\mathbb{Z}$-type mirror topological charge $Q_{\cal M}$. Such a mirror-protected singularity instead takes the form of a 3D fourfold-degenerate Dirac node living on the 1D intersection between the ${\cal S}$-invariant plane and the ${\cal M}$-invariant plane. For example, a ${\cal M}_x$-protected Dirac node can move freely along the $k_y$ axis at either $(k_x, t)=(0,\frac{T}{2})$ or $(k_x, t)=(\pi,\frac{T}{2})$. Notably, such a ${\cal M}$-protected Dirac singularity can be deformed into an exotic nodal loop upon symmetric perturbations, as will be discussed in details in Sec.~\ref{sec:Dn}.

Since the dihedral groups contain both rotation and mirrory symmetries, it is possible that both $Q_J^{(n)}$ and $Q_{\cal M}$ are nontrivial in $D_n$-symmetric systems. For example, it is actually rather common that a phase-band singularity simultaneously carries both rotation and mirror charges. Moreover, when multiple mirror planes are present, a distinct mirror charge $Q_{\cal M'}$ can be defined for each set of inequivalent mirror plane ${\cal M}'$, as long as it commutes with chiral symmetry.
However, not all these topological charges are independent. We find that the total number of independent topological charges sensitively depends on the commutation relations among rotation, mirror, and chiral symmetries. After carefully disentangling the complex relations among various topological charges, we achieve a complete classification of $D_n$-protected phase-band singularities, which is elaborated in Sec. \ref{sec:Dn} and summarized in Table \ref{Table1}.

\subsection{Connection to 4D topological crystalline insulators}

\label{subsection:4D TCI}

We point out that a $C_n$ or $D_n$-protected phase-band topological singularity necessarily features anomalous band connectivity that does not exist in any electronic band structure of 3D static lattice systems. However, such phase-band singularities do resemble exotic surface state physics of 4D TCIs. 

For example, we show in Sec.~\ref{sec:Cn} that a $C_2$-protected DWP exhibits a $C_2$-velocity locking effect. Namely, both right (left) movers along the $t$-axis carry the same $C_2$ eigenvalues.
This obviously violates the energy band connectivity in a static lattice system, where the number of right movers and left movers carrying the same $C_2$ eigenvalue must be the same, due to the periodicity of the Brillouin zone. Such a locking between the velocity and some good quantum number can only occur on the boundary of a static topological system living in one higher dimension, and a well-known example is the spin-momentum locking for the helical edge states circulating around a quantum spin Hall insulator \cite{kane2005z2,bernevig2006QSH}. Indeed, we show explicitly in Appendix~\ref{app:4dTCI} that the surface states of a 4D $C_2$-protected TCI are 3D Weyl nodes exhibiting the same $C_2$-velocity locking effect~\cite{zhang2016topological}.

In fact, the intrinsic relation between a phase-band topological singularity and a 4D TCI surface state is not accidental. Note that if a phase-band singularity can be realized on a static lattice, the fermion doubling theorem always requires the existence of another singularity with exactly the opposite topological characteristics. Then such a pair of singularities can always be eliminated as a whole, while preserving both the symmetries and the bulk Floquet gaps. Therefore, a phase-band singularity {\it should violate the fermion doubling theorem} to guarantee its robustness. Such a violation is possible for the phase band since it does not correspond to the energy spectrum of a local Hamiltonian, and the phases $\widetilde{\phi}({\bf k}, t)$ are periodic in $2\pi$.
Alternatively, in a static system, one can go around the fermion doubling theorem by considering the 3D boundary of a topological system living in four dimensions.


\begin{table}
	\begin{tabular}{c  c  c  c  c  c}  
		\hline \hline
		Rotation Group & $C_1$ & $C_2$ & $C_3$ & $C_4$ & $C_6$ \T\B  \\    \hline      
		$C_{n,-}^+$   & 0 & 0 & 0 & 0 & 0  \T \B \\      
		$C_{n,-}^-$   & 0 & 0 & 0 & 0 & 0  \T \B \\
		$C_{n,+}^+$   & 0 & $\mathbb{Z}$ & 0 & $\mathbb{Z}^2$ & $\mathbb{Z}^3$  \T \B \\      
		$C_{n,+}^-$   & 0 & $\mathbb{Z}$ & 0 & $\mathbb{Z}^2$ & $\mathbb{Z}^3$  \T \B \\  \hline \hline
		Dihedral Group & $D_1$ & $D_2$ & $D_3$ & $D_4$ & $D_6$ \T\B  \\    \hline  
		$D_{n,-,-}^+$   & $\mathbb{Z}$ & 0 & $\mathbb{Z}$ & 0 & 0  \T \B \\  
		$D_{n,-,+}^+$   & 0 & 0 & 0 & 0 & 0  \T \B \\  
		$D_{n,+,-}^+$   & / & $\mathbb{Z}\times 2\mathbb{Z}$ & / & $\mathbb{Z}$ & $\mathbb{Z}^2\times 2\mathbb{Z}$  \T \B \\  
		$D_{n,+,+}^+$   & / & $\mathbb{Z}\times 2\mathbb{Z}$ & / & $\mathbb{Z}$ & $\mathbb{Z}^2\times 2\mathbb{Z}$  \T \B \\  
		$D_{n,-,-}^-$   & $\mathbb{Z}$ & $\mathbb{Z}\times 2\mathbb{Z}$ & $\mathbb{Z}$ & $\mathbb{Z}\times 2\mathbb{Z}$ & $\mathbb{Z}\times 2\mathbb{Z}$  \T \B \\  
		$D_{n,-,+}^-$   & 0 & 0 & 0 & 0 & 0  \T \B \\  
		$D_{n,+,-}^-$   & / & 0 & / & $\mathbb{Z}\times 2\mathbb{Z}$ & $\mathbb{Z}$  \T \B \\  
		$D_{n,+,+}^-$   & / & 0 & / & $\mathbb{Z}\times 2\mathbb{Z}$ & $\mathbb{Z}$  \T \B \\  
		\hline \hline
	\end{tabular}
	\caption{Classification of phase-band topological singularities protected by 2D point groups. We label different classes of $C_n$ and $D_n$ groups as $C_{n,\eta_{C_n}}^s$ and $D_{n,\eta_{C_n},\eta_{M_x}}^s$, respectively. $s=+(-)$ when $(C_n)^n = +1 (-1)$, i.e. single versus double group; $\eta_{C_n}=-(+)$ when $C_n$ commutes (anticommutes) with chiral symmetry; $\eta_{M_x}=-(+)$ when $M_x$ commutes (anticommutes) with chiral symmetry.
	}
	\label{Table1}
\end{table}

\begin{table}
	\begin{tabular}{ c c c c c c }  
		\hline \hline
		Rotation Group & $C_1$ & $C_2$ & $C_3$ & $C_4$ & $C_6$ \T\B  \\    \hline      
		$C_{n,-}^+$   & 0 & 0 & 0 & 0 & 0  \T \B \\      
		$C_{n,-}^-$   & 0 & 0 & 0 & 0 & 0  \T \B \\
		$C_{n,+}^+$   & 0 & $\mathbb{Z}_2$ & 0 & $\mathbb{Z}_2$ & $\mathbb{Z}_2$  \T \B \\      
		$C_{n,+}^-$   & 0 & $\mathbb{Z}_2$ & 0 & $\mathbb{Z}_2$ & $\mathbb{Z}_2$  \T \B \\  \hline \hline
		Dihedral Group & $D_1$ & $D_2$ & $D_3$ & $D_4$ & $D_6$ \T\B  \\    \hline  
		$D_{n,-,-}^+$   & $\mathbb{Z}$ & 0 & $\mathbb{Z}$ & 0 & 0  \T \B \\  
		$D_{n,-,+}^+$   & 0 & 0 & 0 & 0 & 0  \T \B \\  
		$D_{n,+,-}^+$   & / & $\mathbb{Z}$ & / & 0 & $\mathbb{Z}$  \T \B \\  
		$D_{n,+,+}^+$   & / & $\mathbb{Z}$ & / & 0 & $\mathbb{Z}$  \T \B \\  
		$D_{n,-,-}^-$   & $\mathbb{Z}$ & $\mathbb{Z}\times 2\mathbb{Z}$ & $\mathbb{Z}$ & $\mathbb{Z}\times 2\mathbb{Z}$ & $\mathbb{Z}\times 2\mathbb{Z}$  \T \B \\  
		$D_{n,-,+}^-$   & 0 & 0 & 0 & 0 & 0  \T \B \\  
		$D_{n,+,-}^-$   & / & 0 & / & $\mathbb{Z}$ & 0  \T \B \\  
		$D_{n,+,+}^-$   & / & 0 & / & $\mathbb{Z}$ & 0  \T \B \\
		\hline \hline
	\end{tabular}
	\caption{Classification of anomalous Floquet higher-order topology protected by 2D point groups. The definitions of different classes labeled as $C_{n,\eta_{C_n}}^s$ and $D_{n,\eta_{C_n},\eta_{M_x}}^s$ are the same as Table~\ref{Table1}.}
	\label{Table2}
\end{table}

\subsection{From phase-band singularities to higher-order topology}
\label{sec:preliminary_reduction}

As we have mentioned, the anomalous chiral edge mode of an AFTI in class A is indicated by the bulk Weyl singularities in its return map operator $\widetilde{U}({\bf k}, t)$. This is simply because the net phase-band Weyl charge exactly equals to the topological winding number for $\widetilde{U}$, which counts the total number of anomalous chiral edge modes~\cite{rudner2013anomalous}.
Therefore, it is natural to expect that the $C_n$ or $D_n$-protected phase-band singularities dictate the number of robust corner modes in AFHOTIs. However, lacking a topological invariant characterization of AFHOTIs as well as a higher-order bulk-boundary correspondence based on which, new approaches are needed to relate the phase-band singularities to higher-order topology.

\begin{figure}[t]
	\includegraphics[width=0.49\textwidth]{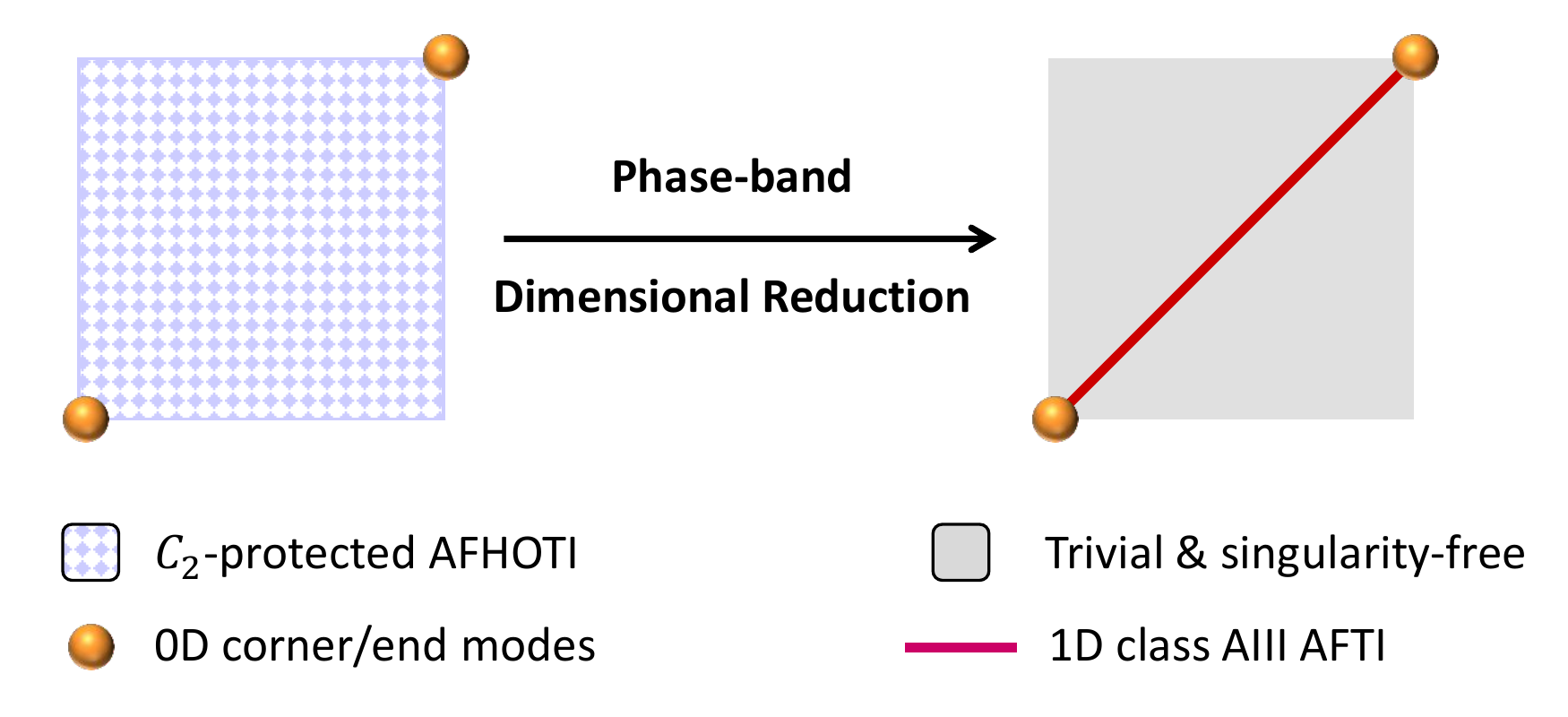}
	\caption{An example of dimensional reduction applied to $C_2$-protected AFHOTI. The dimensional reduction procedure adiabatically connects a 2D $C_2$-protected AFHOTI  harboring corner modes with a 1D AFTI in class AIII harboring end modes. }
	\label{Fig:building block}
\end{figure}

In Sec.~\ref{sec:reduction}, we establish such a higher-order bulk-boundary correspondence and identify the corresponding higher-order topological indices for all $C_n$ and $D_n$-protected AFHOTIs, by developing a {\it phase-band dimensional reduction} (PBDR) method. Such a method was inspired by the dimensional reduction approach that has been applied to classifying TCIs by reducing the problem to classifying TIs in lower dimensions \cite{song2017topological,isobe2015theory}. Similarly, our PBDR approach reduces the problem of classifying 2D AFHOTIs to that of 1D AFTIs in class AIII (see Fig.~\ref{Fig:building block}), for which the classification and bulk-boundary correspondence have been fully established.
	While 2D AFHOTIs host corner localized $0$ and $\pi$ modes, a 1D AFTI with chiral symmetry hosts 0 and $\pi$ modes localized at its endpoints. Such end modes in 1D class AIII AFTIs can be classified by the chiral winding number defined for $\widetilde{U}$ [see Eq.~(\ref{eq:winding})], which we prove is equivalent to the phase-band singularity characterization [see Eq.~(\ref{eq:top_dirac})]. Different from the 3D Weyl/Dirac singularities in 2D AFHOTIs, the phase-band singularities for 1D AFTIs are 2D Dirac nodes in $(k,t)$ space.
	
More precisely, we divide our target $C_n$/$D_n$-symmetric system into symmetry-related patches, as well as a ``skeleton" sandwiched in between. In Fig.~\ref{Fig:building block}, we give a schematic example for an AFHOTI with $C_2$ symmetry. The singularities living on the patches can be adiabatically eliminated by coupling them to a mass term for the phase band. To preserve the crystalline symmetries, such a mass term cannot be spatially uniform, but rather takes the form of a set of symmetry-preserving mass domain walls, which exactly sits on top of the skeleton to respect symmetries. This necessarily leads to chiral-protected 2D Dirac modes localized around each domain wall, which are exactly the defining singularities for 1D AFTIs. Since the singularity-free patches are topologically trivial for our purpose, we have thereby mapped the original 2D system to a collection of symmetry-related 1D AFTIs, completing the PBDR procedure.



After the PBDR, some end modes of the resulting 1D AFTIs may be able to gap out one another and get trivialized. The survivors, however, correspond to robust corner modes in the original 2D system. This allows us to establish the higher-order bulk-boundary correspondence for 2D AFHOTIs protected by point group symmetries. After a careful analysis, we conclude that
\begin{itemize}
	\item a 2D $C_n$-protected AFHOTI is characterized by a $\mathbb{Z}_2$ topological index $\nu_n$, where
	\begin{equation}
	\nu_n \equiv \sum_J |Q_J^{(n)}| \ \ \ \text{mod 2};
	\end{equation}
	\item a 2D $D_n$-protected AFHOTI is characterized by the set of inequivalent mirror topological charges $\{Q_{\cal M}\}$, which admits a higher-order topological classification of $\mathbb{Z}$ when there is only one nontrivial mirror charge, or $\mathbb{Z}\times 2\mathbb{Z}$ when there are two inequivalent mirror charges.
\end{itemize}

We now briefly summarize the boundary features of AFHOTIs indicated by the above classification. First of all, $D_n$-protected corner modes can only live on mirror-invariant corners of the system.
In particular, AFHOTIs with a $\mathbb{Z}$ classification is characterized by only one nontrivial mirror charge $Q_{M_x}$. The value of $|Q_{M_x}|$ corresponds to the number of corner modes on each $M_x$-preserving corner. Notably,  systems with $Q_{M_x}=N$ and $Q_{M_x}=-N$ ($N\in\mathbb{Z}$) are topologically distinct from one another, since their corner modes carry different mirror and chiral eigenvalues.

Secondly, a $\mathbb{Z}\times 2\mathbb{Z}$ classification occurs when a $D_n$-symmetric system has two inequivalent mirror topological charges $Q_M$ and $Q_{M'}$. We find that in this case they must obey: $Q_M = Q_{M'}$ (mod 2). Therefore, the number of corner modes at $M$-invariant corners must equal to that at $M'$-invariant corners, modulo 2.

In contrast, $C_n$-protected corner modes can exist on an arbitrary collection of $C_n$-related corners. However, the $\mathbb{Z}_2$ classification indicates that only an odd number of corner modes are topologically stable in this case.

In Table \ref{Table2}, we summarize our topological classification of 2D AFHOTIs protected by point group symmetries. This classification, along with the associated higher-order topological indices and the bulk-boundary correspondence, constitutes the key result of this work.

\section{Revisiting 1D anomalous Floquet topology}
\label{sec:ssh}

As a warm up, we start by revisiting the theory of 1D anomalous Floquet topological insulator (AFTI) in class AIII, the defining boundary feature of which is the robust end modes pinned at both quasienergies $0$ and $\pi/T$. 
Although such systems are not higher-order topological, they demonstrate in a simple way the general idea which carries along to 2D AFHOTIs, that their anomalous Floquet topology is indicated by phase-band singularities. Moreover, 1D AFTIs provide an elementary building block for establishing the higher-order bulk-boundary correspondence in 2D AFHOTIs via phase-band dimensional reduction (see Sec.~\ref{sec:reduction}).

1D AFTIs in class AIII are known to be characterized by a topological invariant, namely, the chiral winding number \cite{fruchart2016complex}. Below we shall first recast this winding number in terms of phase-band singularities (in this case 2D Dirac points), which clarifies the physical meaning of the topological invariant from a new perspective. As we have mentioned earlier, such a 2D Dirac singularity, when appearing at $t=T$, is precisely the topological critical point between the AFTI phase and the trivial phase. We also illustrate our singularity-based topological diagnostic using a simple concrete model, which is a quantum walk analog of the Su-Schrieffer-Heeger model.

\subsection{Winding number and phase-band singularity}

Let us first introduce the winding number invariant for characterizing class AIII AFTIs in one spatial dimension. Recall from Eq.~(\ref{eq:chiral}) that chiral symmetry acts as a time-reflection symmetry on the return map $\widetilde{U}(k, t)$. This leads to
\begin{equation}
\left[\mathcal{S}, \  \widetilde{U}\left(k, \frac{T}{2}\right) \right]=0,
\end{equation}
which further implies that the return map $\widetilde{U}(k,t)$ at the chiral-invariant point $t_0=\frac{T}{2}$ block diagonalizes according to the chiral eigenvalues $\mathcal{S}=\pm 1$. The topological invariant is thus given by the difference of the winding numbers of the $\mathcal{S}=\pm 1$ blocks of the return map $\widetilde{U}(k, \frac{T}{2})$:
\begin{equation}
{\cal N_S} = \int_{-\pi}^{\pi} \frac{d k}{4\pi} \ {\rm Tr} \left[ \mathcal{S} \widetilde{U}^\dagger\left(k, \frac{T}{2}\right) i \partial_k \widetilde{U}\left(k, \frac{T}{2} \right) \right],
\label{eq:winding}
\end{equation}
where the trace is taken over all eigenstates of $\widetilde{U}$. The chiral winding number ${\cal N_S}$ also classifies the number of edge-localized modes at both quasienergies $0$ and $\pi$ via bulk-boundary correspondence.

Written explicitly in the eigenbasis of $\tilde{U}(k, T/2)$ with phase-band eigenvalues $\{\widetilde{\phi}_n(k, T/2)\}$, the chiral winding number in Eq.~(\ref{eq:winding}) takes the following form:
\begin{eqnarray}
	{\cal N_S} &=& \frac{{\cal W}_+-{\cal W}_-}{2},
	\label{eq:winding2}
	\end{eqnarray}
	where for each chiral sector we have
	\begin{equation}
	{\cal W}_{\pm} = \int_{-\pi}^{\pi} \frac{dk}{2\pi} \sum_{n_\pm} \frac{\partial \widetilde{\phi}_{n_\pm}(k, T/2)}{\partial k}.
	\end{equation}
	Here $n_{\pm}$ labels phase-band eigenstates with chiral eigenvalues $\pm 1$, respectively. Physically, ${\cal W}_{\pm}$ indicates the winding of phase bands at $t=T/2$ around the principle zone for each chiral sector ${\cal S}=\pm$. As pointed out in Ref. \cite{fruchart2016complex}, there exists a compatibility condition that forces the net winding number to vanish:
	\begin{equation}
		{\cal W}_++{\cal W}_-=0.
		\label{eq:chiral compatibility}
	\end{equation}
	We therefore arrive at
	\begin{equation}
	{\cal N_S} = {\cal W}_+.
	\end{equation}
Apparently, a non-zero winding number implies the existence of irremovable phase-band singularities across the principal zone boundary. 
In 2D $({\bf k}, t)$ space, such a zone-boundary singularity is essentially a {\it 2D Dirac point} living on the chiral-invariant line (i.e. $t=\frac{T}{2}$).

\begin{figure}[t]
	\includegraphics[width=0.48\textwidth]{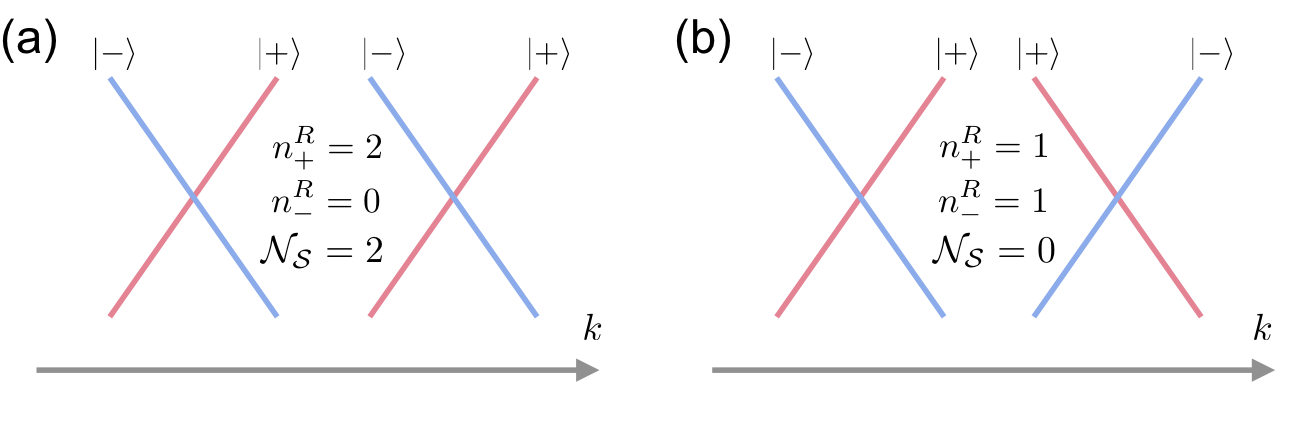}
	\caption{A pair of 2D Dirac nodes on the chiral symmetric line $t_0=\frac{T}{2}$. The chiral eigenvalue of each band is labeled as $|\pm \rangle$. (a) The two right (left) movers have the same chiral eigenvalue (“chiral-velocity locking”), hence the two Dirac nodes are stable. The corresponding $\mathcal{N}_{\mathcal S}=2$. (b) The two right (left) movers have opposite chiral eigenvalues, hence the two Dirac nodes can be gapped out. The corresponding $\mathcal{N}_{\mathcal S}=0$.}
	\label{fig:mirror_dirac}
\end{figure} 

The robustness of a single ${\cal N_S}$-indicated 2D Dirac point is guaranteed by chiral symmetry ${\cal S}$. Since ${\cal W}_+=-{\cal W}_-$, a 2D Dirac point with ${\cal N_S}=1$ is by definition a phase-band crossing between a right-moving channel with ${\cal S}=1$ and a left-moving channel with ${\cal S}=-1$, along the chiral-invariant axis. It is then impossible to couple and gap out these two counterpropagating branches without violating the chiral symmetry.

The stability of multiple Dirac points is then determined by the chiral winding pattern of each singularity, since two counterpropagating modes along $k$ can only gap out each other when carrying the same chiral index. This inspires us to define $n^R_\pm$ as the number of right-movers carrying a chiral index ${\cal S}=\pm$, which will contribute positively (negatively) to the chiral winding number ${\cal N_S}$, respectively. We can then rewrite ${\cal N_S}$ in a physically more intuitive way
\begin{equation}
	\mathcal{N}_{\cal S} = n_+^R - n_-^R \in \mathbb{Z},
	\label{eq:top_dirac}
\end{equation}	
which exactly indicates the net number of robust Dirac points on the zone boundary. For example, when the right-moving modes (along $k$) of two Dirac points are both within the ``$+$" chiral sector, such a pair of Dirac points carry ${\cal N_S}=2$ and cannot annihilate each other without breaking ${\cal S}$ [see Fig.~\ref{fig:mirror_dirac}(a)]. Nevertheless, two Dirac points with opposite chiral winding patterns (i.e. ${\cal N_S}=0$) are always unstable, as shown in Fig.~\ref{fig:mirror_dirac}(b).

\begin{figure*}[t]
	\includegraphics[width=0.75\textwidth]{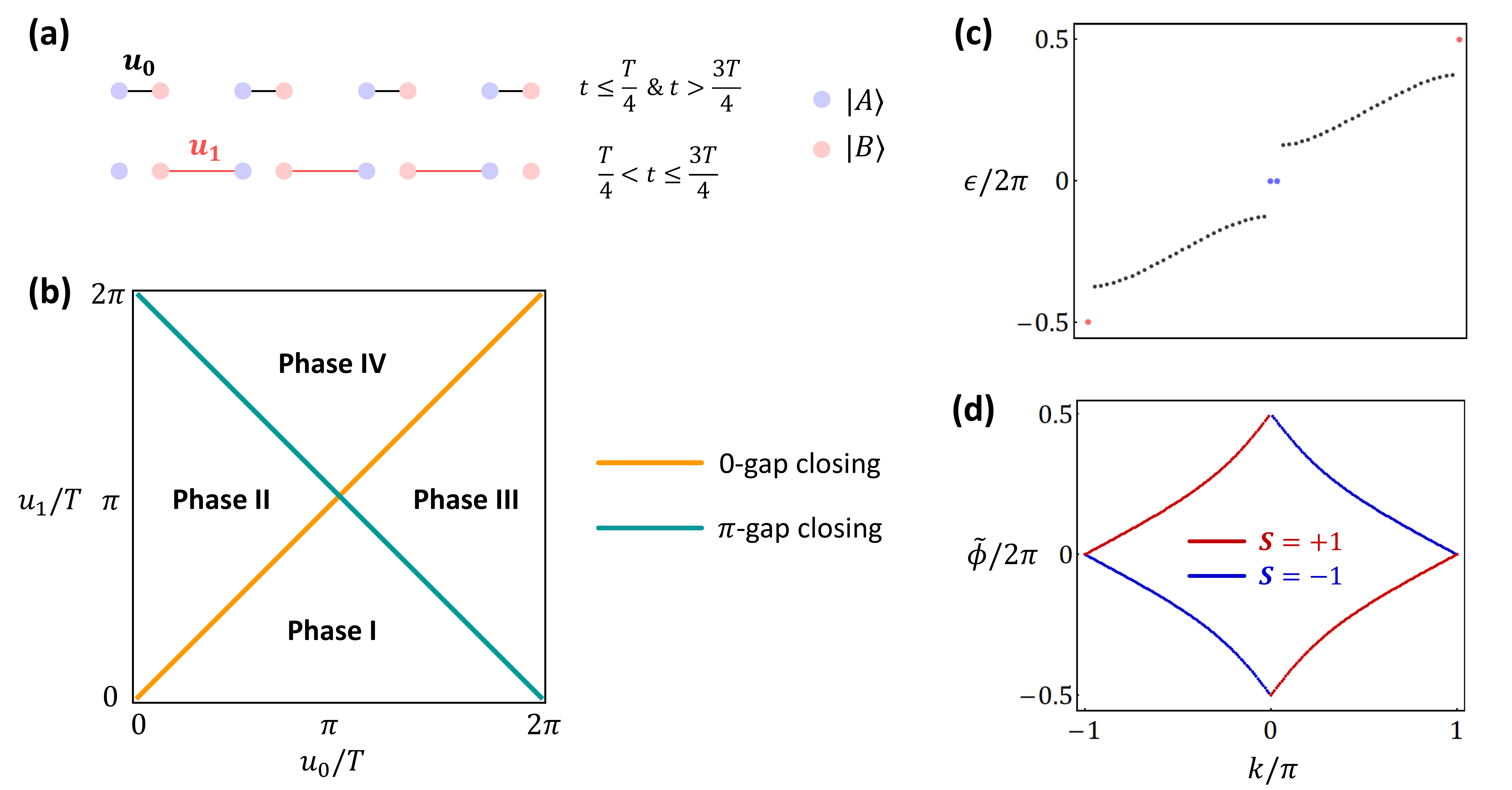}
	\caption{(a) Schematics of the intra-cell $u_0$ and inter-cell $u_1$ couplings for the Hamiltonian $H_{1D}$ in Eq.~(\ref{eq:ssh}). Each unit cell contains two sublattices labeled by $A$ and $B$. (b) Phase diagram of $H_{1D}$. (c) In an open chain geometry, the Floquet spectrum for the phase IV of $H_{1D}$ displays both $0$ and $\pi$ modes at each end of the chain, confirming the system as a 1D AFTI. We have chosen $u_0=\pi/T$ and $u_1=3\pi/(2T)$. (d) Phase bands of the AFTI phase at $t=\frac{T}{2}$ show a nontrivial winding pattern along $k$, confirming the nontrivial chiral winding number ${\cal N_S}=1$. This agrees with the end-mode physics shown in (c).}
	\label{Fig: SSH}
\end{figure*}

One may wonder whether Dirac nodes are still robust singularities if they live away from the chiral-invariant line. We note that for a Dirac node at $(k_0, t_0)$ with $t_0\neq \frac{T}{2}$, ${\cal S}$ enforces the existence of another Dirac point at $(k_0, T-t_0)$. When bringing such a chiral-related Dirac pair to meet at $t_0=\frac{T}{2}$, it is easy to show that they necessarily carry ${\cal N_S}=0$ as a whole. To be specific, since ${\cal S}$ switches between the two right-moving modes of the Dirac points, we can always superpose the two channels to form a bonding (symmetric) state (with ${\cal S}=1$) and an anti-bonding (anti-symmetric) state (with ${\cal S}=-1$), leading to a trivial chiral winding pattern. Therefore,the ``off-line" Dirac nodes can always annihilate with their chiral partners and are thus not robust topological singularities.

Finally, we emphasize on the \textit{chiral-velocity locking} effect of the ${\cal N_S}$-indicated Dirac nodes, where all right (left) movers carry the same chiral eigenvalue, as shown in Fig.~\ref{fig:mirror_dirac}(a). This is similar to the well-known spin-momentum locking effect on the edge of a 2D quantum spin Hall insulator~\cite{hasan2010colloquium,qi2011topological}, implying such Dirac nodes cannot be realized in a 2D static lattice system. However, since ${\cal S}$ behaves as a mirror symmetry for time, the phase-band Dirac nodes are exactly the same as those on the surface of a 3D mirror-protected topological crystalline insulator~\cite{hsieh2012topological}, if we view $t$ as a crystal momentum orthogonal to $k_x$. Such a correspondence between a phase-band topological singularity and the boundary mode of a higher-dimensional topological state is a {\it generic feature for anomalous Floquet topological phases}, as we will show in Sec. \ref{sec:Cn} and \ref{sec:Dn}.

\subsection{Model}
\label{subsec:SSH model}

We shall now give a concrete example to illustrate our general discussions above. Consider a 1D chain subject to the following time-periodic drive:
\begin{widetext}
\begin{equation}
H_{1D}(k,t) =
\begin{cases}
u_0\sum_{{\bf r}} c^{\dagger}_{{\bf r}, A} c_{{\bf r}, B} + \text{h.c.}   &   0<t\leq \frac{T}{4} \quad \text{and} \quad \frac{3T}{4}<t\leq T; \\
\\
u_1\sum_{{\bf r}} c^{\dagger}_{{\bf r + a}_x, A} c_{{\bf r}, B} + \rm{h.c.}  &  \frac{T}{4} < t \leq \frac{3T}{4},
\end{cases}
\label{eq:ssh}
\end{equation}
\end{widetext}
where $A, B$ denote the two sublattices within a unit cell, as shown in Fig.~\ref{Fig: SSH} (a). Under the basis $(|A\rangle, |B\rangle)^T$, chiral symmetry is implemented by $\mathcal{S}=\sigma_z$, and Hamiltonian~(\ref{eq:ssh}) is invariant under chiral symmetry. Notice that this model is in fact a Floquet version of the well-known Su-Schrieffer-Heeger model~\cite{su1979soliton,su1980soliton,liu2018chiral}, where the intra-cell and inter-cell hoppings, i.e. $u_0$ and $u_1$, are periodically switched on and off. 

The phase diagram of Hamiltonian~(\ref{eq:ssh}) is analytically tractable by solving for the vanishing conditions of Floquet bulk gap at quasienergies $0$ and $\pi$. We find that the phase boundaries are determined by:
\begin{equation}
u_0 \pm u_1 = \frac{2m\pi}{T}, \quad  m \in \mathbb{Z},
\end{equation}
which separate four topologically distinct phases, as shown in Fig.~\ref{Fig: SSH}(b). By numerically calculating the Floquet quasienergy spectrum on a finite-size open chain, we find that the four distinct phases can be characterized by the presence of $0$ or $\pi$ end modes: (i) phase I has no boundary modes and is thus topologically trivial; (ii) phase II has robust $0$ end mode; (iii) phase III has robust $\pi$ end mode; (iv) phase IV has both $0$ and $\pi$ end modes [see Fig.~\ref{Fig: SSH}(c)], and thus realizes our expected AFTI phase.

To calculate ${\cal N_S}$ in Eq.~(\ref{eq:top_dirac}), we plot the phase band dispersion at $t=\frac{T}{2}$ for phase IV. As shown in Fig. \ref{Fig: SSH} (d), the phase band manifold does host a topological singularity crossing the principle-zone boundary. In particular, this singularity features our expected chiral-velocity locking effect. Namely, the ${\cal S}=+1$ ($-1$) branch serves as the right (left) mover along the $k$ axis, which directly leads to ${\cal N_S}=1$. The nontrivial value of the chiral winding number thus explains the existence of end modes in Fig.~\ref{Fig: SSH}(c).

\section{$C_n$-protected phase-band singularity}
\label{sec:Cn}
Armed with our discussions of 1D driven class AIII systems, we now proceed to consider phase-band singularities in 2D class AIII Floquet systems that are invariant under a $n$-fold rotation symmetry $C_n$. Since the generalized Brillouin zone $({\bf k}, t)$ for phase bands is three dimensional, it is natural to consider dynamical Weyl nodes as the elementary building block for understanding singular behaviors in the time-evolution process. As we have pointed out in Sec.~\ref{sec:summary_singularity}, the chiral symmetry ${\cal S}$, a unitary time-reflection operation, guarantees that Weyl nodes must come in pairs (i.e. dynamical Weyl pairs or DWPs for short), leading to a vanishing net Weyl charge. This is in contrast with the 2D dynamical Dirac points discussed in Sec.~\ref{sec:ssh}, where there can be an odd number of Dirac points in total. The DWPs classified in this section are inherently the quantum criticality separating a $C_n$-protected dynamical phase from a static limit. 

Let us start with a dynamical Weyl node at $({\bf k}_0, t_0)$. If $C_n{\bf k}_0\neq {\bf k}_0$ (i.e. the Weyl node living off an rotation-invariant axis), there will be $(n-1)$ additional Weyl points that are rotation-related. Since there is no symmetry protection, such a group of Weyl points along with their chiral-related partners can always be brought together and adiabatically eliminated as a whole. Therefore, it is sufficient to only consider Weyl points on the $t$-axis intersecting the rotation-invariant momentum ${\bf k}_0$.

\begin{figure*}[t]
	\includegraphics[width=0.7\textwidth]{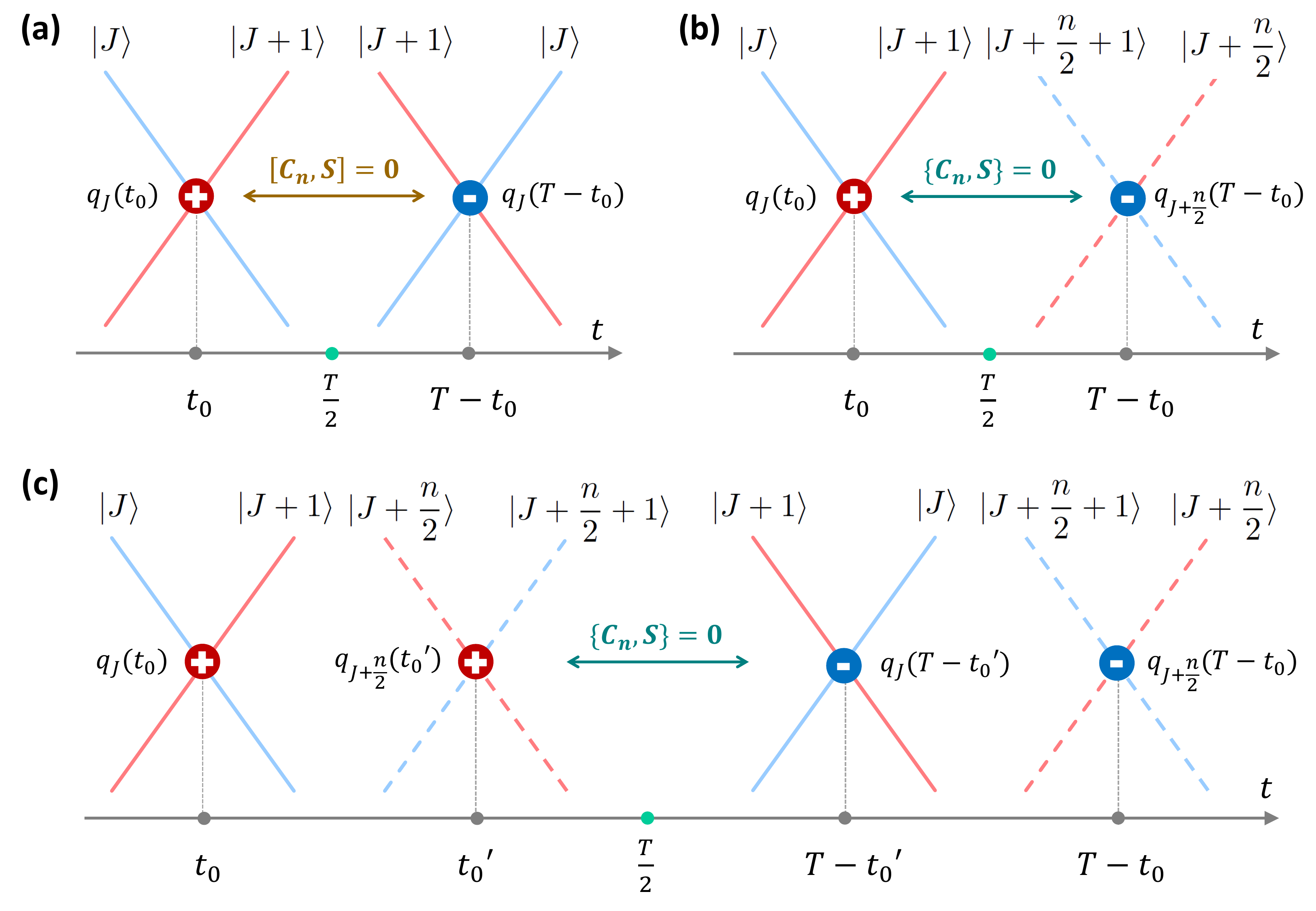}
	\caption{(a) Pair annihilation of dynamical Weyl nodes when $[C_n,{\cal S}]=0$. (b) Stable dynamical Weyl pair when $\{C_n, {\cal S} \}=0$. (c) Annihilation of two DWPs when $\{C_n, {\cal S} \}=0$.}
	\label{Fig2}
\end{figure*} 

To describe a Weyl node at $({\bf k}_0, t_0)$, we start by defining the Weyl basis  
\begin{equation}
\psi_J = (|J\rangle, |J+1\rangle)^T,
\label{eq:basis}
\end{equation}
under which the corresponding Weyl charge is labeled as $q_J({\bf k}_0, t_0)$.
Here $j_z=J$ or $J+1$ is the $z$-component angular momentum
\begin{equation}
C_n |J\rangle = e^{i \frac{2\pi}{n} J} |J \rangle,
\end{equation}
which is defined mod $n$ and takes integer (half-integer) values for spinless (spinful) fermions. The effective Weyl Hamiltonian near $({\bf k}_0, t_0)$ can be written as
\begin{equation}
	h_{w}(t_0) = \pi \mathbb{1}_2 + v_x k_x\sigma_x - v_y k_y\sigma_y + v_t t \sigma_z,
	\end{equation}
	where $v_x=v_y$ is required for $n>2$. The Weyl monopole charge $q_J$ of $h_w$ is thus defined as 
	\begin{equation}
	q_J = -{\rm sgn}(v_x v_y v_t).
	\end{equation}
	and is further simplified as $q_J = -{\rm sgn}(v_t)$ when $n>2$.
We emphasize that our definition of $q_J$ as well as its $J$ label depends on our choice of basis in Eq.~(\ref{eq:basis}), where the $j_z$ eigenvalues are rank-ordered in ascending order. Such a notation is helpful in defining the annihilation rule of Weyl points under $C_n$. For example, a Weyl node with $q_{J_1}$ can annihilate with another Weyl node with $q_{J_2}$ {\it only if} (i) $J_1=J_2$ and (ii) $q_{J_1}=-q_{J_2}$. \\

When $[C_n, {\cal S}]=0$, we have
\begin{eqnarray}
C_n \left( \mathcal{S} |J, {\bf k}_0, t_0\rangle \right) &=& e^{i\frac{2\pi}{n}J} \left( \mathcal{S} |J, {\bf k}_0, t_0\rangle \right) \nonumber  \\
& = & e^{i\frac{2\pi}{n}J}  |J, {\bf k}_0, T-t_0\rangle,
\end{eqnarray}
up to an unimportant global phase factor.
Thus, chiral symmetry maps a phase-band state with $j_z=J$ at $({\bf k}_0, t_0)$ to another state at $({\bf k}_0, T-t_0)$ with the same $j_z$.
Therefore, the chiral-related partner of $h_{w}(t_0)$ is also formed by the crossing of two phase bands with $j_z=J$ and $J+1$ at $T-t_0$, as schematically shown in Fig.~\ref{Fig2}(a). Since this Weyl pair now has Weyl charges
\begin{equation}
 q_J({\bf k}_0, t_0) = -q_J({\bf k}_0, T-t_0),
\end{equation}
they can nnihilate one another. To see this, we move this pair of Weyl nodes along $t$-axis while respecting chiral symmetry, such that they coincide at $t_0=\frac{T}{2}$ and form a fourfold degenerate Dirac point. Under the basis
\begin{equation} 
\Psi_{d,-} = (|J\rangle, |J+1\rangle, |J\rangle, |J+1\rangle)^T, 
\end{equation}
the effective Hamiltonian for this Dirac point near $({\bf k}_0, \frac{T}{2})$ is 
\begin{equation}
	h_d = \pi \mathbb{1}_4 + v_x k_x\gamma_1 - v_y k_y\gamma_2 + v_t t \gamma_5.
\label{eq:dirac}
\end{equation}
Again, $v_x=v_y$ is required for $n>2$. For convenience, we define the $4\times 4$ $\gamma$ matrices as 
$\gamma_1 = \tau_0 \otimes \sigma_x,\ \gamma_2 = \tau_0 \otimes \sigma_y,\ \gamma_3=\tau_x\otimes \sigma_z,\ \gamma_4=\tau_y\otimes \sigma_z,\ \gamma_5 = \tau_z \otimes \sigma_z$, based on which $\gamma_{jl}=[\gamma_j, \gamma_l]/(2i)$ can be generated for $j,l\in\{1,2,3,4,5\}$ and $j<l$. We find that
\begin{equation} 
{\cal S} = \tau_x \otimes \sigma_0 \equiv \gamma_{45}, \quad C_n = e^{i\frac{2\pi}{n}{\cal J}_z} 
\end{equation}
with ${\cal J}_z = \text{diag}(J, J+1, J, J+1)$.
Clearly, we can introduce a mass term proportional to $\gamma_3$ that gaps out this DWP without breaking $C_n$ and $\mathcal{S}$, since 
$[\gamma_3, C_n] = [\gamma_3, {\cal S}] = 0$.
We thus conclude that
\begin{itemize}
	\item there is no stable DWP when $[C_n,{\cal S}]=0$.
\end{itemize}

When $\{C_n, {\cal S}\}=0$, we have
\begin{eqnarray}
C_n \left( \mathcal{S} |J, {\bf k}_0, t_0\rangle \right) &=& -e^{i\frac{2\pi}{n}J} \left( \mathcal{S} |J, {\bf k}_0, t_0\rangle \right) \nonumber  \\
& = & e^{i\frac{2\pi}{n}(J+\frac{n}{2})}  |J, {\bf k}_0, T-t_0\rangle,
\end{eqnarray}
up to a phase factor.
Thus, ${\cal S}$ maps a phase-band state with $j_z=J$ at $({\bf k}_0, t_0)$ to another state at $({\bf k}_0, T-t_0)$ with a \textit{different} $j_z=J+\frac{n}{2}$.
Since the change of angular momentum $\frac{n}{2}$ can only be an integer, $n$ must be even. This immediately implies that
\begin{itemize}
	\item $C_3$ symmetry cannot protect a DWP.
\end{itemize}

For $n=2,4,6$, the chiral partner of $h_w(t_0)$ is now a Weyl node formed by two phase bands with $j_z=J+\frac{n}{2}$ and $J+\frac{n}{2}+1$, which carries a Weyl charge 
\begin{equation}
q_{J+\frac{n}{2}}({\bf k}_0, T-t_0)=-q_J({\bf k}_0, t_0),
\end{equation}
as shown in Fig.~\ref{Fig2}(b). To check the stability of such DWP, we again make them coincide at $\frac{T}{2}$ and form a Dirac point described by Eq.~(\ref{eq:dirac}).
However, the new basis is now 
\begin{equation}
\Psi_{d,+} = \left(|J\rangle, |J+1\rangle, |J+\frac{n}{2}\rangle, |J+\frac{n}{2}+1\rangle \right)^T, 
\end{equation}
which is distinct from $\Psi_{d,-}$. When $n=4, 6$, 
the four phase bands of this 3D Dirac point belong to four different one dimensional irreducible representations (irreps) of $C_{4,6}$. Therefore, this Dirac point cannot be gapped out along the rotation-invariant $t$-axis, since $C_{4,6}$ symmetry strictly prohibits any coupling between the DWP. The robustness of this DWP is also encoded in their distinct $J$ indices of Weyl charges $q_J$ and $q_{J+\frac{n}{2}}$ [see Fig.~\ref{Fig2}(c)].   

$C_2$ symmetry, however, has only two distinct irreps and thus the above protection mechanism does not seem to apply. Interestingly, in this case we find
\begin{equation} 
C_2= e^{i\pi J} \gamma_5
\end{equation} 
under the Dirac-point basis
\begin{equation} 
\Psi_{d,+}=(|J\rangle, |J+1\rangle, |J+1\rangle, |J\rangle)^T, 
\end{equation}
which is proportional to the $t$-linear term of $h_d$ in Eq.~(\ref{eq:dirac}). Therefore, any $C_2$-preserving perturbation, which must commute with the $t$-linear term in $h_d$, can only split the Weyl nodes along the $t$-axis, but is incapable of gapping them out.
Physically, the two phase bands with $C_2=(-1)^J$ propagate along the same direction on $t$-axis, while the other two with $C_2=(-1)^{J+1}$ propagate along the opposite direction. Similar to the chiral-velocity locking effect discussed in Sec.~\ref{sec:ssh}, here the stable DWP features a locking between $C_2$ eigenvalue and the velocity along $t$ direction. As we have mentioned before, this locking phenomena between the band dispersion and certain quantum numbers are generally anomalous in the band theory, and cannot be realized in any static 3D lattice system. Nevertheless, we show in Appendix \ref{app:4dTCI} that such $C_2$-velocity locking effect in DWP is exactly the defining characteristic for the surface physics of a 4D $C_2$-protected TCI. In fact, {\it a $C_n$-protected phase-band singularity is generally related to the surface state of a 4D $C_n$-protected TCI characterized by a rotation Chern number} \cite{zhang2016topological}. To conclude, we have proved in general that 
\begin{itemize}
	\item A pair of chiral-related dynamical Weyl points cannot annihilate each other {\it iff} $\{C_n, {\cal S}\}=0$.
\end{itemize}

When multiple DWPs coexist on the ${\bf k}_0$-axis, Weyl nodes from {\it different} chiral-related pairs may annihilate one another, although each pair by itself is stable. Fig.~\ref{Fig2}(c) gives an example where two DWPs can be gapped out altogether. This inspires us to define 
\begin{equation}
q_J^{(n)}({\bf k}_0)=\sum_{t_i\in (0,T)}q_J({\bf k}_0,t_i),
\end{equation} 
which counts the net Weyl charge along the $t$-axis at momentum ${\bf k}_0$ labeled by $J$. Crucially, since the Weyl nodes come in chiral-related pairs, these $q_J^{(n)}$ satisfy the following constraint:
\begin{equation}
q_J^{(n)}({\bf k}_0) = -q^{(n)}_{J+\frac{n}{2}}({\bf k}_0). 
\end{equation}
Then for each rotation-invariant axis at momentum ${\bf k}_0$, we find the following set of independent net monopole charges for each $J$,
\begin{eqnarray}
	\{q_{\frac{\alpha}{2}}^{(n)}({\bf k}_0), q_{\frac{\alpha}{2}+1}^{(n)}({\bf k}_0),..., q_{\frac{n+\alpha}{2}-1}^{(n)}({\bf k}_0) \} \in \mathbb{Z}^{\frac{n}{2}},
\end{eqnarray}
which captures all stable DWPs. Here $\alpha=0$ for spinless fermion with $(C_n)^n=1$ and $\alpha=1$ for spinful fermion with $(C_n)^n=-1$. One can easily check that the example shown in Fig.~\ref{Fig2}(c) precisely has vanishing $J$-dependent net monopole charges.

We notice that the definition of $q_J^{(n)}({\bf k}_0)$ relies on well-defined translation symmetries. For example, a Weyl node with $q_J({\bf k}_0, t_i)=1$ is generally incapable of annihilating with another Weyl node with $q_J({\bf k}_0', t_i)=-1$ if ${\bf k}_0 \neq {\bf k}_0'$. However, when translation symmetry is broken, ${\bf k}_0$ and ${\bf k}_0'$ are folded to the same point and thus the two Weyl nodes can be gapped out. For our purpose, we shall focus on the ``strong classification" of topological singularities that does not rely any translation symmetry. Such a classification for DWPs is characterized by the {\it $C_n$-topological charge},
\begin{equation}
	Q^{(n)}_J = \sum_{{\bf k}_0} q^{(n)}_J({\bf k}_0),
	\label{Eq: Topo Charge Q_n}
\end{equation}
where $J \in\{\frac{\alpha}{2}, \frac{\alpha}{2}+1,..., \frac{n+\alpha}{2}-1\}$ for $(C_n)^n=(-1)^{\alpha}$, and ${\bf k}_0$ runs over all rotation-invariant momenta. Systems with different $Q_J^{(n)}$s cannot be adiabatically connected to each other without breaking the symmetries, and are hence topologically distinct.

\section{$D_n$-protected phase-band singularity} 
\label{sec:Dn}

We now proceed to classify phase-band singularities protected by dihedral group symmetries in 2D class AIII Floquet systems.
Besides $n$-fold rotations, a dihedral group $D_n$ has $n$ additional in-plane mirror symmetries. The physical significance of a mirror symmetry ${\cal M}$ is twofold: 
\begin{enumerate}
	\item[(1)] when $\{C_n, {\cal S}\}=0$, ${\cal M}$ imposes additional constraints on the relationship among various rotation topological charges $Q_J^{(n)}$; 
	\item[(2)] ${\cal M}$ by itself is capable of protecting new types of topological singularity that are indicated by a mirror topological charge $Q_{\cal M}$, distinct from $Q_J^{(n)}$.
\end{enumerate}
As will be shown later, it is both the rotation and mirror topological charges, as well as the interplay among them, that together determines the classification of $D_n$-protected phase-band singularities.

The structure of this section is organized as follows. In Sec.~\ref{subsec:mirror constraint} and Sec.~\ref{subsec:dirac point doublet}, we first discuss the constraints on the rotation topological charges $Q_J^{(n)}$ imposed by mirror symmetry. In Sec.~\ref{subsec:mirror}, we explore the topological singularities protected by mirror symmetry alone (i.e. $D_1$ group) and define the corresponding mirror topological charge. This allows us to list all possible topological charges for a general $D_n$ group in Sec.~\ref{subsec:all possible D_n charges}. However, through a comprehensive case study of $D_2$ group in Sec.~\ref{subsec:D_2 singularity}, we find that there exists a complex interplay among rotation and mirror topological charges. Namely, not all charges are independent from one another. In Sec.~\ref{subsec:classification of D_n singularity}, we outline the general rules that determine a complete set of independent topological charges for all $D_n$ groups, which finalizes the classification of topological phase-band singularities protected by two-dimensional point groups.

\subsection{Constraining $C_n$ topological charges with a mirror symmetry}
\label{subsec:mirror constraint}

The constraints on $Q_J^{(n)}$ imposed by mirror symmetry can be understood as follows. When $[C_n, M_x]\neq 0$, the mirror symmetry will ``glue" certain one-dimensional irreps of $C_n$ into two-dimensional irreps, thereby imposing additional relations that rotation topological charges with different $J$ must satisfy. 
In particular, since $C_nM_x=M_xC_n^{-1}$, we have
\begin{eqnarray}
C_n \left( M_x |J, {\bf k}_0, t_0 \rangle \right)  &=& e^{-i \frac{2\pi}{n} J} \left( M_x |J, {\bf k}_0, t_0 \rangle \right)  \nonumber  \\
&=& e^{-i \frac{2\pi}{n} J}  |-J, {\bf k}_0, t_0 \rangle,
\end{eqnarray}
up to a phase factor, where ${\bf k}_0$ is a rotation and mirror symmetric momentum.
Namely, $M_x$ maps a phase-band state with $j_z=J$ at $({\bf k}_0, t_0)$ to another state at the same location with $j_z=-J$. In what follows, we shall focus on $D_n$ double groups with $(C_n)^n=-1$, which only has 2d irreps. 
A discussion for $D_n$ single groups with $(C_n)^n=1$ is detailed in Appendix~\ref{app:dn}. 

Consider a phase-band Weyl node at $({\bf k}_0, t_0)$ under the basis $\psi_J = (|J\rangle, |J+1\rangle)^T$ that carries a charge $q_J({\bf k}_0, t_0)$. $M_x$ will enforce the existence of another Weyl node at the same location with a charge 
\begin{equation}
q_{-J-1}({\bf k}_0, t_0) = -q_J({\bf k}_0, t_0)
\end{equation}
under the basis $\psi_{-J-1}=(|-J-1\rangle, |-J\rangle)^T$. This mirror-enforced Weyl pair forms a 4-fold degenerate {\it dynamical Dirac point} (DDP). Such a DDP is  unstable {\it only if} $J+1 = -J$ (mod $n$), where the two mirror-related Weyl nodes carry opposite monopole charges with exactly the same $J$ label. In other words, there is no protected singularity if $J=-\frac{1}{2}$ or $\frac{n-1}{2}$, leading to
\begin{equation} 
Q_{\frac{n-1}{2}}^{(n)} = 0  \quad \text{for any} \ D_n \  \text{group}. 
\end{equation}
In particular, the $D_2$ double group always has trivial topological charges and hence no stable topological singularity. For $n=4,6$, this indicates that $Q^{(4)}_{\frac{3}{2}}=0$, and $Q^{(6)}_{\frac{5}{2}}=0$.
Furthermore, we notice that ${\cal S}$ and $M_x$ together impose
\begin{equation} 
Q^{(n)}_{J} = - Q^{(n)}_{-J-1} = - Q^{(n)}_{-J-1+n} = Q^{(n)}_{-J-1+\frac{n}{2}}, 
\end{equation}
which requires 
\begin{equation}
Q_{\frac{1}{2}}^{(6)} = Q_{\frac{3}{2}}^{(6)}.
\end{equation} 
Taken together, the nontrivial rotation topological charges under the constraint of mirror symmetry for each $D_n$ double group are:
\begin{eqnarray}
&D_2:&  0,  \nonumber \\
&D_4:&  Q^{(4)}_{\frac{1}{2}} \in \mathbb{Z},   \nonumber  \\
&D_6:&  Q^{(6)}_{\frac{1}{2}} \in \mathbb{Z}.   
\label{eq:rotation_double_dn}
\end{eqnarray}
As shown in Appendix \ref{app:dn}, the mirror-constrained rotation topological charges for $D_n$ single groups are
\begin{eqnarray}
&D_2:&  Q_0^{(2)} \in \mathbb{Z},  \nonumber \\
&D_4:&  Q^{(4)}_0 \in \mathbb{Z},   \nonumber  \\
&D_6:&  \{Q^{(6)}_0, Q^{(6)}_1\} \in \mathbb{Z}^2.   
\label{eq:rotation_single_dn}
\end{eqnarray}

\begin{figure}[!t]
	\includegraphics[width=0.45\textwidth]{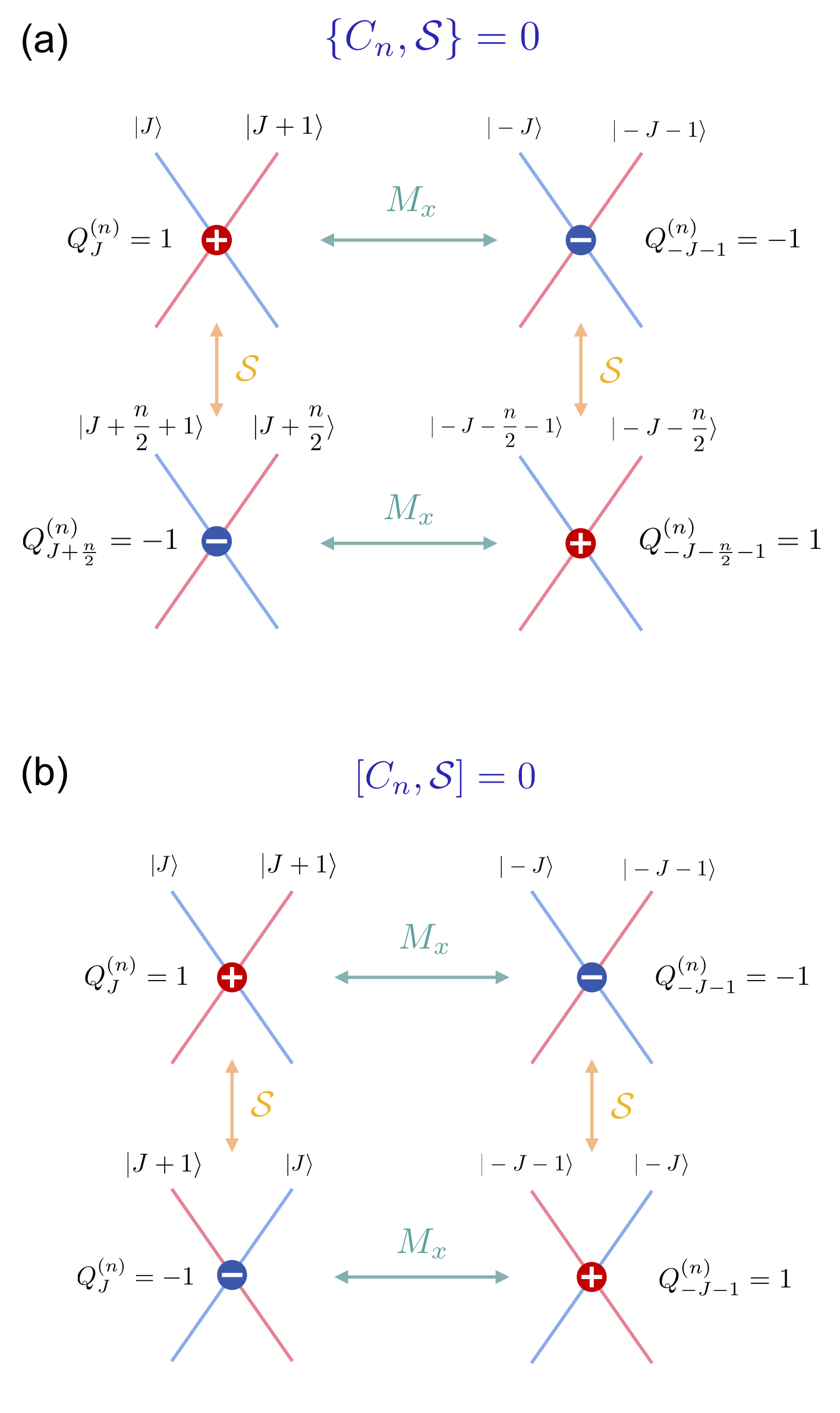}
	\caption{Commutative diagrams for the Weyl node partners under the action of $M_x$ and $\mathcal{S}$. (a) $\{C_n, \mathcal{S}\}=0$. A single DDP is possible when $2J+1=\frac{n}{2}$ (mod $n$). (b) $[C_n, \mathcal{S}]=0$. A single DDP is possible when $2J+1=0$ (mod $n$). When the above conditions are not satisfied, one will instead have a pair of DDPs (i.e. a DDP doublet).}
	\label{fig:doublet}
\end{figure} 

\subsection{Dynamical Dirac point and its doublet}
\label{subsec:dirac point doublet}

In this part, we will further elaborate on the properties of DDPs defined in the previous subsection. First of all, we emphasize that the existence of DDPs does not rely on the rotation topological charges. Namely, even when $[C_n, {\cal S}]=0$ with vanishing $Q_J^{(n)}$, it is possible to have similar DDPs that are protected by mirror symmetries themselves. Nevertheless, in the current subsection, we will only derive the conditions for such DDPs to exist for various $D_n$ groups when the rotation charges are trivial, and leave their explicit topological characterizations (i.e. the mirror topological charges) to the coming subsections.

Secondly, unlike the case for $C_n$ groups, the mirror-constrained $C_n$ topological charges in Eqs.~(\ref{eq:rotation_double_dn}) and~(\ref{eq:rotation_single_dn}) indicate either the number of DDPs or that of {\it DDP doublets} (i.e. chiral-related pairs of DDPs). The formation of a DDP doublet versus that of a single DDP can be understood as follows. Starting from a dynamical Weyl node $W_1$ and its mirror partner $W_2$ at $({\bf k}_0, t_0)$, we denote their chiral partners at $({\bf k}_0, T-t_0)$ as $W_3$ and $W_4$, respectively. Generally, $W_{1,2}$ and $W_{3,4}$ will form two chiral-related DDPs, which we call a DDP doublet. However, in some special cases, the mirror partner of a Weyl node is identical to its chiral partner, e.g. $W_1=W_4$ and $W_2=W_3$, leading to a single DDP pinned at $t_0=T/2$. As we will see in the next subsection, differentiating between a DDP and a DDP doublet is essential for our general classification of $D_n$-protected singularities. Below, we shall systematically clarify for which $D_n$ groups can a single DDP exist.

The situation where $\{C_n, \mathcal{S}\}=0$ is generally captured by the commutative diagram shown in Fig.~\ref{fig:doublet}(a). We start from a single Weyl node under the basis $\psi_J$, with rotation charge $Q_J^{(n)}=1$. In Fig.~\ref{fig:doublet}(a) we explicitly show its partners under the action of $M_x$ and $\mathcal{S}$. Indeed, generically $M_x$ and $\mathcal{S}$ will give rise to a doublet of Dirac cones, unless the mirror and chiral partners of the original Weyl node coincide. When this happens, it is possible to have a single DDP sitting at $t_0=\frac{T}{2}$. From Fig.~\ref{fig:doublet}(a), one can immediately see that the condition for this to happen is:
\begin{equation}
J+\frac{n}{2}+1=-J \quad {\rm mod} \ n, \quad {\rm i.e.},  \quad 2J+1=\frac{n}{2} \quad {\rm mod} \ n.
\end{equation}
More specifically, the above condition can be satisfied when
\begin{enumerate}
\item $(C_n)^n=1$ (spinless):  $n=2,6$;
\item $(C_n)^n = -1$ (spinful): $n=4$.
\end{enumerate}
Recall that $C_3$ cannot anticommute with $\mathcal{S}$ and hence does not appear on the list above. Therefore, we conclude that 
{\it when $\{C_n, \mathcal{S}\}=0$, a single DDP is possible only for spinless $D_2$ and $D_6$ groups, and spinful $D_4$ group.}
Otherwise, we generally expect  DDP doublets as the phase-band singularities.

Similarly, the situation where $[C_n, \mathcal{S}]=0$ is shown in Fig.~\ref{fig:doublet}(b). One can read off from this commutative diagram that the condition for the existence of a single DDP in this case is given by:
\begin{equation}
J+1=-J \quad {\rm mod} \ n, \quad {\rm i.e.},  \quad 2J+1=0 \quad {\rm mod} \ n.
\end{equation}
We first note that a single DDP can always exist in a $D_1$ group, where all angular momentum irreps are equivalent. In addition, the above condition can be satisfied when
\begin{enumerate}
\item $(C_n)^n=1$ (spinless): $n=3$;
\item $(C_n)^n=-1$ (spinful): all $n$.
\end{enumerate}
Therefore, we conclude that {\it when $[C_n, \mathcal{S}]=0$, a single DDP is possible for spinless and spinful $D_1$ groups, spinless $D_3$ group and all spinful $D_n$ groups.}
Fig.~\ref{fig:doublet}(b) also shows that the rotation charge in this case must be trivial, as expected from our discussions in the previous section. Nonetheless, it is mirror symmetry that is responsible for protecting the above DDPs, which we will show next.

\subsection{Topological singularity protected by a single mirror symmetry: $D_1$ group}
\label{subsec:mirror}

Besides constraining rotation topological charges, mirror symmetry by itself can protect robust dynamical singularities.
Let us consider a single mirror symmetry $M_x$, which constitutes the $D_1$ group. We will show that the topological singularity protected by $M_x$ generically takes the form of a nodal loop, which can be shrunk to a single DDP. Furthermore, we will also demonstrate that a DDP doublet related by both chiral and mirror symmetries always has a trivial mirror charge. Hence, the condition for the existence of a single DDP derived previously will become important when we discuss mirror charges for $D_n$ groups in general.

Combining with chiral symmetry, there are now two mirror planes in the 3D $({\bf k}, t)$ space: the $k_x=0$ plane and the $t_0=\frac{T}{2}$ plane. The line along $k_y$ direction with $k_x=0$ and $t_0=\frac{T}{2}$ is thus invariant under both mirror symmetries. Therefore, topological singularities residing on the $k_y$-axis can appear as a single DDP that is invariant under $M_x$ and $\mathcal{S}$ symmetries. Alternatively, a DDP doublet on the $t$-axis (or $k_x$-axis) is also allowed by symmetry.

Let us first consider a DDP on the mirror symmetric line. If there is only one mirror plane $M_x$, the DDP can appear anywhere on the mirror symmetric line, and we can thus always place it at the origin (we have redefined $t-\frac{T}{2}$ as $t$). The DDP can again be described by the effective Hamiltonian~(\ref{eq:dirac}). However, the action of chiral and mirror symmetry remains to be determined. By a direct enumeration, we find that there are two possible representations for the chiral and mirror symmetries that are consistent with their actions on Hamiltonian~(\ref{eq:dirac}):
\begin{eqnarray}
\mathcal{S} &=& \gamma_{35}, \quad \text{or} \quad \mathcal{S} = \gamma_{45};  \nonumber  \\
M_x &=& \gamma_{13}, \quad \text{or} \quad M_x = \gamma_{14}.
\end{eqnarray}
One can easily check that out of the 4 combinations of $\mathcal{S}$ and $M_x$, two of which satisfy $[\mathcal{S}, M_x]=0$, and the other two satisfy instead $\{\mathcal{S}, M_x\}=0$. We shall now examine the stability of the DDP for each situation separately; namely, whether or not such a DDP can be gapped out by a symmetric mass term.

When $\{\mathcal{S}, M_x\}=0$, we have $\mathcal{S}=\gamma_{35}, \ M_x=\gamma_{13}$; or $\mathcal{S}=\gamma_{45}, \ M_x=\gamma_{14}$. Apparently a mass term $h_m \propto \gamma_4$ in the former case, or $h_m \propto \gamma_3$ in the latter case gaps out the DDP while respecting both symmetries: $[\mathcal{S}, h_m]=0$, $[M_x, h_m]=0$. We thus conclude that 
\begin{itemize}
	\item there is no stable mirror-protected DDP when $\{ \mathcal{S}, M_x\}=0$.
\end{itemize}

\begin{figure}[t]
	\includegraphics[width=0.48\textwidth]{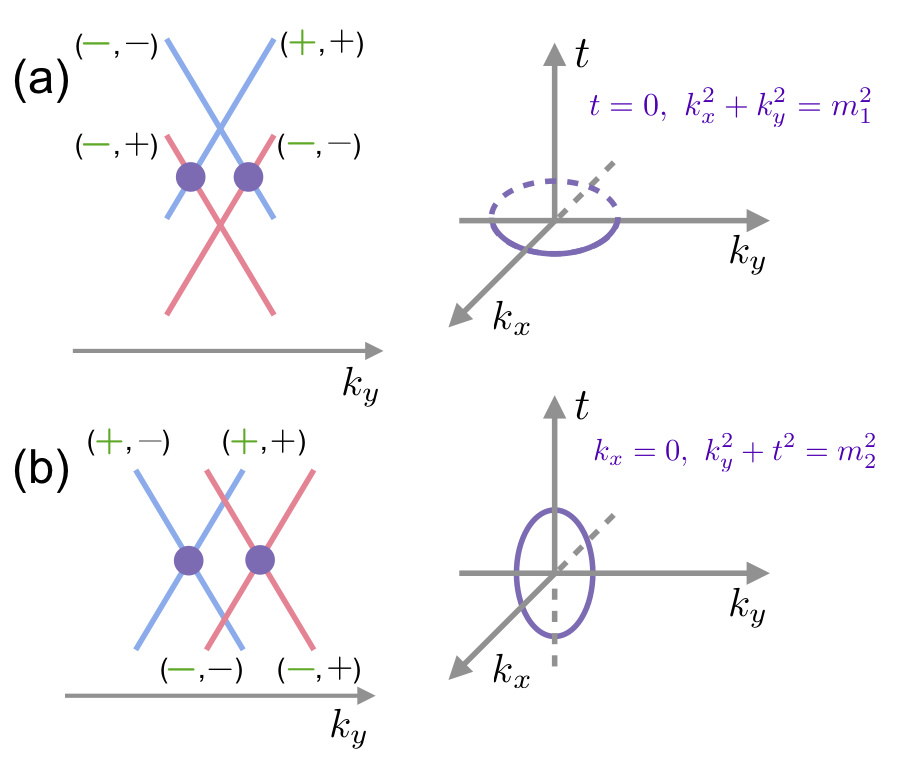}
	\caption{A DDP in the presence of perturbations respecting both $\mathcal{S}$ and $M_x$ symmetries. (a) The first allowed perturbation moves the Weyl nodes up and down symmetrically in energy, yileding a nodal loop as indicated by the purple dots and line. (b) The second allowed perturbation moves the location of the Weyld nodes symmetrically along $k_y$ axis, yielding a nodal loop as indicated by the purple dots and line. The green and black “$\pm$” label the $\mathcal{S}$ and $M_x$ eigenvalues of each band: $(\mathcal{S}, M_x)$, respectively.}
	\label{fig:mirror_nodal}
\end{figure}

When $[\mathcal{S}, M_x]=0$, we have $\mathcal{S}=\gamma_{45}, \ M_x=\gamma_{13}$; or $\mathcal{S} = \gamma_{35}, \ M_x=\gamma_{14}$. Let us take the second case as an example. Similarly to our previous discussions, we are again looking for perturbations that commute with both $\mathcal{S}$ and $M_x$. In the current situation, there are only two such perturbations: $h_m \propto \gamma_{35}$, or $h_m \propto \gamma_{14}$. However, neither of them is capable of gapping out the DDP in this case, since neither $\gamma_{35}$ nor $\gamma_{14}$ anticommutes with the full kinetic terms in Hamiltonian~(\ref{eq:dirac}). For example, the phase band spectrum of Hamiltonian~(\ref{eq:dirac}) in the presence of $h_m = m_1\gamma_{35}$ is given by:
\begin{equation}
\widetilde{\phi}({\bf k}, t) = \pi \pm \sqrt{t^2 + \left(\sqrt{k_x^2+k_y^2}\pm m_1 \right)^2},
\end{equation}
which simply corresponds to shifting the two Weyl nodes up and down symmetrically in energy, yielding a nodal loop at:
\begin{equation}
t = 0,  \quad  k_x^2+k_y^2 = m_1^2,
\end{equation}
as depicted in Fig.~\ref{fig:mirror_nodal}(a).
On the other hand, in the presence of $h_m = m_2 \gamma_{14}$, the phase band spectrum takes the form:
\begin{equation}
\widetilde{\phi}({\bf k}, t) = \pi \pm \sqrt{k_x^2 + \left( \sqrt{k_y^2 + t^2} \pm m_2 \right)^2},
\end{equation}
which now corresponds to shifting the two Weyl nodes symmetrically along the $k_y$ axis, yielding a nodal loop at:
\begin{equation}
k_x = 0, \quad k_y^2 + t^2 = m_2^2,
\end{equation}
as depicted in Fig.~\ref{fig:mirror_nodal}(b). However, neither of the perturbations is able to remove the topological singularity in the phase band, due to the combination of $\mathcal{S}$ and $M_x$ symmetries. 

To understand the protection mechanism for the above DDP, we first label every phase band dispersing along the $k_y$ axis with both its chiral and mirror eigenvalues as $({\cal S}, M_x)$. We then find that a general mirror-preserving DDP with $[\mathcal{S}, M_x]=0$ always features the following properties:
\begin{enumerate}
	\item[(i)] it consists of four branches along the $k_y$ axis that are labeled by $({\cal S}, M_x) = (+,+), (+,-), (-,+), (-,-)$;
	\item[(ii)] the $(+,+)$ and $(-,-)$ branches disperse along the same direction, while the $(+,-)$ and $(-,+),$ branches disperse along the other direction.
\end{enumerate}

With this pattern of symmetry indices, one immediately finds that all crossings of the DDP are protected by both chiral and mirror symmetries. Namely, any two bands that cross necessarily have either opposite $\mathcal{S}$ or $M_x$ eigenvalues. For example, as shown in Fig.~\ref{fig:mirror_nodal}(a), the perturbation $m_1\gamma_{35}$ shifts the two Weyl nodes up and down in phase band energy, yielding two crossings labeled by purple dots in Fig.~\ref{fig:mirror_nodal}(a). While the two bands crossings at the purple dots have identical $M_x$ eigenvalues, they cannot be gapped out for sharing the opposite $\mathcal{S}$ eigenvalues. An analysis of all other cases can be carried out in a similar fashion, and leads to the same result. We thus conclude that 
\begin{itemize}
\item A single mirror symmetry $M_x$ can protect stable phase band topological singularities \textit{iff} $[\mathcal{S}, M_x]=0$. The generic topological singularities in this case are nodal loops, which can be shrunk to DDPs but can never be adiabatically eliminated while preserving $M_x$.
\end{itemize}

To characterize such mirror-protected loop-like or point-like singularities, we make use of their unique symmetry pattern and define the following {\it mirror topological charge}:
\begin{equation}
Q_{M_x} = \sum_{k_y=0,\pi}(n_{++}^R - n_{++}^L) \in \mathbb{Z},
\label{eq:mirror_charge}
\end{equation}
where $n_{++}^{R/L}$ counts the number of left and right movers with $\mathcal{S}=M_x=+1$ along the $k_y$ axis. The summation over all high-symmetry $k_y$ restricts our classification to singularities that do not rely on translation symmetry.
For a single DDP, this definition implies that $Q_{M_x} = {\rm sgn}(v_y)$, as the sign of $Q_{M_x}$ depends on whether the branch with $\mathcal{S}=M_x=+1$ is a left or right mover (see Fig.~\ref{fig:mirror_nodal}).

\begin{figure}[!t]
	\includegraphics[width=0.4\textwidth]{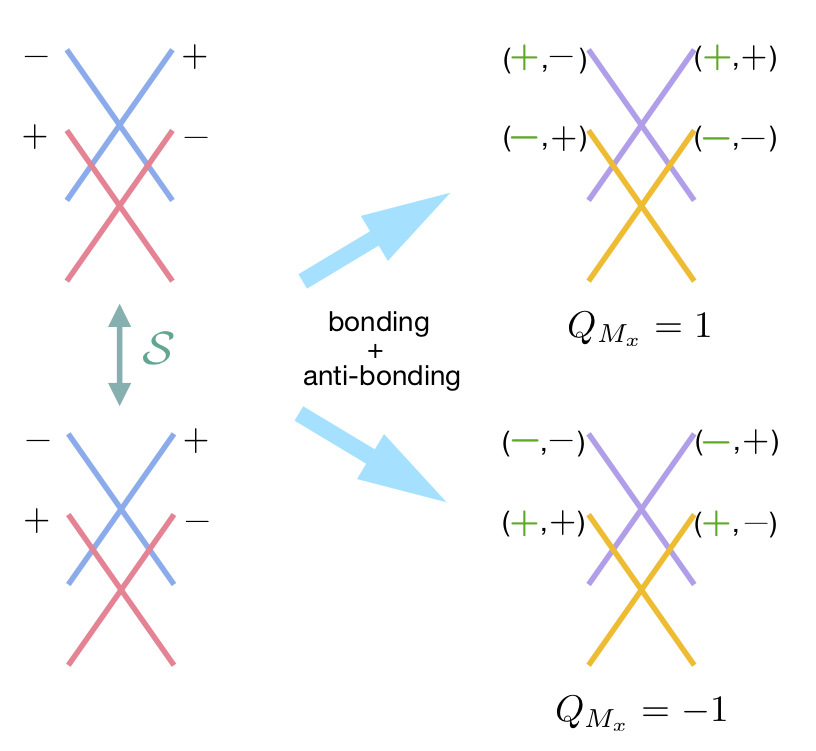}
	\caption{Mirror charge of a DDP doublet. A DDP doublet consists of two Dirac nodes related by chiral symmetry, as shown on the left. The black $\pm$ labels the $M_x$ eigenvalue of each branch. Chiral eigenstates are bonding (symmetric) and anti-bonding (antisymmetric) superpositions of the two left / right movers. The resultant chiral eigenvalues of each branch are labeled by green $\pm $ on the right. Such a DDP doublet apparently has $Q_{M_x}=0$.}
	\label{fig:doublet_charge}
\end{figure} 

Next, we explore the $Q_{M_x}$ of a DDP doublet. Consider a single Dirac node within a DDP doublet, which consists of a pair of mirror-related Weyl nodes. Such a DDP can be effectively described by Hamiltonian~(\ref{eq:dirac}), and carries a $M_x$ index in the $M_x$-invariant plane. From Fig.~\ref{fig:mirror_nodal}, along $k_y$-axis, the $M_x$ eigenvalues of each branch for this single DDP are: $+v_k k_y, \ M_x=1;\ +v_k k_y, \ M_x=-1; \ -v_k k_y, \ M_x=1; \ - v_k k_y, \ M_x=-1$. The same is also true for its chiral partner since $[{\cal S}, M_x]=0$, and we depict both DDPs in Fig.~\ref{fig:doublet_charge}(a).

The next step is to determine the chiral index for each branch of the DDP doublet. As shown in Fig.~\ref{fig:doublet_charge}(a), the chiral symmetry maps one branch of a DDP to a branch in the other DDP with the same $M_x$ eigenvalue and $k_y$ dispersion. Therefore, eigenstates of $\mathcal{S}$ are symmetric and antisymmetric superpositions of these two chiral-related branches, with the same $M_x$ eigenvalue. Namely, two chiral-related branches with $M_x=\pm$ will contribute to two branches with $(+,\pm)$ and $(-,\pm)$. Then it is easy to see that such a DDP doublet always consists of a DDP with $Q_{M_x}=1$ and another with $Q_{M_x}=-1$. Therefore, we have proved that
\begin{itemize}
	\item a DDP doublet always carries $Q_{M_x}=0$.
\end{itemize}

\subsection{A list of possible topological charges for $D_n$}
\label{subsec:all possible D_n charges}

\begin{figure}[!t]
	\includegraphics[width=0.45\textwidth]{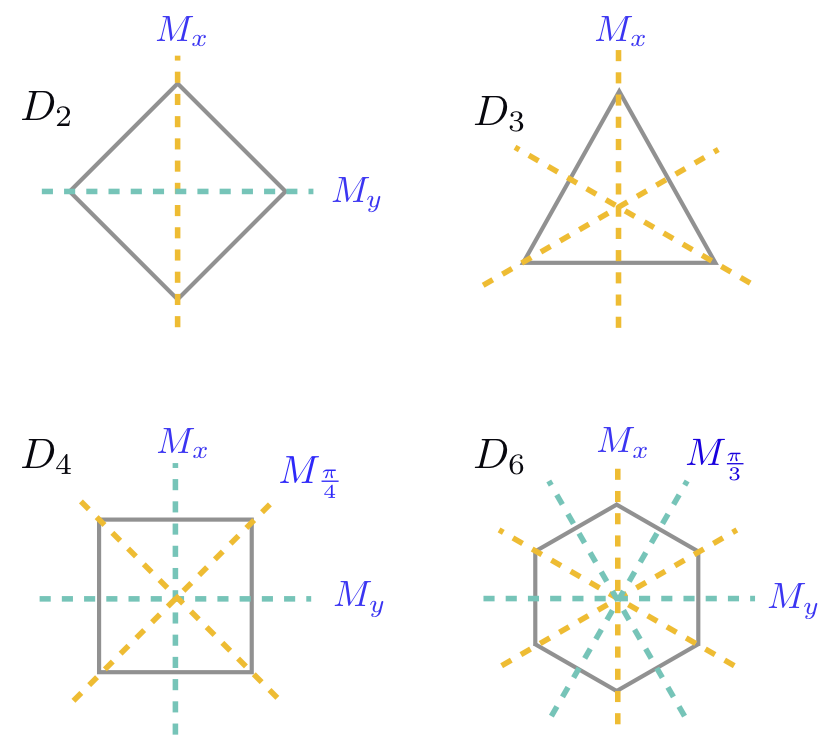}
	\caption{Mirror planes of the dihedral groups $D_2$, $D_3$, $D_4$ and $D_6$. Inequivalent mirror planes are drawn in different colors.}
	\label{fig:mirror}
\end{figure} 

We are now ready to present the complete list of possible topological charges for a general $D_n$ group. Summarizing our previous discussions of rotation and mirror charges, the set of nontrivial topological charges will depend on three factors:
\begin{enumerate}
\item the commutation relation between $M_x$ and $\mathcal{S}$; 
\item the commutation relation between the inequivalent mirror $M' = C_n M_x$ and $\mathcal{S}$; or, equivalently, the commutation relation between $C_n$ and $\mathcal{S}$;
\item $(C_n)^n=+1$ or $-1$, i.e. single or double group.
\end{enumerate}
Here, (1)-(3) together lead to $2^3=8$ possibilities. We shall label them as $D_{n, \eta_{C_n}, \eta_{M_x}}^{s}$ for a dihedral group $D_n$ in the following discussions, where $\eta=- (+)$ when $M_x$ or $C_n$ commutes (anticommutes) with $\mathcal{S}$, and $s=+(-)$ when $(C_n)^n=+1 (-1)$. Besides, we will often denote the mirror $M'$ inequivalent to $M_x$ as $M_{\theta_n}$. Here for $D_n$ group, we have defined
\begin{equation}
	\theta_n=\frac{\pi}{2}-\frac{\pi}{n}
\end{equation}	
as the polar angle of the corresponding mirror axis in 2D real space. For example, $M_{\theta_2}$ for $D_2$ group is exactly $M_y$, since the $M_y$-axis is labeled by a zero polar angle.

Of all eight possibilities, $D_{n, -, +}^{\pm}$ contains no nontrivial topological charge, since in this case $\{\mathcal{S}, M_x\}=[C_n, \mathcal{S}] = \{ M_{\theta_n}, \mathcal{S} \}=0$, and hence all rotational and mirror topological charges are trivial. Furthermore, $D_{n, +, +}^{\pm}$ and $D_{n, +, -}^{\pm}$ are in fact equivalent, since the former corresponds to $\{\mathcal{S}, M_x\} = [\mathcal{S}, M_{\theta_n}]=\{C_n, \mathcal{S}\}=0$ and the latter corresponds to $[\mathcal{S}, M_x] = \{ \mathcal{S}, M_{\theta_n} \} = \{C_n, \mathcal{S} \}=0$; hence, they are equivalent upon switching the definition of $M_x$ and $M_{\theta_n}$. Therefore, we are left with only four distinct cases, which we choose to be: $D^+_{n, +, -}$ ($D^+_{n,+,+}$), $D^-_{n, +, -}$ ($D^-_{n,+,+}$), $D^+_{n,-,-}$, and $D^-_{n,-,-}$. Notice that for the $D_3$ group, there exist only two nontrivial classes $D^{\pm}_{3,-,-}$, since all of its mirrors are equivalent under $C_3$ (see Fig.~\ref{fig:mirror}).

Now we have the following rules for determining the nontrivial topological charges for each $D^s_{n,\eta_{C_n}, \eta_{M_x}}$ class: 
\begin{enumerate}
\item rotation charges $Q_J^{(n)}$: nontrivial if $\eta_{C_n}=+$ and the $D_n$ mirror-constrained rotation charge is nontrivial;

\item mirror charge $Q_{M_x}$: nontrivial if $\eta_{M_x}=-$ and the existence of a single DDP is possible for the corresponding $D_n$ group;

\item both $\{Q_{M_x}, Q_{M_{\theta_n}}\}$ are nontrivial if $\eta_{C_n}=\eta_{M_x}=-$, and a single DDP is possible.

\item If two mirror symmetries $M_x$ and $\widetilde{M}$ are equivalent under $C_n$ rotation (e.g. $\widetilde{M}=C_n M_x C_n^{-1}$), we necessarily have $Q_{M_x}=Q_{\widetilde{M}}$ for $\eta_{M_x}=-$, and only $Q_{M_x}$ will be included in our classification to avoid double counting. 
\end{enumerate}
Based on the above rules, we can list all possible nontrivial topological charges for each $D^s_{n,\eta_{C_n}, \eta_{M_x}}$ class as follows:
\begin{itemize}
\item $D_2$ group:
\begin{enumerate}
\item $D^+_{2,+,\pm}$: $\{Q_0^{(2)}, Q_{M_x}\}$
\item $D^-_{2,+,\pm}$: trivial
\item $D^+_{2,-,-}$: trivial
\item $D^-_{2,-,-}$: $\{Q_{M_x}, Q_{M_y}\}$
\end{enumerate}

\item $D_3$ group: $D^{\pm}_{3,-,-}$: $Q_{M_x}$

\item $D_4$ group:
\begin{enumerate}
\item $D^+_{4,+,\pm}$: $Q_0^{(4)}$
\item $D^-_{4,+,\pm}$: $\{Q_{\frac{1}{2}}^{(4)}, Q_{M_x}\}$
\item $D^+_{4,-,-}$: trivial
\item $D^-_{4,-,-}$: $\{Q_{M_x}, Q_{M_{\frac{\pi}{4}}}\}$
\end{enumerate}

\item $D_6$ group:
\begin{enumerate}
\item $D^+_{6,+,\pm}$: $\{Q_0^{(6)}, Q_1^{(6)}, Q_{M_x}\}$
\item $D^-_{6,+,\pm}$: $Q_{\frac{1}{2}}^{(6)}$
\item $D^+_{6,-,-}$: trivial
\item $D^-_{6,-,-}$: $\{Q_{M_x}, Q_{M_{\frac{\pi}{3}}}\}$
\end{enumerate}
\end{itemize}

While the above list exhausts all possible topological charges, there remains one important question: are all the mirror and rotation topological charges for a given $D_n$ group completely independent from one another? To address this question, below we shall first use the $D_2$ group as a case study. Then we will provide a general prescription allowing us to classify all other $D_n$ groups in a similar fashion.

\subsection{A case study: $D_2$ group}
\label{subsec:D_2 singularity}

The $D_2$ group has two inequivalent mirror planes $M_x$ and $M_y$ that are $C_2$-related: $C_2=M_x M_y$, as shown in Fig.~\ref{fig:mirror}. According to our list of topological charges, there are only two nontrivial classes: $D^+_{2,+,\pm}$ and $D^-_{2,-,-}$. While this follows from our general rules, for $D_2$ group we can alternatively reach this conclusion from a straightforward analysis, which we shall present first.

If there exists only one DDP in $D_2$ group, it must be invariant under $M_y$ by itself and is thus pinned at a high symmetry position intersected by $k_x,k_y$ and $t$ axes (e.g. the origin of phase space). For DDPs that are not $M_y$ invariant, they must come in $M_y$-related pairs, forming DDP doublets with a trivial mirror charge.
As it turns out, which one of these two scenarios actually happens is completely determined by the commutation relations between $M_x$, $M_y$, and $\mathcal{S}$. Let us take a single DDP at the origin. Since $[\mathcal{S}, M_x]=0$ is required for a stable DDP, we are left with four possibilities:
\begin{enumerate}
\item $[M_x, M_y]=0$, $[\mathcal{S}, M_y]=0$. In this case, the right mover of the DDP with $\mathcal{S}=M_x=+1$ (see Fig.~\ref{fig:mirror_nodal}) is mapped to a left mover with $\mathcal{S}=M_x=+1$ under $M_y$. This implies the existence of a mirror-enforced DDP doublet, and the total mirror charge according to Eq.~(\ref{eq:mirror_charge}) must be $Q_{M_x}=0$.

\item $[M_x, M_y]=0$, $\{\mathcal{S}, M_y \}=0$. In this case, the right mover of the DDP with $\mathcal{S}=M_x=+1$ is mapped to a left mover with $\mathcal{S}=-1$, $M_x=+1$, which is precisely the left mover of the other Weyl node of the same DDP. One can readily check for each left and right mover, and confirm that the DDP is mapped to itself under $M_y$. Therefore, the DDP is this case is stable.

\item $\{M_x, M_y\}=0$, $[\mathcal{S}, M_y]=0$. In this case, the right mover of the DDP with $\mathcal{S}=M_x=+1$ is mapped to a left mover with $\mathcal{S}=+1$, $M_x=-1$, which is precisely the left mover of the same Weyl node. One can easily check that the DDP is also mapped to itself under $M_y$. Therefore, the DDP in this case is also stable.

\item $\{M_x, M_y\}=0$, $\{\mathcal{S}, M_y \}=0$. In this case, the right mover of the DDP with $\mathcal{S}=M_x=-1$ is mapped to a left mover with $\mathcal{S}=M_x=+1$. This implies the existence of a mirror-enforced DDP doublet with trivial mirror charges.

\end{enumerate}

From the above discussions, we find that, out of the four possible scenarios, only the following two cases give rise to a single stable DDP:
\begin{enumerate}
\item $[M_x, M_y]=0$, $[M_x, \mathcal{S}]=\{M_y, \mathcal{S}\}=0$, $\{C_2, \mathcal{S}\}=0$, and $(C_2)^2=1$;
\item $\{M_x, M_y\}=0$, $[M_x, \mathcal{S}]=[M_y, \mathcal{S}]=0$, $[C_2, \mathcal{S}]=0$, and $(C_2)^2=-1$.
\end{enumerate}
These two situations precisely correspond to the two nontrivial classes $D^+_{2,+,\pm}$ and $D^-_{2,-,-}$, respectively.
We remark that the above conclusion can also be obtained using an explicit representation of the DDP, as in Hamiltonian~(\ref{eq:dirac}). As discussed in the previous subsection, there are two different choices of $\mathcal{S}$ and $M_x$ satisfying $[M_x, \mathcal{S}]=0$. Again, we can take $\mathcal{S}=\gamma_{35}$ and $M_x=\gamma_{14}$ as an example. Given the DDP in Eq.~(\ref{eq:dirac}), there are two allowed representations for $M_y$: $M_y=\gamma_{23}$ or $M_y = \gamma_{24}$. One can directly check that these two allowed $M_y$ operations precisely correspond to the above two situations with stable DDPs. In what follows, we will elaborate on the two nontrivial classes and carefully disentangle the relations among the mirror and $C_2$ topological charges.

\subsubsection{$D_{2, +, -}^{+}$}

\label{subsubsection:D_{2+-}}

In this case, we have
\begin{equation}
[\mathcal{S}, M_x]=\{\mathcal{S}, M_y\}=0, \quad  \{C_2, \mathcal{S}\}=0.
\end{equation}
From the above relations, we conclude that $M_y$ cannot protect stable singularity. Moreover, since $(C_2)^2=1$, this corresponds to the $D_2$ single group with $Q_{0}^{(2)} \in \mathbb{Z}$. Together with the mirror topological charge $Q_{M_x} \in \mathbb{Z}$, the topological singularities in this case are classified by two integers: $\{Q_0^{(2)}, Q_{M_x}\}$. As it turns out, these two charges are not necessarily equal, but they are not completely independent either. 

\begin{figure}[!t]
	\includegraphics[width=0.48\textwidth]{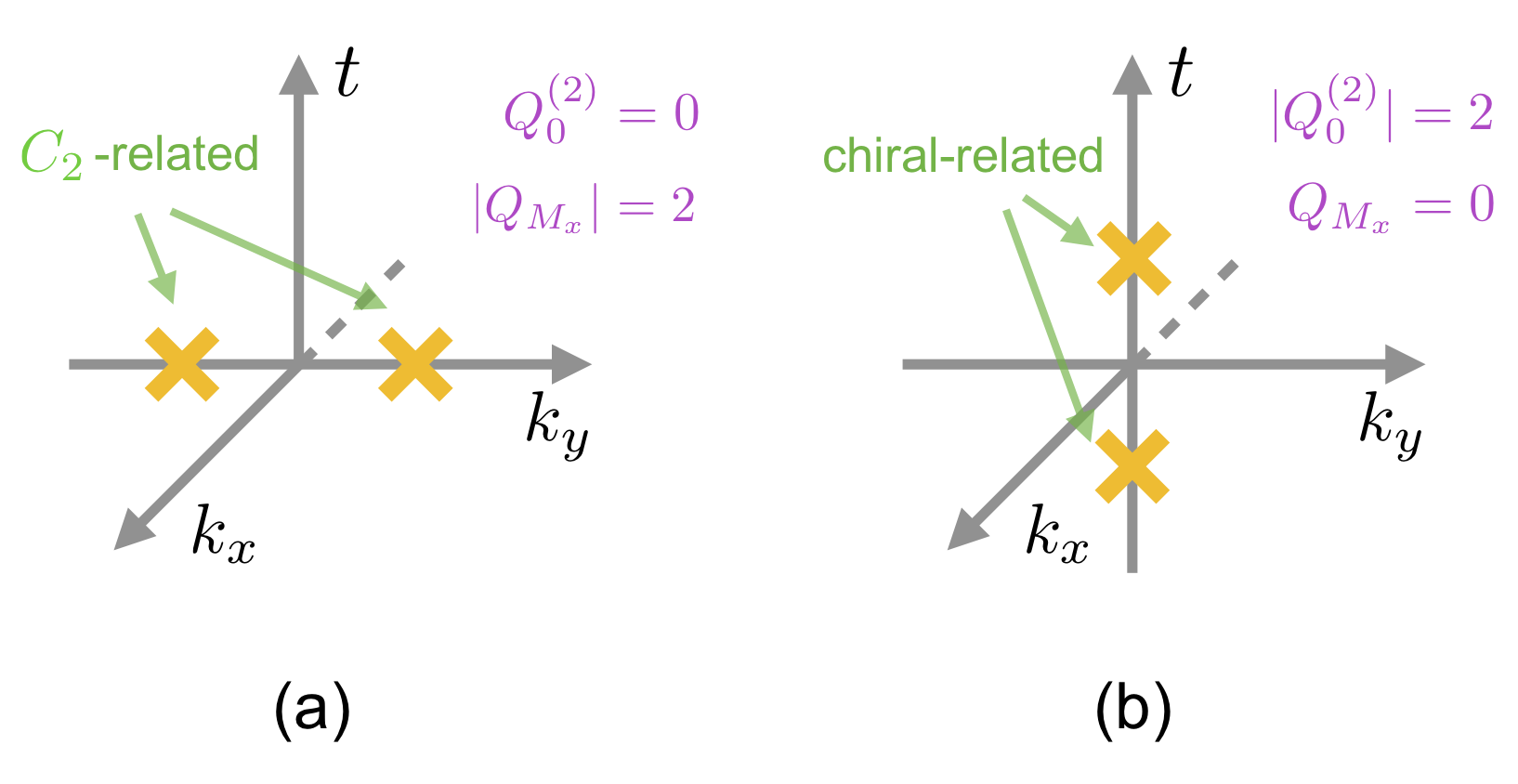}
	\caption{Relation between $Q^{(2)}_0$ and $Q_{M_x}$ can be seen from inspecting the mobility of a pair of DDPs. (a) A DDP doublet is pinned on the $k_y$ axis by $M_x$ and $\mathcal{S}$ symmetries, but can split along the $k_y$ axis in a $C_2$ symmetric way. This corresponds to $Q^{(2)}_0=0$ and $|Q_{M_x}|=2$. (b) A DDP doublet is pinned on the $C_2$-invariant axis, but can split along the $t$ axis in a chiral symmetric way. This corresponds to $|Q^{(2)}_0|=2$ and $Q_{M_x}=0$.}
	\label{fig:D2_obstruction}
\end{figure}

To see this, it is instructive to consider two DDPs. While a single DDP must be pinned at the origin by $M_x$, $\mathcal{S}$ and $C_2$ symmetries, a pair of DDPs can appear at $M_x$, chiral, or $C_2$-related positions forming a DDP doublet, as shown in Fig.~\ref{fig:D2_obstruction}. If a pair of DDPs can be symmetrically split along the $k_y$-axis and yet cannot move along the $t$-axis as illustrated in Fig.~\ref{fig:D2_obstruction}(a), it is a $C_2$-related DDP doublet. From our previous discussion in Sec.~\ref{sec:Cn}, such a group of off-axis DWPs must have $Q^{(2)}_0 = 0$. In the meantime, the total mirror charge is nontrivial: $|Q_{M_x}|=2$, since we essentially have two identical copies of DDP on the $k_y$ axis that are related by $C_2$. Similarly, when a pair of DDPs are pinned on the $C_2$-invariant axis $k_x=k_y=0$ and yet can be symmetrically split along the $t$ axis, it forms a chiral-related DDP doublet as illustrated in Fig.~\ref{fig:D2_obstruction}(b). We instead have $Q_{M_x}=0$ and $|Q^{(2)}_0|=2$. Finally, if both DDPs are pinned at the origin, one has $|Q_{M_x}|=|Q^{(2)}_0|=2$. 

Therefore, if a pair of DDPs can be split in some way, we always have $|Q^{(2)}_0+Q_{M_x}|=2$ and $Q^{(2)}_0Q_{M_x}=0$. Otherwise, we have $|Q^{(2)}_0|=|Q_{M_x}|=2$ if no splitting is allowed. This directly leads to the conclusion that
\begin{equation}
Q_{M_x} \equiv Q^{(2)}_0  \quad  {\rm mod} \ 2.
\end{equation}
Consequently, the set of independent topological charges in this symmetry class is given by
\begin{equation}
\{Q_{M_x}, Q_{M_x}-Q^{(2)}_0\} \in \mathbb{Z} \times 2\mathbb{Z}.
\end{equation}

\subsubsection{$D_{2, -, -}^{-}$}

In this case, we have
\begin{equation}
[\mathcal{S}, M_x]=[\mathcal{S}, M_y]=0, \quad  [C_2, \mathcal{S}]=0.
\end{equation}
This immediately implies that the rotation topological charge is trivial, and that both $M_x$ and $M_y$ can protect stable topological singularities. One can thus assign a pair of mirror topological charges: $\{Q_{M_x}, Q_{M_y}\}$ to characterize the current situation. Moreover, since $(C_2)^2=-1$, this corresponds to the $D_2$ double group.

To investigate the relation between $Q_{M_x}$ and $Q_{M_y}$, let us first notice that for a single DDP, $|Q_{M_x}|=|Q_{M_y}|=1$, since $Q_{M_x} = {\rm sgn}(v_y)$ and $Q_{M_y}={\rm sgn}(v_x)$ for a single DDP taking the form of Eq.~(\ref{eq:dirac}), as discussed in Sec. \ref{subsec:mirror}. For a pair of DDPs, using a similar rationale as in Fig.~\ref{fig:D2_obstruction}, we find that $Q_{M_x}$ ($Q_{M_y}$) is trivial if the two DDPs can be split symmetrically along the $k_x$ ($k_y$) axis. Following a similar argument in Sec. \ref{subsubsection:D_{2+-}}, we therefore conclude that
\begin{equation}
Q_{M_x} \equiv Q_{M_y} \quad {\rm mod} \ 2.
\end{equation}
Once again, upon redefining the topological charges as $\{Q_{M_x}, Q_{M_x}-Q_{M_y}\}$, we arrive at the following classification:
\begin{equation}
\{ Q_{M_x}, Q_{M_x}-Q_{M_y} \} \in \mathbb{Z} \times 2\mathbb{Z}.
\end{equation}

\subsection{Classification of $D_n$-protected topological singularities}
\label{subsec:classification of D_n singularity}

Based on the results for $D_2$ group, we now lay down the following rules for topological charges:
\begin{enumerate}
	\item For $D^{\pm}_{n,+,-}$, we always have 
	\begin{equation}
	Q^{(n)}_J \equiv Q_{M_x} \quad {\rm mod} \ 2,
	\end{equation}
	if $|Q^{(n)}_J|$ indicates the number of DDPs instead of DDP doublets. A similar rule holds for $D_{n,+,+}^{\pm}$ if we replace $M_x$ with $M_{\theta_n}$.
	\item For $D^-_{n,-,-}$, we always have
	\begin{equation}
	Q_{M_{\theta_n}} \equiv Q_{M_x}  \ {\rm mod} \ 2
	\label{eq:mirror charge mod 2}
	\end{equation}
\end{enumerate}

The first rule is based on the observation that a single DDP in class $D^{\pm}_{n,+,-}$ simultaneously carries rotation and mirror charges with $|Q_{M_x}|=|Q_J^{(n)}|=1$. This relation can also be understood through a similar argument in Sec. \ref{subsubsection:D_{2+-}}. Therefore, any phase-band singularities that are built from DDPs should always follow the first rule listed above.

The second rule can be intuitively understood for $D_{2,-,-}^-$, since $Q_{M_x}$ and $Q_{M_y}$ of a single DDP explicitly depend on the sign of $v_x$ and $v_y$, respectively. For $n>2$, one might naturally expect that $Q_{M_{\theta_n}}$ should exactly equal to $Q_{M_x}$, given that $C_n$ requires $v_x=v_y$. However, we will argue below that this naive expectation is not true. The simple underlying reason is that for DDPs, the representation of a mirror symmetry sensitively depends on the angular momentum basis.

To see this, we recall that a DDP protected by $D_{n,-,-}^-$ generally admits a phase-band description under the basis $\Psi_J=(|J\rangle, |J+1\rangle, |-J\rangle, |-J-1\rangle)^T$, as shown in Fig.~\ref{fig:doublet}(b). For convenience, we now label its rotation and mirror operations as $C_n(J)$ and $M_x(J)$, following the subscript of $\Psi_J$. In particular, $C_n(J)=\text{diag}(e^{i\frac{2\pi J}{n}},e^{i\frac{2\pi (J+1)}{n}},e^{-i\frac{2\pi J}{n}},e^{-i\frac{2\pi (J+1)}{n}})$. As pointed out in Sec. \ref{subsec:dirac point doublet}, a single DDP with $\Psi_J$ can exist in $D_{n,-,-}^-$ only when $J=-\frac{1}{2}$ or $J=\frac{n-1}{2}$. As we show in Appendix~\ref{app:mirror_angular momentum}, the corresponding mirror symmetries $M_x(-\frac{1}{2})$ and $M_x(\frac{n-1}{2})$ are related by
\begin{equation}
	M_x\left(\frac{n-1}{2}\right) = (-1)^{\frac{n}{2}} M_x\left(-\frac{1}{2}\right),
	\label{eq:mirror_angular momentum}
\end{equation}
On the other hand, since $M_{\theta_n} = C_n M_x$, we have
\begin{eqnarray}
M_{\theta_n}\left(\frac{n-1}{2}\right) &=& C_n\left(\frac{n-1}{2}\right) M_x\left(\frac{n-1}{2}\right) \nonumber \\
&=& -(-1)^{\frac{n}{2}} C_n\left(-\frac{1}{2}\right) M_x\left(-\frac{1}{2}\right)  \nonumber \\
&=& -(-1)^{\frac{n}{2}} M_{\theta_n}\left(-\frac{1}{2}\right),
\end{eqnarray}
where we have used the fact that $C_n(\frac{n-1}{2})=-C_n(-\frac{1}{2})$. This directly leads to
\begin{eqnarray}
M_x\left(\frac{n-1}{2}\right) &=& (-1)^{\frac{n}{2}} M_x\left(-\frac{1}{2}\right),  \nonumber \\
M_\theta\left(\frac{n-1}{2}\right)&=&-(-1)^{\frac{n}{2}} M_\theta \left(-\frac{1}{2}\right).
\label{eq:mirror relation}
\end{eqnarray}

Now let us consider two DDPs described by the same effective Hamiltonian~(\ref{eq:dirac}), under bases $\Psi_{-\frac{1}{2}}$ and $\Psi_{\frac{n-1}{2}}$ respectively. Suppose the DDP under $\Psi_{-\frac{1}{2}}$ carries $Q_{M_x}=Q_{M_{\theta_n}}=1$. Then the DDP under $\Psi_{\frac{n-1}{2}}$ will instead carry $Q_{M_x}=-Q_{M_{\theta_n}}=(-1)^{\frac{n}{2}}$, based on Eq.~(\ref{eq:mirror relation}).
This is because the sign of the mirror topological charge as defined in Eq.~(\ref{eq:mirror_charge}) relies on the explicit representation of the mirror symmetry. Namely, if we add a minus sign to the mirror representation, we also need to flip the corresponding mirror charge. This is why we have the ``modulo 2" relation for the mirror charges in Eq.~(\ref{eq:mirror charge mod 2}), instead of a naive equality.

Equipped with the above principles, it is quite straightforward to read off the classification for phase-band singularities from our list in Sec.~\ref{subsec:all possible D_n charges}. In particular, whenever two charges $Q$ and $Q'$ satisfy $Q\equiv Q'$ mod 2, we denote their contribution to the classification as $\{Q,Q-Q'\}\in\mathbb{Z}\times 2\mathbb{Z}$.
For example, class $D^-_{4,+,-}$ has a classification: $\{Q_{M_x}, Q_{M_x}-Q^{(4)}_{\frac{1}{2}}\} \in \mathbb{Z} \times 2\mathbb{Z}$, and class $D^+_{6,+,-}$ has a classification: $\{Q_{M_x}, Q_0^{(6)}, Q_{M_x}-Q_1^{(6)}\} \in \mathbb{Z}^2 \times 2\mathbb{Z}$. The reason we only include $Q^{(6)}_1$ in the subtraction is that $Q_0^{(6)}$ only accounts for DDP doublets with a trivial mirror charge. Similarly, the second rule requires that class $D^-_{4,-,-}$ has a classification: $\{Q_{M_x}, Q_{M_x}-Q_{M_{\frac{\pi}{4}}}\} \in \mathbb{Z} \times 2\mathbb{Z}$, and class $D^-_{6,-,-}$ likewise has a classification: $\{Q_{M_x}, Q_{M_x}-Q_{M_{\frac{\pi}{3}}}\} \in \mathbb{Z} \times 2\mathbb{Z}$. 

We are now finally in the position to summarize our full classification of topological singularities for all two-dimensional point groups. We summarize our results in Table~\ref{Table1}.

\section{Bulk-boundary correspondence \& higher-order topological indices}
\label{sec:reduction}

In this section, we establish a higher-order bulk-boundary correspondence, by relating the bulk phase-band singularities to symmetry-protected corner modes in systems with open boundary, which is a defining boundary signature of 2D AFHOTIs. Such a PBDR procedure extends the idea of dimensional reduction, which was previously developed for classifying static crystalline symmetry-protected topological states. As explained in Sec.~\ref{sec:preliminary_reduction}, the key step in our PBDR approach is to reduce the original 3D phase-band singularities to 2D singularities characterizing the end modes of the resultant 1D chains after the reduction. This is achieved by introducing globally symmetry-preserving mass domain walls for the original 3D phase-band singularities. The 2D phase-band singularities will thus manifest themselves as domain wall modes after PBDR. The end modes of the resulting 1D AFTI that survive will correspond to robust corner modes in the original 2D system. This allows us to establish the higher-order bulk-boundary correspondence for 2D AFHOTIs protected by point group symmetries.

In what follows, we shall start by classifying general $C_n$-protected AFHOTIs with rotation topological charges to illustrate our PBDR procedure. This also allows us to construct the corresponding higher-order topological indices. A similar analysis is then performed to $D_1$-protected AFHOTIs with a single mirror symmetry. Next, we proceed to consider $D_2$ group as a case study for the more complicated $D_n$ groups, where the interplay between rotation and mirror charges needs to be carefully disentangled in order to predict the correct boundary features and construct the topological indices.
	Finally, we provide a general discussion on the classification of AFHOTIs for all $D_n$ groups, as well as the required boundary geometry in order to observe the desired boundary features.

\subsection{Classification of $C_n$-protected AFHOTIs}
\label{subsection: Cn classification}

\begin{figure*}[t]
	\includegraphics[width=0.9\textwidth]{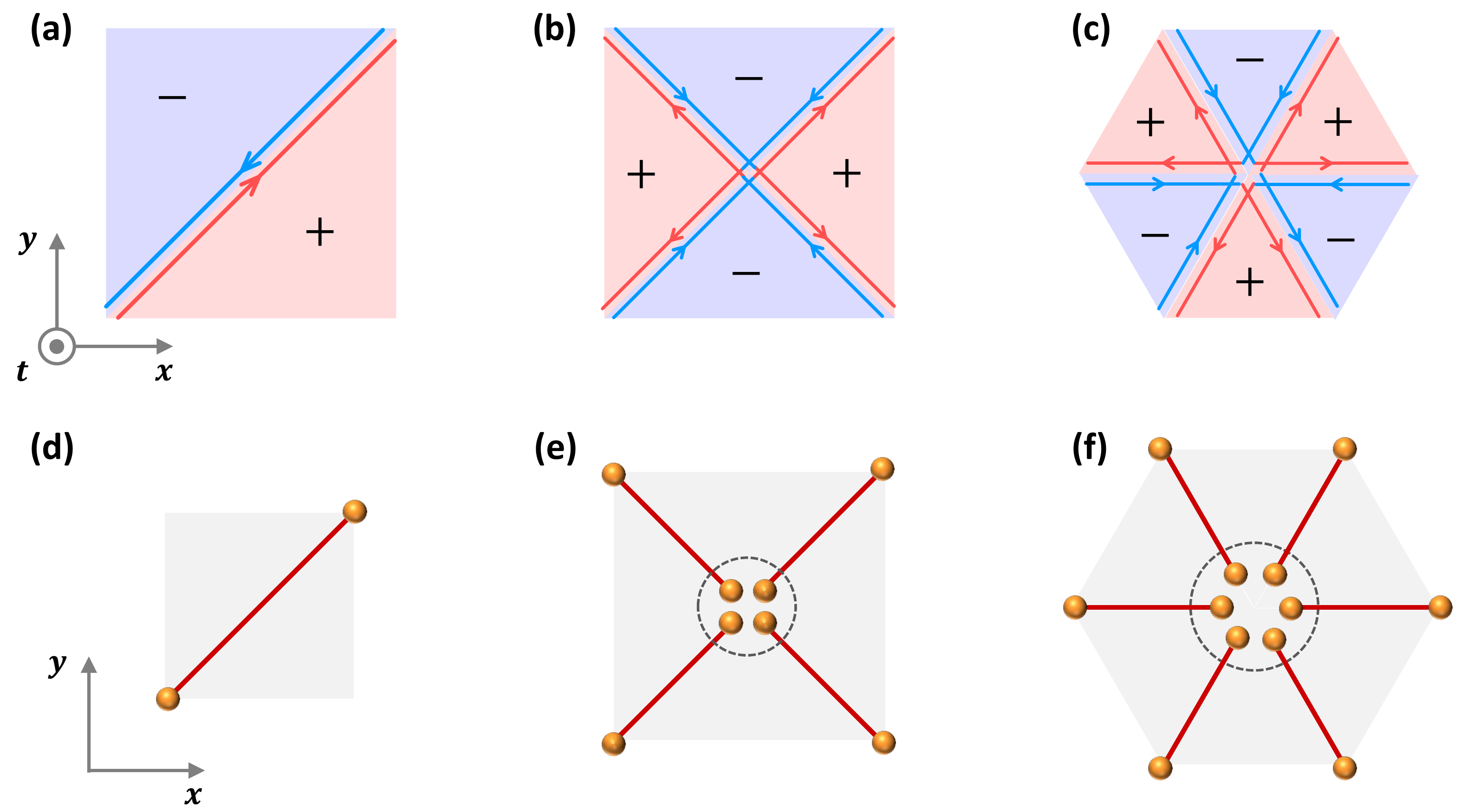}
	\caption{(a), (b), and (c) show the pattern of mass domain walls to achieve the PBDR for $C_2$, $C_4$, and $C_6$-symmetric systems, respectively. The ``$\pm$" labels the sign of the mass term $m(r,\theta)$ in Eq.~(\ref{Eq: mass domain wall_Cn}). The colored arrows along the mass domain walls represent the chiral-protected domain wall modes, which are 2D Dirac nodes. (d)-(f) Accordingly, the original 2D Floquet systems are reduced to a collection of symmetry related 1D AFTIs with end modes (depicted as golden dots). End modes residing in the bulk of the system encircled by black dashed lines can be adiabatically eliminated.}
	\label{Fig: reduction_Cn}
\end{figure*}

\subsubsection{Dimensional reduction}

As discussed in Sec.~\ref{sec:Cn}, the topological singularities protected by $C_n$ symmetry are DWPs carrying nontrivial rotation charges. We shall now exemplify our PBDR scheme by coupling a 3D DWP to a spatially-varying mass term that respects $C_n$ symmetry globally, and explicitly solving for the resultant 2D domain wall modes.

Let us recall that a single $C_n$-protected DWP  is described by the following Dirac Hamiltonian: 
\begin{equation}
h_d = v_x k_x \gamma_1 - v_y k_y\gamma_2 + v_t t \gamma_5,
\label{eq:dirac2}
\end{equation}
where the chiral symmetry ${\cal S}=\gamma_{45}$, $n$-fold rotation symmetry $C_n = e^{i\frac{2\pi}{n}{\mathcal J}_z}$ with ${\mathcal J}_z=\text{diag}(J,J+1,J+\frac{n}{2},J+\frac{n}{2}+1)$, and $v_x=v_y\equiv v_k$ is required for $n>2$. Under this basis, the rotation charge of this DWP is defined as $Q_J^{(n)} = -\text{sgn}(v_t)$. To gap out this DWP, the only ${\cal S}$-preserving constant mass term is $h_m = m \gamma_3$, which necessarily breaks rotation symmetry since $\{C_n, \gamma_3\} = 0$. Nevertheless, it is possible to add a spatially-varying mass term $h_m({\bf r})=m({\bf r}) \gamma_3$, such that $C_n$ symmetry is preserved globally. For example, we can take $m({\bf r})$ to have the following spatial dependence:
\begin{equation}
m(r, \theta) = m_0 \  \text{sgn} \left[\cos \left(\frac{n}{2}(\theta+\theta_0) \right) \right],
\label{Eq: mass domain wall_Cn}
\end{equation}      
where $\theta = \tan^{-1} (y/x)$ and $r=\sqrt{x^2+y^2}$ are the polar coordinates, and $\theta_0$ is an arbitrary constant parameter. In particular, $m(r, \theta)$ flips its sign under a $C_n$ operation: $m(r, \theta+\frac{2\pi}{n}) = -m(r, \theta)$, which directly cancels the additional sign flip from $\{C_n, \gamma_3\} = 0$ and leads to $[C_n, h_m({\bf r})]=0$. Thus, $h_m({\bf r})$ globally preserves $C_n$ symmetry [see Fig.~\ref{Fig: reduction_Cn}(a)-(c)].

However, the singularity in the phase band is not completely eliminated by $h_m({\bf r})$, since the mass term is forced to vanish when
\begin{equation}
\theta = \theta_m = \frac{(2m+1)\pi}{n}-\theta_0,\ \forall\ m\in \mathbb{Z}.
\end{equation}
From a different perspective, $h_m({\bf r})$ flips its sign across $\theta=\theta_m$, forming a mass domain wall for the DWP at every $\theta_m$, which necessarily binds 2D domain wall modes.

As an example, we consider a single mass domain wall at $\theta=0$ (i.e. parallel to $x$-axis) by choosing $\theta_0 = \pi/n$. In the presence of such a domain wall, the Hamiltonian for the DWP now becomes
\begin{equation}
\widetilde{h}_d = v_x k_x \gamma_1 - v_y k_y\gamma_2 + v_t t \gamma_5 + m_0 \text{sgn}(y) \gamma_3.
\end{equation}
The domain wall modes can be analytically solved from the zero-mode equation $\widetilde{h}_d |\psi\rangle = 0$ at $k_x= t =0$. Substituting $k_y \rightarrow -i \partial_y$, this becomes a differential equation:
\begin{equation}
\left[v_y \partial_y + m_0 \text{sgn}(y) \gamma_{23}\right]\psi = 0.
\end{equation}
Consider an ansatz solution $\psi = f(y) \xi_s$, where $\xi_s$ is an eigenstate of $\gamma_{23}$ with eigenvalue $s=\pm 1$: $\gamma_{23}\xi_s = s \xi_s$. Then, $f(y)$, the spatial part of $\psi$, can be easily solved as 
\begin{equation}
f(y) = {\cal N} e^{-s\frac{m_0}{v_y}|y|},
\end{equation}
where ${\cal N}$ is the normalization factor. To obtain a normalizable solution, we require $m_0v_ys>0$, or equivalently, $s = \text{sgn} (m_0 v_y)$. In other words, the choice of eigenstate $\xi_{\pm}$ is determined by the sign of $m_0 v_y$.

By projecting $\widetilde{h}_d$ onto the two eigenbases of $\xi_{s}$, we obtain the effective Hamiltonian for the domain wall modes
\begin{equation}
h_{\text{dw},s} = v_x k_x \sigma_z + v_t t \sigma_x,
\label{eq:domain}
\end{equation}
which describes a massless 2D Dirac node in $(k_x, t)$-space, as anticipated. The projected chiral symmetry is given by
\begin{equation}
\widetilde{{\cal S}}_{s} = s \sigma_z.
\end{equation}
This indicates that for whichever choice of $s=\pm 1$, the left and right movers of the domain wall mode~(\ref{eq:domain}) always carry opposite chiral eigenvalues.
Therefore, this domain wall mode is protected by chiral symmetry and carries a chiral topological charge
\begin{equation}
{\cal N_S} = \text{sgn}(m_0 v_x v_y).
\end{equation}
As discussed in Sec.~\ref{sec:ssh}, such a nonzero ${\cal N_S}$ for the chiral-protected 2D Dirac point is precisely the topological invariant for 1D AFTI in class AIII. This also implies that the resultant 1D chain must carry a single chiral-protected edge mode at each endpoint, when the original 2D system has a single DWP.
This result directly applies to other $C_n$-related domain walls, which indicates that a 2D class AIII Floquet system with $C_n$-protected phase-band singularities is adiabatically equivalent to $C_n$-symmetric arrays of 1D class AIII AFTIs.

\subsubsection{Bulk-boundary correspondence \& topological indices}

\begin{figure*}[t]
	\includegraphics[width=0.98\textwidth]{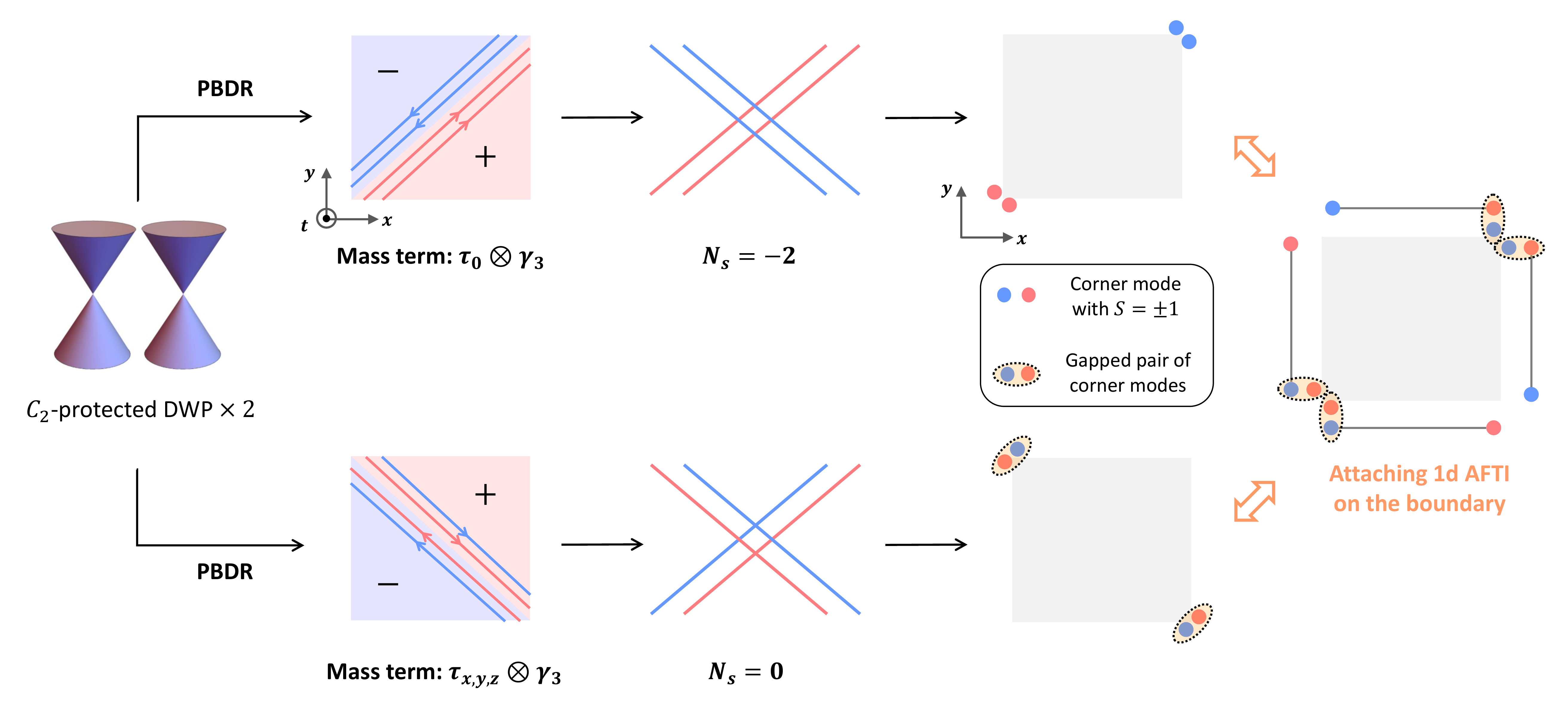}
	\caption{Starting from two $C_2$-protected DWPs, there exist two classes of mass terms in the PBDR procedure giving rise to 2D Dirac nodes carrying different chiral charges $\mathcal{N}_{\mathcal S}$. This further leads to two seemingly distinct corner mode patterns: zero or two corner modes at each corner. These two corner mode patterns correspond to the same bulk phase, and hence can be transformed to one another by symmetrically attaching 1D AFTIs on the boundary, without closing the bulk gap. 
		This demonstrates the $\mathbb{Z}_2$ classification of general $C_n$-protected AFHOTIs.}
	\label{Fig: C2_2Dirac}
\end{figure*}

Next, we shall study the number of robust end modes of the 1D chains after dimensional reduction using the chiral winding number $\mathcal{N}_{\mathcal S}$, and establish a bulk-boundary correspondence by relating the corner mode counting to the bulk $C_n$ topological charges. Fig.~\ref{Fig: reduction_Cn}(d)-(f) shows the resultant $C_n$-invariant arrays after dimensional reduction, along with their end modes. End modes that are close together in the bulk of the system can be adiabatically eliminated due to their finite spatial overlaps in the wavefunctions, whereas those that reside at $C_n$-related corners correspond to the corner modes of the original 2D systems. Since these corner modes are protected by a nonzero $\mathcal{N}_{\mathcal S}$ that arises from bulk singularities and are further distributed spatially in a $C_n$-symmetric way, they cannot be eliminated by any $C_n$ symmetric perturbations without closing the bulk Floquet gap. We thus conclude that a 2D Floquet system with $C_n$ and chiral symmetry must be an AFHOTI with $n$ $C_n$-related corner modes, if there is a single phase band DWP.

To examine the robustness of the corners modes when there are multiple DWPs in the system, we now proceed to consider the boundary physics in the presence of two DWPs. We shall work out an explicit example of $C_2$-protected AFHOTIs, and the arguments therein apply to $C_4$ and $C_6$ symmetries as well. Since the resultant 2D singularity no longer carries any information about the angular momentum $J$ of the rotation charge $Q^{(n)}_J$, such two DWPs can carry rotaion charges $Q^{(n)}_{J_1}$ and $Q^{(n)}_{J_2}$ with either $J_1=J_2$ or $J_1\neq J_2$, as long as they do not annihilate one another. Such a double-DWP can be described by the phase-band Hamiltonian
\begin{equation}
h_{d\times 2} = \mu_0 \otimes h_d,
\end{equation}
where $h_d$ follows the definition in Eq.~(\ref{eq:dirac2}), and $\mu_{0,x,y,z}$ are Pauli matrices for the valley degrees of freedom associated with the two DWPs. The symmetries for this double-DWP system are given by ${\cal S} = \mu_0\otimes \gamma_{45}$, and $C_2 = \mu_0\otimes \gamma_5$, accordingly.
We find four distinct $C_2$ and chiral symmetric mass terms for the DWPs:
\begin{equation}
h_{m\times 2} = m({\bf r})\ \mu_{0,x,y,z} \otimes \gamma_3 \equiv m({\bf r})M,
\end{equation}
where $m({\bf r})= m_0\ \text{sgn}[{\rm cos}(\theta+\theta_0)]$ describes a mass domain wall at $\theta=\frac{\pi}{2}-\theta_0$. However, as shown in Fig.~\ref{Fig: C2_2Dirac}, different choices of the mass matrix $M$ in $h_{m\times 2}$ above lead to distinct domain wall physics, which eventually leads to distinct corner-mode configurations:
\begin{enumerate}
	\item[(i)] $M=\mu_0\otimes \gamma_3$: The domain wall modes carry ${\cal N_S}=\pm 2$. The sign of ${\cal N_S}$ again depends on ${\rm sgn}(m_0 v_x v_y)$. The 2D system hosts two corner modes at each of the two $C_2$-related corners;
	\item[(ii)] $M=\mu_{x,y,z}\otimes \gamma_3$: The domain wall modes have ${\cal N_S}=0$ and thus can be gapped out. The 2D system has no protected corner mode. 
\end{enumerate}
Therefore, by two adiabatic deformations of the same Floquet system, we end up with scenarios with either zero or two corner modes. However, these two reduction schemes are both adiabatically connected to the same original 2D system, hence the two seemingly distinct boundary features must correspond to the same bulk phase. In other words, these two different boundary configurations can be converted from one to the other without closing the bulk gap. Indeed, as we demonstrate in Fig.~\ref{Fig: C2_2Dirac} explicitly, the two corner modes in case (i) above can be removed by symmetrically attaching 1D AFTIs on the boundary. This implies that the number of $C_2$-protected corner-modes is only well-defined modulo 2 at a single corner.
Similar arguments can be directly applied to $C_4$ and $C_6$ Floquet systems with double DWPs. Therefore, we have shown that
\begin{itemize}
	\item 2d $C_n$-protected AFHOTIs admit a $\mathbb{Z}_2$ topological classification.
\end{itemize}

The final step is to construct a topological index that counts the number of robust corner modes. As we have seen from the above examples, the chiral charge ${\mathcal N}_{\mathcal S}$ and hence the number of corner modes only cares about the number of net DWPs, but not the explicit values of their topological charges $Q_J^{(n)}$ or the angular momentum $J$. For example, a spinless $C_4$-symmetric system having three DWPs with $Q_0^{(4)}=-1$ and $Q_1^{(4)}=2$ has the same boundary feature as any other system with $(Q_0^{(4)},Q_1^{(4)})=(l_1,l_2)$, where $l_{1,2}\in \mathbb{Z}$ and $l_1+l_2 \equiv 1$ mod 2. Therefore, we define the following $\mathbb{Z}_2$ corner mode index for $C_n$-symmetric AFHOTIs as the higher-order topological index:
\begin{equation}
\nu_n \equiv \sum_J |Q_J^{(n)}| \ \ \ (\text{mod 2}),
\end{equation} 
The value of $\nu_n$ directly indicates the presence ($\nu_n=1$) or absence ($\nu_n=0$) of robust corner mode at each corner. We will provide more examples to illustrate the $\mathbb{Z}_2$ corner mode index in Sec.~\ref{sec:model}.  

\subsection{Classification of $D_1$-protected AFHOTIs}

In Sec.~\ref{subsec:mirror}, we have shown that the phase-band singularity protected by $M_x$ symmetry alone (i.e. $D_1$ group) is essentially a single Dirac node that can move freely along $k_y$ axis at $(k_x,t)=(0/\pi, \frac{T}{2})$. We show in this subsection how mirror-protected corner modes are related the mirror topological charge $Q_{M_x}$, via our PBDR analysis.

\begin{figure*}[t]
	\includegraphics[width=0.98\textwidth]{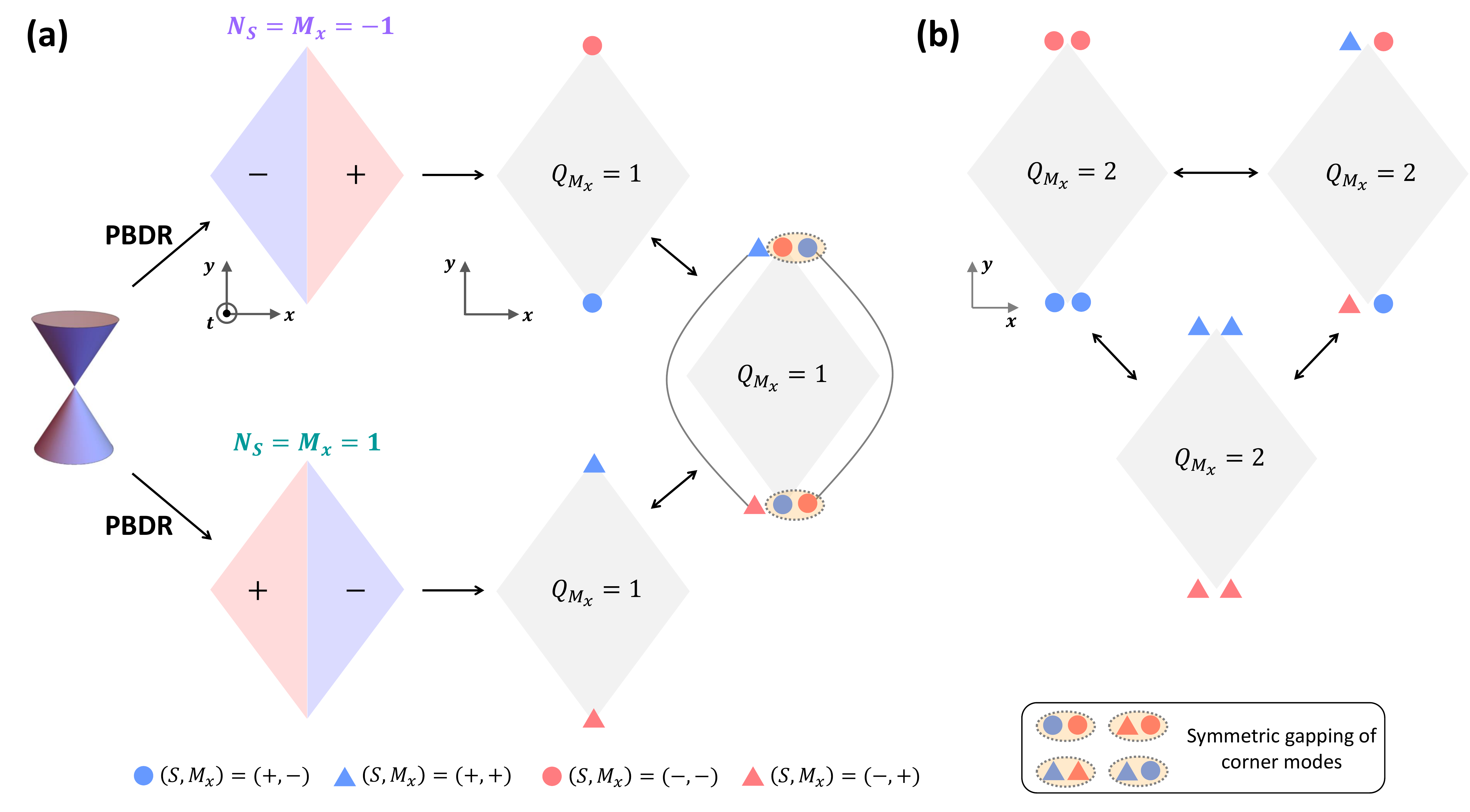}
	\caption{(a) Starting from a single DDP with $Q_{M_x}=1$, there exist two distinct corner mode patterns via PBDR (we take $v_xv_y>0$), which correspond to the same bulk phase and can be connected to one another by symmetrically attaching 1D AFTIs at the boundary. The colors of the corner modes indicate their chiral eigenvalues: ${\mathcal S} = +1$ (blue), and ${\mathcal S}=-1$ (red). The shapes of the corner modes indicate their mirror eigenvalues: $\widetilde{M}_x = +1$ (triangle), and $\widetilde{M}_x=-1$ (circle). Pairs of corner modes can annihilate one another if their net mirror charge $Q_{M_x}$ is zero. Since $Q_{M_x} = \mathcal{N}_{\mathcal S} \widetilde{M}_x$, there are four such possibilities, which we circle in dashed lines. (b) Three topologically equivalent corner mode configurations for $Q_{M_x}=2$, which can be transformed to one another by symmetrically attaching 1D AFTIs at the boundary.}
	\label{Fig: mirror_corner}
\end{figure*}

\subsubsection{Dimensional reduction}

Consider again the Dirac Hamiltonian~(\ref{eq:dirac2}) now describing a DDP protected by mirror symmetry. As in Sec.~\ref{subsec:mirror}, we take the chiral symmetry ${\cal S} = \gamma_{45}$ and $M_x = \gamma_{13}$, so that $[M_x, {\cal S}] =0$ is satisfied and $Q_{M_x}$ is nontrivial. Since the $({\cal S}, M_x)=(+,+)$ branch propagates along the $\text{sgn}(v_y)\hat{y}$ direction as shown in Fig.~\ref{fig:mirror_nodal}, we have $Q_{M_x} = \text{sgn}(v_y)$.

To perform the PBDR, we choose the following $M_x$-symmetric mass term:
\begin{equation}
h_m ({\bf r}) = m_0 \ \text{sgn}(x) \gamma_3,
\end{equation}
which creates a mass domain wall along the $y$-axis. At $k_y=t=0$, the zero-mode equation is given by
\begin{equation}
[v_x \partial_x - m_0 \text{sgn}(x) \gamma_{13}]\psi = 0.
\end{equation}
Consider an ansatz solution for the zero-mode wavefunction $\psi = f(x) \xi_s$, where $\xi_s$ is an eigenstate of $\gamma_{13}$ with eigenvalue $s=\pm 1$: $\gamma_{13}\xi_s = s \xi_s$. Then the spatial part can be solved as $f(x) = {\cal N} e^{s\frac{m_0}{v_x}|x|}$, where ${\cal N}$ is the normalization factor. The condition for a normalizable wavefunction is given by $m_0 v_x s <0$, which can be rewritten as
\begin{equation}
s=-\text{sgn}(m_0 v_x).
\end{equation}
Upon projecting onto the $\xi_s$ basis, we obtain the effective Hamiltonian for the domain wall mode
\begin{equation}
\widetilde{h}_{\text{dw},s} = -v_y k_y \sigma_z - v_t t \sigma_x,
\end{equation}
and the projected symmetries:
\begin{equation}
\widetilde{{\cal S}}_s = -s \sigma_z,\quad  \widetilde{M}_{x} = s \sigma_0.
\end{equation}
Then the chiral charge for the domain wall fermion is given by
\begin{equation}
{\cal N_S} = \text{sgn}(s v_y) = -{\rm sgn}(m_0 v_x v_y)= s Q_{M_x},
\end{equation}
However, a key difference from the previously discussed $C_n$ case is that, the domain wall mode here carries an additional mirror index. The projected mirror eigenvalue $\widetilde{M}_{x}$ satisfies
\begin{equation}
\widetilde{M}_{x} = s = {\cal N_S} Q_{M_x},
\label{Eq: mirror_BBC}
\end{equation}
which relates the bulk mirror charge to the topological indices of the domain wall mode. This relation is the key to understand the bulk-boundary correspondence for $D_1$-protected AFHOTIs.

\subsubsection{Bulk-boundary correspondence \& topological indices}

According to Eq.~(\ref{Eq: mirror_BBC}), starting from a $M_x$-protected DDP with $Q_{M_x}=1$, there are two phenomenologically distinct scenarios for the domain wall modes characterized by different $(\mathcal{N}_{\mathcal S}, \widetilde{M}_x)$ indices, namely, $(\mathcal{N}_{\mathcal S}, \widetilde{M}_x)=(+1,+1)$ or $(\mathcal{N}_{\mathcal S}, \widetilde{M}_x)=(-1,-1)$. Although both cases lead to a single corner mode located at each $M_x$-invariant corner, the corner modes in these two cases carry distinct $\mathcal S$ and  $\widetilde{M}_x$ indices, as shown in Fig.~\ref{Fig: mirror_corner}(a). However, since these two cases are both adiabatically connected to the same original 2D system, they correspond to the same bulk phase. As expected, these two corner mode configurations can be transformed from one to the other by symmetrically attaching 1D AFTIs on the boundary without closing the bulk gap, as shown in Fig.~\ref{Fig: mirror_corner}(a).   

Similarly, for two identical $M_x$-protected Dirac points (i.e. $Q_{M_x} = \pm 2$), there exist three topologically equivalent corner mode configurations that can be generated via the PBDR procedure, as shown in Fig.~\ref{Fig: mirror_corner}(b). These three configurations correspond to all possible combinations of corner mode configurations shown in Fig.~\ref{Fig: mirror_corner}(a) for a single DDP.
We emphasize that, unlike $C_n$-protected AFHOTIs, it is not possible to adiabatically eliminate the two $M_x$-protected corner modes when $Q_{M_x}=\pm 2$. Instead, attaching a pair of mirror-related 1D AFTIs to the system's boundary can only switch among the three tcorner mode configurations shown in Fig.~\ref{Fig: mirror_corner}(b).

We have shown that for both single and double DDPs, $|Q_{M_x}|$ directly indicates the number of robust corner modes at each mirror-invariant corner. Systems with $\pm Q_{M_x}$, although harboring the same number of corner modes, are topologically distinct. This is because their corner modes carry distinct ${\mathcal S}$ and $\widetilde{M}_x$ indices, and hence cannot be adiabatically connected to one another. Therefore, we conclude that 
\begin{itemize}
	\item $Q_{M_x}\in\mathbb{Z}$ is the topological index for 2D $D_1$-protected AFHOTIs.
\end{itemize}

\subsection{$D_2$-protected AFHOTIs}

\label{subsection:D2 topo class}

\begin{figure*}[t]
	\includegraphics[width=0.98\textwidth]{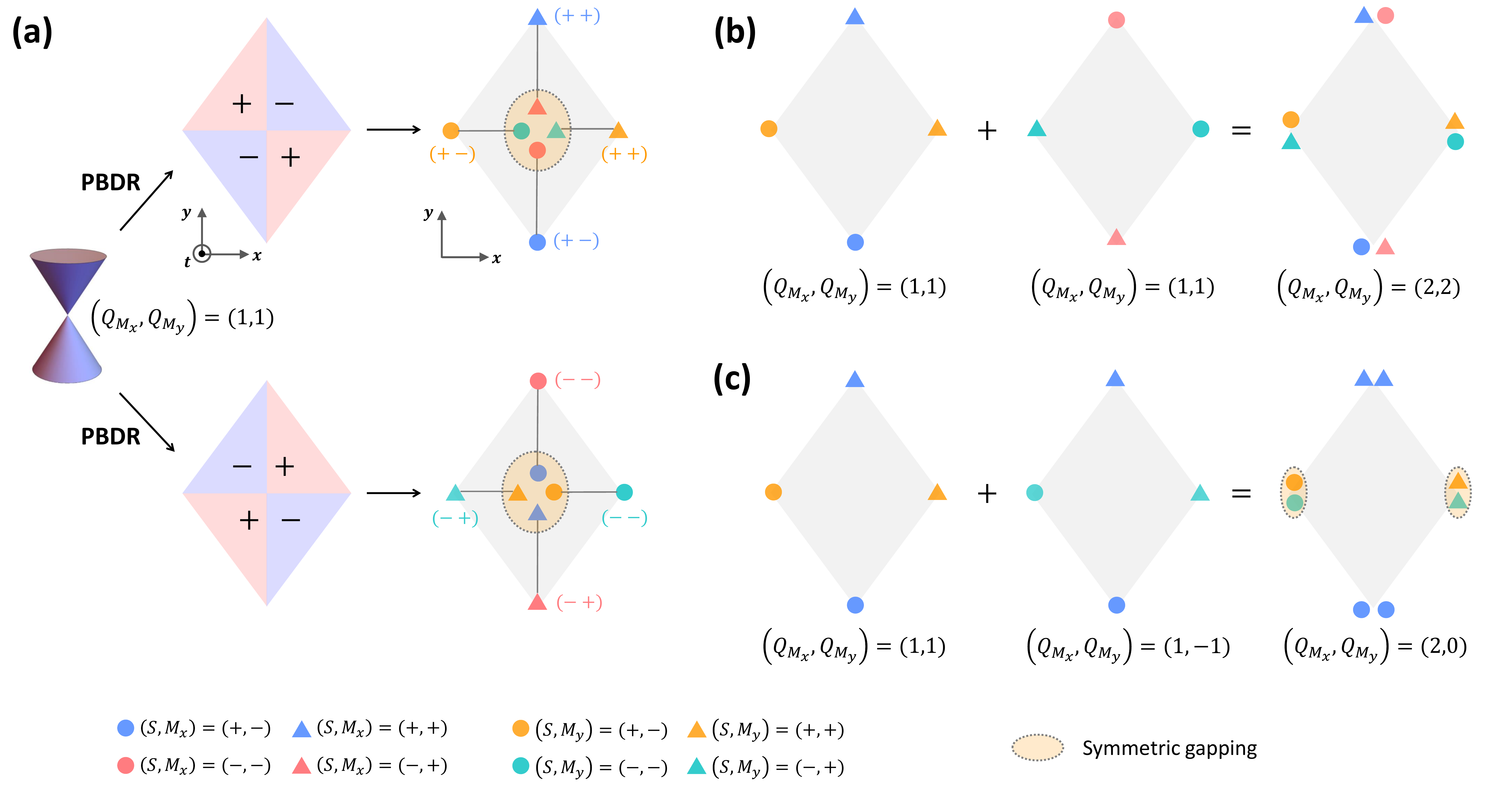}
	\caption{(a) Two possible domain wall and corner mode patterns generated from a $D_2$-protected DDP with $(Q_{M_x}, Q_{M_y})=(1,1)$ under PBDR, where we take $v_xv_y>0$ for definiteness. The chiral and mirror eigenvalues can be read off from the results shown in Fig.~\ref{Fig: mirror_corner}, and we use the same convention for the shape and color of each corner mode as in Fig.~\ref{Fig: mirror_corner}.
		(b) Coupling two systems with $(Q_{M_x}, Q_{M_y})=(1,1)$ and $(Q_{M_x}, Q_{M_y})=(1,1)$ leads to a net mirror charge of $(Q_{M_x}, Q_{M_y})=(2,2)$, which agrees with the presence of two stable corner modes at each $M_x$ or $M_y$-invariant corner. (c) Coupling two systems with $(Q_{M_x}, Q_{M_y})=(1,1)$ and $(Q_{M_x}, Q_{M_y})=(1,-1)$ leads to a net mirror charge of $(Q_{M_x}, Q_{M_y})=(2,0)$. In this case, two robust corner modes only show up at $M_x$-invariant corners, which is also consistent with the net mirror charges. Similarly to Fig.~\ref{Fig: mirror_corner}, two corner modes that are close together can be symmetrically gapped out only if they have the same color (chiral eigenvalues) but different shapes (mirror eigenvalues), or vice versa.} 
	\label{Fig:D2_2Dirac}
\end{figure*}

In this subsection, we shall use $D_2$-protected AFHOTIs as a case study to demonstrate the general principles for characterizing and classifying $D_n$-protected AFHOTIs. According to our previous discussions, there are two nontrivial classes with $D_2$ symmetry: $D_{2,+,\pm}^+$ and $D_{2,-,-}^-$, both having a $\mathbb{Z}\times 2\mathbb{Z}$ classified phase-band singularities.

\subsubsection{$D_{2,+,\pm}^+$} 

In Sec.~\ref{subsubsection:D_{2+-}}, we have shown that the phase-band singularities for $D_{2,+,-}^+$ are characterized by a pair of topological charges $\{Q_{M_x}, Q_0^{(2)}\}$ satisfying $Q_0^{(2)} \equiv Q_{M_x}$ (mod 2). Therefore, the bulk-boundary correspondence here can be directly understood based our results in Sec.~\ref{sec:Cn} and Sec.~\ref{subsec:mirror} for systems with $C_2$ and $M_x$ symmetries, respectively. 

First of all, {\it any corner mode at a generic, locally $M_x$-breaking corner can always be gapped out}. To see this, let us consider a system on a rhombus geometry with two $x$-corners and two $y$-corners. Assuming the existence of a corner mode with chiral eigenvalue ${\cal S}=1$ at the left $x$-corner, its mirror partner at the right $x$-corner must also carry ${\cal S}=1$, since $[M_x, {\cal S}]=0$. However, since $\{C_2,{\cal S}\}=0$, the $C_2$-partner of the left $x$-corner mode gives rise to another corner mode carrying ${\cal S}=-1$ at the right $x$-corner. Therefore, corner modes at every generic corner must come in pairs with opposite ${\cal S}$ eigenvalues, which can always be eliminated together.

As a result, the only robust corner modes must reside at $M_x$-invariant corners, which are solely accounted for by the mirror topological charge, instead of the rotation charges. We conclude that
\begin{itemize}
	\item the topological index for class $D_{2,+,-}^+$ AFHOTIs is $Q_{M_x}\in \mathbb{Z}$. 
\end{itemize}

The above conclusion can be understood in a formal yet simpler way. Symmetry protected corner modes can only arise from the following topological indices: the $C_2$ corner mode index $\nu_2\in \mathbb{Z}_2$ and the mirror charge $Q_{M_x}\in \mathbb{Z}$. Since $\nu_2\equiv |Q_0^{(2)}| \equiv Q_{M_x}$ (mod 2), the resulting higher-order topological classification is then
\begin{equation}
(\mathbb{Z}_2\times \mathbb{Z})/\mathbb{Z}_2 = \mathbb{Z}, 
\end{equation}  
which agrees with our previous physical argument. Notably, if we explicitly break $M_x$ while preserving $C_2$, the system is now characterized by the $C_2$ corner mode index $\nu_2$ alone. Therefore, if there were originally an odd number of corner modes before $M_x$ was broken, a single corner mode at each corner can still survive after breaking $M_x$. This corner mode can then be symmetrically shifted to generic $M_x$ breaking and $C_2$ related corners, and remains protected by $\nu_2$.

Similarly, class $D_{2,+,+}^+$ is equivalent to $D_{2,+,-}^+$ with $M_x$ replaced by $M_y$, and hence is also classified by $Q_{M_y} \in \mathbb{Z}$, hosting robust corner modes at $M_y$ invariant corners.

\subsubsection{$D_{2,-,-}^-$}

Class $D_{2,-,-}^-$ has two nontrivial mirror charges $\{Q_{M_x}, Q_{M_y}\}$ satisfying $Q_{M_x}\equiv Q_{M_y}$ (mod 2). Therefore, both mirrors can protect their own corner modes at the corresponding mirror-invariant corners. Based on our previous discussions, it is straightforward to conclude that
\begin{itemize}
	\item AFHOTIs belonging to class $D_{2,-,-}^-$ are characterized by $(Q_{M_x}, Q_{M_x}-Q_{M_y})\in \mathbb{Z}\times 2\mathbb{Z}$. 
\end{itemize}  
The physical meaning of the $\mathbb{Z}\times 2\mathbb{Z}$ classification is that {\it the number of corner modes at each $M_x$-invariant corner must equal to that at each $M_y$-invariant corner, modulo 2.} We shall now demonstrate this classification explicitly following the PBDR procedure.

In Fig.~\ref{Fig:D2_2Dirac}(a), we show the possible mass domain wall patterns for a single DDP described by Hamiltonian~(\ref{eq:dirac2}) with $(Q_{M_x}, Q_{M_y})=(1, 1)$. For example, we again take $\mathcal{S}=\gamma_{45}$, $M_x=\gamma_{13}$, and $M_y=\gamma_{23}$, consistent with $D_{2,-,-}^-$. Then, a globally mirror symmetric mass domain wall can be generated by coupling to a mass term $h_m ({\bf r}) = m_0 \ \text{sgn}(xy) \gamma_3$. We obtain two topologically equivalent corner mode configurations (with $m_0>0$ and $m_0<0$), hosting one mode at each corner with different $\mathcal{S}$, $M_x / M_y$ eigenvalues. To further demonstrate that the number of modes at each $M_x$-invariant corner is equal to that at each $M_y$-invariant corner modulo two, in Fig.~\ref{Fig:D2_2Dirac}(b)\&(c) we stack our original system $(Q_{M_x}, Q_{M_y})=(1,1)$ with another system with $(Q_{M_x}, Q_{M_y})=(1,\pm 1)$ respectively. Indeed, we arrive at two different corner mode configurations with two modes at each $M_x$-invariant corner, which corresponds to $Q_{M_x}=2$. The number of corner modes at each $M_y$-invariant corner is either two, indicating $Q_{M_y}=2$; or zero, indicating $Q_{M_y}=0$.
Therefore, we have demonstrated the $\mathbb{Z}\times 2\mathbb{Z}$ classification for class $D_{2,-,-}^-$ AFHOTIs with two inequivalent mirror charges.

\subsection{Classifying general $D_n$-protected AFHOTIs}

\subsubsection{Guidelines for classifying $D_n$-protected AFHOTIs}
Now we are ready to summarize the general rules for classifying higher-order topology as well as establishing the bulk-boundary correspondence for all $D_n$-protected AFHOTIs:

\begin{enumerate}
	\item[(i)] If a $C_n$ topological charge $Q_J^{(n)}$ leads to DDP doublets, it does not contribute to $C_n$-protected corner modes;
	
	\item[(ii)] If a $C_n$ topological charge $Q_J^{(n)}$ and a mirror topological charge $Q_M$ coexist and satisfy
	\begin{equation}
	Q_J^{(n)}\equiv Q_M \ \ (\text{mod }2),
	\end{equation}
	they together contribute $\mathbb{Z}$ to the classification, which is solely determined by the value of $Q_M$;
	
	\item[(iii)] If two inequivalent mirror topological charges $Q_M$ and $Q_{M'}$ coexist and satisfy
	\begin{equation}
	Q_M\equiv Q_{M'} \ \ (\text{mod }2),
	\end{equation}
	they together contribute $\mathbb{Z}\times 2\mathbb{Z}$ to the topological classification.
\end{enumerate}
Let us briefly explain the above rules. Based on our discussions of $C_n$-symmetric AFHOTIs in Sec.~\ref{subsection: Cn classification}, it is straightforward to see that a DDP doublet will always lead to an even number of corner modes at each corner, which is trivial due to the $\mathbb{Z}_2$ classification.
For the second rule, one can use the argument in Sec.~\ref{subsection:D2 topo class} to show that the net contribution from $\{Q_J^{(n)}$ and $Q_{M_x}$ to the higher-order topology is $(\mathbb{Z}\times \mathbb{Z}_2)/\mathbb{Z}_2 = \mathbb{Z}$. The third rule directly follows from our results for $D^-_{2,-,-}$ in Sec.~\ref{subsection:D2 topo class}.

\begin{figure*}[t]
	\includegraphics[width=0.98\textwidth]{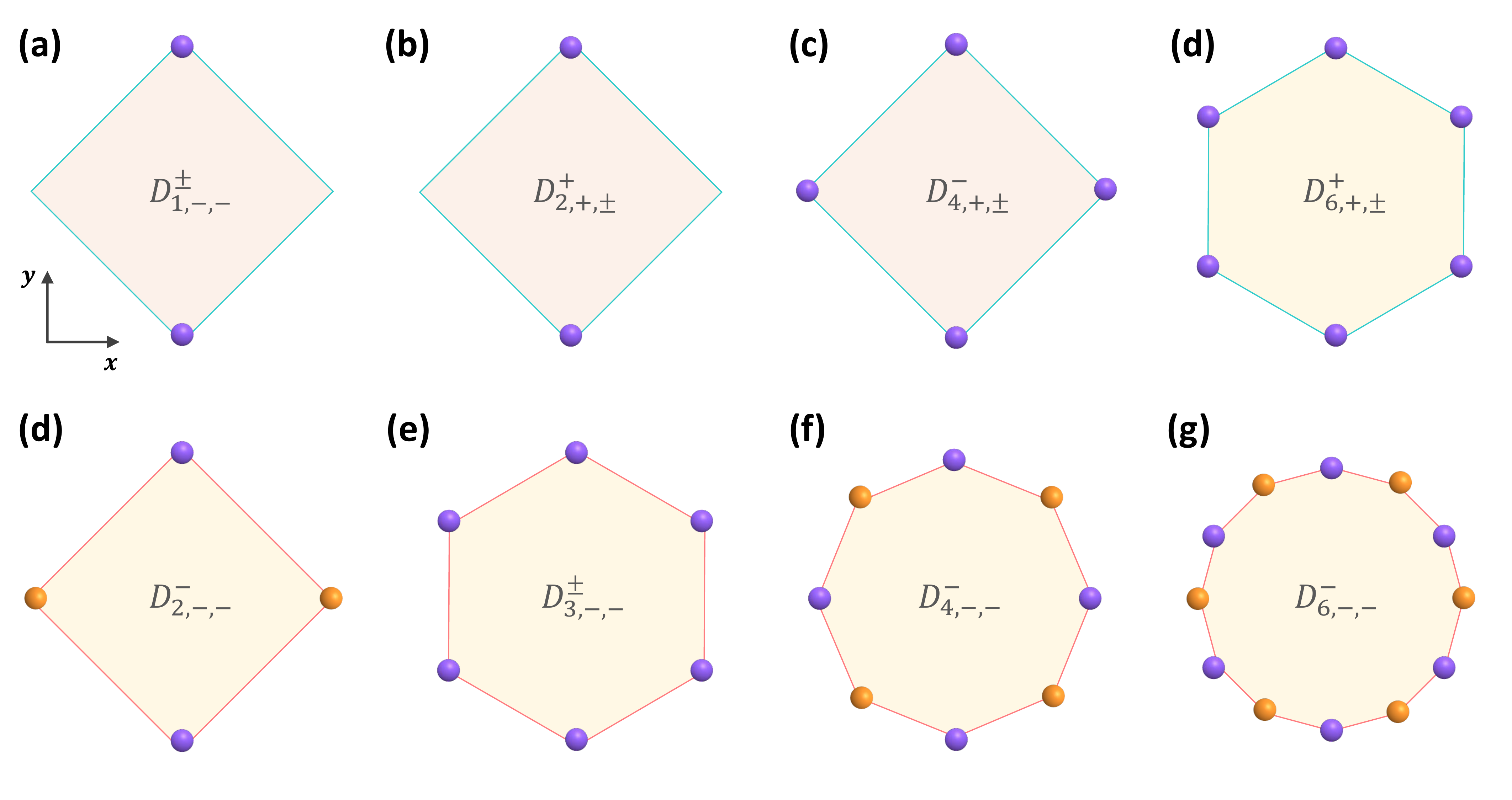}
	\caption{Schematics of corner mode patterns for all nontrivial $D_n$ classes on appropriately chosen boundary geometries. Corner modes are denoted by spheres, and the different colors denote corner modes protected by two inequivalent mirrors $M_x$ and $M'$ respectively.}
	\label{Fig: Dn_corner}
\end{figure*}

\subsubsection{Higher-order topological indices for $D_n$-protected AFHOTIs}

Combining the above general rules  and the classfication of phase-band singularities in Sec.~\ref{sec:Dn}, we are now in place to list our classification results for all $D_n$-protected AFHOTIs, along with the corresponding topological indices as follows:
\begin{itemize}
	\item $D_1$ group: $D^{\pm}_{1,-,-}$: $Q_{M_x} \in \mathbb{Z}$;
	\item $D_2$ group:
	\begin{enumerate}
		\item $D^+_{2,+,\pm}$: $Q_{M_x} \in \mathbb{Z}$;
		\item $D^-_{2,+,\pm}$: trivial;
		\item $D^+_{2,-,-}$: trivial;
		\item $D^-_{2,-,-}$: $\{Q_{M_x}, Q_{M_x}-Q_{M_y}\}\in \mathbb{Z}\times 2\mathbb{Z}$;
	\end{enumerate}
	
	\item $D_3$ group: $D^{\pm}_{3,-,-}$: $Q_{M_x} \in \mathbb{Z}$;
	
	\item $D_4$ group:
	\begin{enumerate}
		\item $D^+_{4,+,\pm}$: trivial;
		\item $D^-_{4,+,\pm}$: $Q_{M_x}\in \mathbb{Z}$;
		\item $D^+_{4,-,-}$: trivial;
		\item $D^-_{4,-,-}$: $\{Q_{M_x}, Q_{M_x}-Q_{M_{\frac{\pi}{4}}}\}\in \mathbb{Z}\times 2\mathbb{Z}$;
	\end{enumerate}
	
	\item $D_6$ group:
	\begin{enumerate}
		\item $D^+_{6,+,\pm}$: $Q_{M_x}\in \mathbb{Z}$;
		\item $D^-_{6,+,\pm}$: trivial;
		\item $D^+_{6,-,-}$: trivial;
		\item $D^-_{6,-,-}$: $\{Q_{M_x}, Q_{M_x}-Q_{M_{\frac{\pi}{3}}}\}\in \mathbb{Z}\times 2\mathbb{Z}$.
	\end{enumerate}
\end{itemize}
Here we have ignored classes $D^{\pm}_{n,-,+}$, which do not have any robust topological singularities.

Let us make a few remarks on the above  results. 
First of all, despite having nontrivial rotation charges, $D^{+}_{4,+,\pm}$ and $D^-_{6,+,\pm}$ admit a trivial topological classification since their rotation topological charges always lead to DDP doublets. Secondly, the higher-order topological classification for $D^+_{6,+,\pm}$ is reduced to $\mathbb{Z}$, even though the corresponding topological singularities are characterized by $\{Q^{(6)}_0, Q_{M_x}, Q_{M_x}-Q^{(6)}_1\}\in \mathbb{Z}^2\times 2\mathbb{Z}$. This is simply because (i) $Q^{(6)}_0$ always leads to DDP doublets; (ii) $Q^{(6)}_1 \equiv Q_{M_x}$ (mod 2). Then based on the second rule listed above, the only topological  index in this case is $D^+_{6,+,\pm}$ is $Q_{M_x}\in \mathbb{Z}$. 

We summarize the full  classification of $D_n$-protected AFHOTIs in Table~\ref{Table2}.

\subsubsection{Boundary geometry for observing all symmetry-protected corner modes}

Since all nontrivial $D_n$-protected AFHOTIs are characterized solely by the mirror topological charges, we conclude that robust corner modes can only live on the mirror-invariant corners of a $D_n$-protected AFHOTI. Indeed, following a similar argument for $D_2$ symmetry in Sec.~\ref{subsection:D2 topo class}, one can easily show that corner modes appearing on a generic mirror-breaking corner always come in pairs, and thus are unstable for a general $D_n$ group.

As a result, a $\mathbb{Z}$-classified $D_n$-protected AFHOTI will host robust corner modes on its $M_x$-invariant corners, as well as all other corners that are related to the $M_x$-symmetric corners via $C_n$ rotations. According to our classification in Table \ref{Table2}, this includes classes $D_{1,-,-}^{\pm}$, $D^+_{2,+,\pm}$, $D^{\pm}_{3,-,-}$, $D^-_{4,+,\pm}$, and $D^+_{6,+,\pm}$ in $D_n$ groups. On the other hand, if the corresponding AFHOTIs admit a $\mathbb{Z}\times 2\mathbb{Z}$ classification, corner modes can appear on geometric corners that are invariant under either $M_x$ or $M'=C_nM_x$ symmetry, as well as all other corners that are $C_n$-related. AFHOTIs protected by $D^-_{2,-,-}$, $D^-_{4,-,-}$, and $D^-_{6,-,-}$ fall into this class. Therefore, in order to observe all symmetry protected corner modes, the boundary of the system must be properly chosen so as to possess the appropriate mirror-invariant corners. Below we summarize the geometry needed for each class of $D_n$-protected AFHOTI in order to observe all boundary features:
\begin{itemize}
	\item kite: $D_{1,-,-}^{\pm}$;
	\item rhombus: $D^+_{2,+,\pm}$, $D^-_{2,-,-}$;
	\item square: $D^-_{4,+,\pm}$;
	\item regular hexagon: $D^{\pm}_{3,-,-}$, $D^+_{6,+,\pm}$;
	\item regular octagon: $D^-_{4,-,-}$;
	\item regular dodecagon: $D^-_{6,-,-}$.
\end{itemize} 
One can, of course, choose a geometry with a higher symmetry than listed above to observe the desired corner modes. In Fig.~\ref{Fig: Dn_corner}, we schematically show the boundary geometry listed above for each class of $D_n$-protected AFHOTIs. 

This completes our higher-order topological classification and bulk-boundary correspondence analysis for $D_n$-protected AFHOTIs.

\section{Model realizations and experimental detections}
\label{sec:model}

As a demonstration of our classification scheme, in this section we present two concrete lattice models for $C_2$ and $D_4$-protected AFHOTIs, respectively. We shall analyze the symmetry and topological properties for both models, plot their phase-band spectra to reveal the singularity physics, and explicitly show how the anomalous corner modes are indicated by the corresponding higher-order topological indices defined in Sec. \ref{sec:reduction}. Finally, we will briefly discuss realizations of these AFHOTI models in various experimental platforms and further provide a proposal on accessing the higher-order topological indices in experiment. 

\begin{figure*}[t]
	\includegraphics[width=0.98\textwidth]{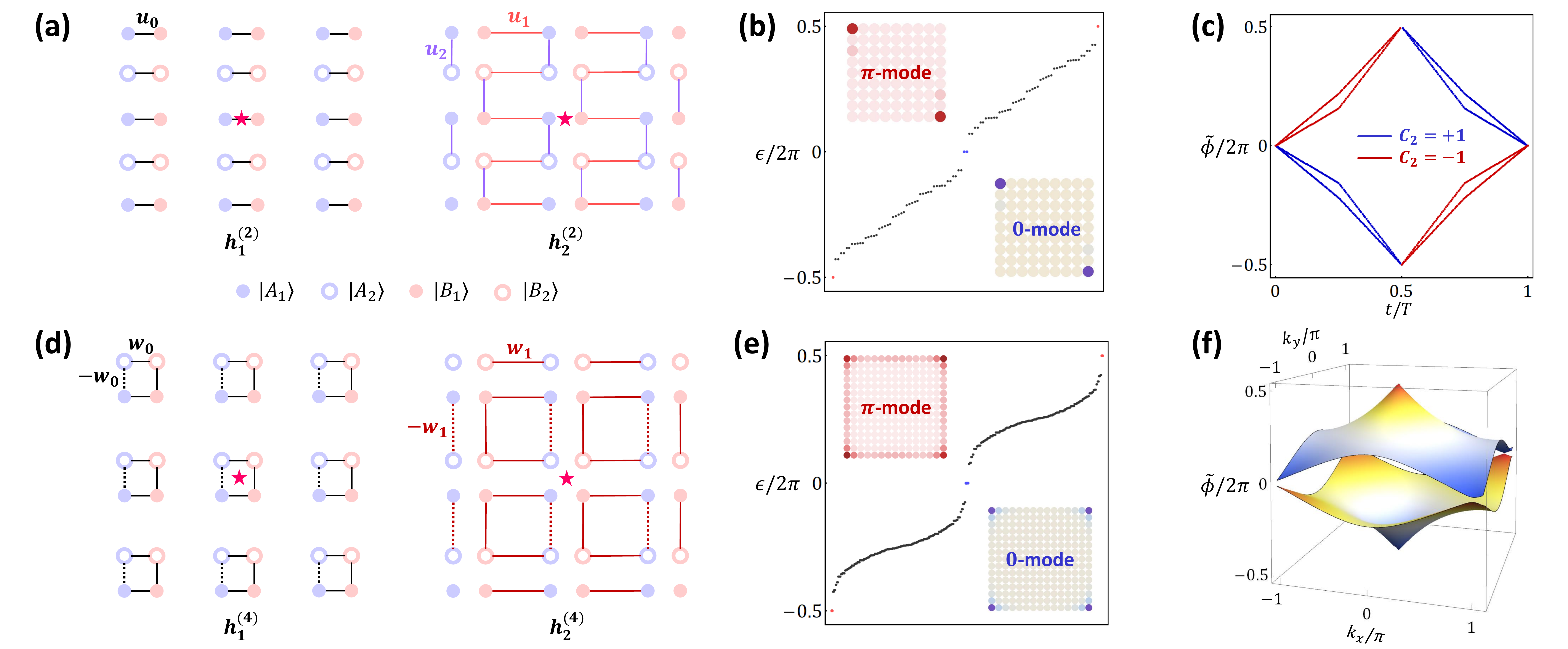}
	\caption{Schematics of $H^{(2)}$ and $H^{(4)}$ are shown in (a) and (d), where the red stars label the rotation center. Anomalous Floquet corner modes and phase band with $C_2$-protected [$D_4$-protected] DWP for $H^{(2)}$ [$H^{(4)}$] are shown in (b) and (c) [(e) and (f)], respectively.}
	\label{Fig3}
\end{figure*} 

\subsection{$C_2$-symmetric AFHOTI} 
Our first model of $C_2$-protected AFHOTI starts from a coupled-chain construction of the 1D AFTI model in Sec.~\ref{subsec:SSH model}, which is further decorated with a set of interchain dimer bonds $u_2$, as shown in Fig.~\ref{Fig3}(a). Each unit cell now contains four inequivalent atoms, labeled by $A_1, A_2, B_1$ and $B_2$. The time-dependent Hamiltonian is given by:
\begin{widetext}
\begin{equation}
H^{(2)}({\bf k},t) =
\begin{cases}
u_0\sum_{{\bf r},i=1,2} c^{\dagger}_{{\bf r}, A_i} c_{{\bf r}, B_i} + \text{h.c.}   &   0<t\leq \frac{T}{4} \quad \text{and} \quad \frac{3T}{4}<t\leq T; \\
\\
u_1\sum_{{\bf r}, i=1,2} c^{\dagger}_{{\bf r + a}_x, A_i} c_{{\bf r}, B_i} + u_2\sum_{{\bf r}} (c^{\dagger}_{{\bf r}, B_1} c_{{\bf r}, B_2} + c^{\dagger}_{{\bf r + a}_y, A_1} c_{{\bf r}, A_2}) + \text{h.c.}  &  \frac{T}{4} < t \leq \frac{3T}{4},
\end{cases}
\end{equation}
\end{widetext}
In Appendix~\ref{app:c2}, we provide the matrix representation of $H^{(2)}({\bf k}, t)$ in momentum space. By construction, $H^{(2)}({\bf k}, t)$ respects both $C_2$ and ${\cal S}$, where
\begin{equation}
	{\cal S} = \begin{pmatrix}
	\sigma_z & 0 \\
	0 & -\sigma_z \\
	\end{pmatrix},\ \ 
	C_2 = \begin{pmatrix}
	\sigma_x & 0 \\
	0 & \sigma_x e^{i k_y}
	\end{pmatrix},
\end{equation}
under the basis $\Phi_2 = (|A_1\rangle, |B_1\rangle, |A_2\rangle, |B_2\rangle)^T$. Note that $(C_2)^2=1$ and the anti-commutation relation $\{C_2, {\cal S}\}=0$ is satisfied. This model thus corresponds to class $C_{2,+}^{+}$ in Table~\ref{Table2}, which has a $\mathbb{Z}_2$ classification.

As discussed in Appendix~\ref{app:c2}, $H^{(2)}({\bf k} ,t)$ has a rather rich phase diagram with the phase boundary determined by 
\begin{equation}
u_0 \pm u_1 + u_2=\frac{2m\pi}{T}, \ \text{and} \ u_0 \pm u_1 - u_2=\frac{2m\pi}{T}, \ m \in \mathbb{Z}. 
\end{equation}
In Fig.~\ref{Fig3}(b), we calculate the Floquet quasienergy spectrum with open boundary in both spatial directions 
for $u_1=2u_0=4u_2=2\pi/T$. As shown in Fig.~\ref{Fig3}(b), there exists a pair of robust corner-localized $0$ and $\pi$ modes. To show that these corner modes are protected by irremovable DWPs, we plot the phase band of the return map $\widetilde{U}({\bf k}, t)$ along $\Gamma$-axis in Fig.~\ref{Fig3}(c). Indeed, we find a DWP in the phase band featuring the anomalous $C_2$-velocity locking effect discussed above and hence is topologically stable. Based on the relation between $C_2$ eigenvalues and $t$-directional velocities, it is easy to find that the $C_2$ corner-mode index $\nu_2 = |Q_{0}^{(2)}| = 1$, in agreement with the number of corner modes. Therefore, the corner modes in our model is indeed $C_2$-protected, and the corresponding higher-order bulk-boundary correspondence also holds. \\

\subsection{$D_4$-symmetric AFHOTI} 

We now give another example of AFHOTI with $D_4$ symmetry. Notably, this model was first proposed in Ref.~\cite{huang2020floquet} as a Floquet quadrupole insulator, similar to that in Ref. \cite{hu2020dynamical}, while the key role of $D_4$ and ${\cal S}$ in protecting bulk higher-order topology was overlooked. Here we revisit this model and fully demonstrate its AFHOTI physics based on our theoretical framework. 

The $D_4$-symmetric AFHOTI model is described by the following time-dependent Hamiltonian
\begin{equation}
H^{(4)}({\bf k},t) =
\begin{cases}
h_1^{(4)}  &   0<t\leq \frac{T}{4} \quad \text{and} \quad \frac{3T}{4}<t\leq T; \\
\\
h_2^{(4)}  &  \frac{T}{4} < t \leq \frac{3T}{4}.
\end{cases}
\end{equation}
Each unit cell for $H^{(4)}$ contains four atoms labeled by $A_{1,2}$ and $B_{1,2}$, following $H^{(2)}$'s convention. Under the basis $\Phi_4 = (|A_1\rangle, |B_1\rangle, |A_2\rangle, |B_2\rangle)^T$, the momentum-space Hamiltonian matrices are 
\begin{eqnarray}
h_1^{(4)} &=& w_0 (\gamma_{45} + \gamma_{14}), \nonumber  \\
h_2^{(4)} &=& w_1(\cos k_x \gamma_{45} + \cos k_y \gamma_{14} - \sin k_x \gamma_4 - \sin k_y \gamma_{24}).  \nonumber \\
\end{eqnarray}
The definition of the $\gamma$ matrices is the same as in Eq.~(\ref{eq:dirac}).
$H^{(4)}$ respects both $D_4$ and chiral symmetries, where
\begin{equation}
	C_4=\begin{pmatrix}
	0 & \sigma_0 \\
	-i \sigma_y & 0 \\
	\end{pmatrix},\ \ M_x=\gamma_3,\ \ {\cal S} = \gamma_{34}.
\end{equation}
Here $(C_4)^4=-1$ and $\{{\cal S}, C_4\}=0$ is again satisfied. Since $\{M_x, \mathcal{S}\}=0$, this model corresponds to class $D_{4,+, +}^{-}$ in Table~\ref{Table2}, which has a $\mathbb{Z}$ classification.

When $w_0=\frac{2}{3}w_1=\frac{\pi}{\sqrt{2}T}$, $H^{(4)}$ hosts a pair of zero and $\pi$ modes at each corner of the system, as shown in Fig.~\ref{Fig3}(e). We further scan the phase band of the return map $\widetilde{U}({\bf k}, t)$ within one period and identify the existence of a DDP located at the $\Gamma$-axis, as shown in Fig.~\ref{Fig3}(f). The $D_4$-symmetric model is analytically tractable, and we derive the effective Hamiltonian of this DDP in appendix~\ref{app:d4}. The effective Hamiltonian has a block-diagonal form up to ${\cal O}(k^2)$, corresponding to the two Weyl points that form this DDP,
\begin{eqnarray}
	H_\text{eff} &=& \begin{pmatrix}
	v_t t & e^{i\frac{\pi}{4}}v_k k_+ & 0 & 0 \\
	e^{-i\frac{\pi}{4}}v_k^* k_- & -v_t t & 0 & 0\\
	0 & 0 & v_t t & e^{i\frac{\pi}{4}} v_k k_+ \\
	0 & 0 & e^{-i\frac{\pi}{4}} v_k^* k_- & -v_t t \\
	\end{pmatrix}, \nonumber \\
\end{eqnarray} 
where $v_t = \frac{w_1 T}{\sqrt{2}} $, $v_l = (-i/\sqrt{2}) e^{i[\frac{w_0 T}{2\sqrt{2}}]} \sin \frac{w_0T}{2\sqrt{2}}$. We find that the $j_z$ eigenvalues of the four bands form the basis $\Psi=\left(|\frac{1}{2}\rangle, |\frac{3}{2}\rangle, |-\frac{1}{2}\rangle, |-\frac{3}{2}\rangle \right)^T$, corresponding to the previous $\Psi_{d,+}$ with $n=4$ and $J=\frac{1}{2}$. Indeed, such a DDP is stable, and can be analytically shown to carry a $C_4$ topological charge $Q_{\frac{1}{2}}^{(4)} = -\text{sgn}(v_t)$. The mirror topological charge can only be defined for $M_{\frac{\pi}{4}} = C_4 M_x$, since $[M_{\frac{\pi}{4}}, {\cal S}]=\{M_x,{\cal S}\}=0$, where we numerically find $|Q_{M_{\frac{\pi}{4}}}|=1$ for our choice of parameters, in agreement with the corner mode configuration depicted in Fig.~\ref{Fig3}(e). This unambiguously demonstrate $D_4$-protected anomalous Floquet higher-order topology in this system. Furthermore, if $M_x$ is broken, the higher-order topology is still $C_4$-protected, as the DDP might split into a DWP yielding $|Q_{\frac{1}{2}}^{(4)}| = 1$.

\subsection{Experimental realization and detection of phase-band singularities}

Besides serving as a proof of concept, both models can be feasibly realized in various state-of-the-art experimental platforms, such as cold atoms \cite{Wintersperger2020}, photonic waveguides \cite{mukherjee2017experimental,maczewsky2017observation,mukherjee2018state,mukherjee2020observation}, ring resonators \cite{adzal2020realization}, and acoustic platforms \cite{peng2016experimental}, where their first-order cousin (i.e. 2D AFTIs) has been successfully demonstrated. In particular, we also note a recent experiment on Floquet SSH model with anomalous $\pi$ mode in photonic waveguides \cite{cheng2019observation}, which, upon some slight modifications, can be directly coupled to achieve our $C_2$ AFHOTI phase.

Besides detecting the corner modes at both quasienergies 0 and $\pi$, one can instead prove the higher-order topological nature by directly measuring the phase-band singularities in the bulk. To achieve this, we propose to design a path shown in Fig.~\ref{Fig:phase_transition} that connects an AFHOTI phase to the high frequency limit by slowly increasing the driving frequency $\omega$, until $\omega$ is much larger than the band width of the static Hamiltonian. During this process, there exists a topological critical point that closes the $\pi$ quasienergy gap and further separates the AFHOTI phase from a static limit, which is exactly our proposed phase-band singularity. Namely, by simply tuning the system to its high frequency limit, one can directly extract our proposed higher-order topological index by measuring the properties of the topological critical point in the Floquet spectrum. It should be noted that a similar measurement of the monopole charge for Weyl singularities has already been achieved for 2D class A AFTIs in a cold atom system \cite{Wintersperger2020}. Therefore, the boundary signal of corner modes, together with the bulk measurement of the higher-order topological index, will demonstrate the anomalous Floquet higher-order topological physics in an unambiguous way.

\section{Discussions} 
\label{sec:discussion}
To conclude, we have established a long-sought theoretical framework to diagnose and classify anomalous Floquet higher-order topological phenomena, which is completely beyond any known topological diagnostic for static crystalline systems. Our approach is based on the concept of symmetry-protected phase-band singularities, which manifest themselves as exotic Dirac/Weyl singularities featuring unconventional dispersion relations that can only occur on the surface of higher-dimensional TCIs. To prove the higher-order bulk-boundary correspondence between phase-band singularities and Floquet corner modes, we develop the method of phase-band dimensional reduction, which provides a direct mapping between 2D AFHOTIs to 1D AFTIs. With the 1D AFTIs as the topological building blocks, we manage to extract the higher-order topological indices for all point-group symmetry protected 2D AFHOTIs, which thus accomplishes both the topological classification and characterization.

We mention that there are still several intriguing open questions for Floque topological matters, waiting to be explored within our proposed framework. For example, we have limited our discussion in the present work to ``strong" topological phenomena. In principle, a finer topological classification with translation symmetries can be achieved similarly, where we expect the weak anomalous Floquet topological phases to be indicated by line-node-like phase-band singularities. Featuring flat-band-like 0 and $\pi$ edge modes, an example of such weak AFTI is simply a ``trivial" stacking of 1D AFTIs, which is already contained in the phase diagram of our $C_2$-AFHOTI model (see Appendix \ref{app:c2} for details).

On the other hand, it is straightforward to apply our framework to classify anomalous Floquet higher-order topological phenomena in other tenfold-way symmetry classes. Nonetheless, switching from the chiral symmetry ${\cal S}$ to other internal symmetries (e.g. particle-hole symmetry) could lead to completely distinct topological classification and boundary physics. For example, we have demonstrated that ${\cal S}$ strictly prohibits any anomalous chiral edge modes, since the net phase-band Weyl charge must vanish. In contrast, such a constraint is relaxed when we consider driven systems with particle-hole symmetry. Namely, a 2D driven superconductor could be 1st-order topological with anomalous chiral Majorana edge modes, or 2nd-order topological with anomalous corner Majorana modes at both quasienergies 0 and $\pi$ \cite{Vu2020floquet}.       

Beyond spatial crystalline symmetries, 2D AFHOTIs can also be protected by nonsymmorphic space-time crystalline symmetries \cite{morimoto2017floquet,peng2019floquet,peng2020floquet}, e.g. a time-glide mirror that combines both spatial mirror reflection and a half-period translation along the time direction. Note that such a time-glide AFHOTI has been proposed to admit a winding-number-based topological index, similar to a 1D class AIII AFTI discussed in Sec.~\ref{sec:ssh}. Therefore, the topological winding number for time-glide AFHOTIs is also expected to indicate the existence of certain topological singularities in the phase bands, which fits well into our classification framework.       

The present classification scheme is also readily generalized to 3D AFHOTIs with either (i) anomalous chiral/helical hinge modes traversing the Floquet band gap or (ii) robust $0$ and $\pi$ corner modes. The phase bands now live in a four-dimensional $({\bf k}, t)$ space, and hence exhibits new types of phase-band topological singularities that are related to the surface states of 5D TCIs.

Last but not least, it is also desirable to reformulate our proposed higher-order topological indices into more accessible topological invariants for $\widetilde{U}$, perhaps similar to those homotopy invariants for 1D and 2D AFTIs. We leave this interesting topic to future works \cite{Yu2020dynamical}.

\section*{Acknowledgments}
We thank Sheng-Jie Huang, DinhDuy Vu, Yi-Ting Hsu, and in particular S. Das Sarma and Jiabin Yu for helpful discussions. We also thank Biao Huang and Haiping Hu for useful exchanges regarding Ref.~\cite{hu2020dynamical}. R.-X. Z. acknowledges support from a JQI Postdoctoral Fellowship. Z.-C. Y. is supported by MURI ONR N00014-20-1-2325, MURI AFOSR, FA9550-19-1-0399, and Simons Foundation. Z.-C. Y. also acknowledges support from the NSF PFCQC program.

\appendix

\section{$C_2$-protected 4D topological crystalline insulators}
\label{app:4dTCI}

In this section, we present a general theory of $C_2$-protected 4D TCIs that are characterized by a new 4D topological invariant, the rotation Chern number. We will construct the corresponding minimal model, define the rotation Chern number, and further solve analytically for the 3D topological surface states. We will show that the surface states of such a $C_2$-protected 4D TCI are precisely Weyl nodes exhibiting the $C_2$-velocity locking as in the $C_2$-protected DWPs.

\subsection{Minimal model and rotation Chern number}

We start by presenting a minimal model of $C_2$-protected 4D TCI, following the construction scheme in Ref. \cite{zhang2016topological}. Consider the following $\gamma$ matrices
\begin{eqnarray}
\gamma_1 &=& \tau_x \otimes \sigma_z,\ \gamma_2 = \tau_y \otimes \sigma_z, \ \gamma_3 = \tau_0 \otimes \sigma_x, \nonumber \\
\gamma_4 &=& \tau_0 \otimes \sigma_y, \ \gamma_5 = \tau_z \otimes \sigma_z,
\end{eqnarray}
with which all other $\gamma$ matrices can be generated as $\gamma_{ij} = [\gamma_i, \gamma_j]/(2i)$.
Then for a general 4D Dirac system with $h= \sum_{i=1}^5 d_i \gamma_i$, its second Chern number is given by \cite{qi2008topological}
\begin{equation}
{\cal C}^{(2)} = \frac{3}{8 \pi ^2} \int d^4 k \epsilon^{abcde} \hat{d}_a \partial_{k_x} \hat{d}_b \partial_{k_y} \hat{d}_c \partial_{k_z} \hat{d}_d \partial_{k_w} \hat{d}_e,
\end{equation}
which is defined in the 4D momentum space $(k_x, k_y, k_z, k_w)$. Here $\hat{d}_i = d_i/\sqrt{\sum_i d_i^2}$. Our minimal model consists of two 4D Chern insulators with opposite ${\cal C}^{(2)}$ values, which can be achieved by the following Hamiltonian, 
\begin{equation}
h_\text{TCI}({\bf k}) = \begin{pmatrix}
h_+({\bf k}) & 0 \\
0 & h_-({\bf k}) \\
\end{pmatrix},
\end{equation}
where
$h_{\pm}({\bf k}) = v [\sin k_x \gamma_1 \pm \sin k_y \gamma_2 + \sin k_z \gamma_3 + \sin k_w \gamma_4] + m({\bf k}) \gamma_5$. 
In particular, $m({\bf k}) = A_0 + A_1 [\cos k_x + \cos k_y + \cos k_z + \cos k_w]$ and the matrix block $h_{\pm}$ hosts $C^{(2)}=\pm 1$ when $-4A_1 < A_0 < -2A_1$.

We now consider a two-fold rotation symmetry $C_2=s_0\otimes \gamma_{12}$ that transforms $h_\text{TCI}$ as,
\begin{eqnarray}
C_2 h_\text{TCI} (k_x, k_y, k_z, k_w) C_2^{-1} = h_\text{TCI} (-k_x, -k_y, k_z, k_w) \nonumber \\
\end{eqnarray}
On the $C_2$-invariant plane with $k_{x,y}=0$ spanned by $(k_z, k_w)$, $h_\text{TCI}$ becomes four decoupled $2\times 2$ matrix blocks,
\begin{equation}
h_\text{TCI} = \bigoplus_{i=1}^4 h_i,
\end{equation} 
Here, we have
\begin{eqnarray}
h_1 &=& h_3 = v (\sin k_z \sigma_x + \sin k_w \sigma_y) + \tilde{m}(k_z, k_w)\sigma_z, \nonumber \\
h_2 &=& h_4 = v (\sin k_z \sigma_x + \sin k_w \sigma_y) - \tilde{m}(k_z, k_w)\sigma_z,
\end{eqnarray}
with $\tilde{m}(k_z, k_w) = A_0+2A_1 + A_1(\cos k_z + \cos k_w)$. Therefore, when $-4A_1 < A_0 < -2A_1$, we find $h_{1,3}$ has a first Chern number ${\cal C}^{(1)}=1$, while the first Chern number for $h_{2,4}$ is $-1$. According to the representation of $C_2$, $h_{1,3}$ and $h_{2,4}$ live in the $C_2=+1$ and $C_2=-1$ subspaces, respectively. Although the net ${\cal C}^{(1)}$ is zero for all $h_i$s, following Ref.~\cite{zhang2016topological}, we can define a new $C_2$-protected topological invariant dubbed the rotation Chern number ${\cal C_R}$, where
\begin{equation}
{\cal C_R} \equiv \frac{{\cal C}^{(1)} (C_2=1) - {\cal C}^{(1)} (C_2=-1)}{2} \in \mathbb{Z}.
\end{equation} 
For $h_\text{TCI}$ with $-4A_1 < A_0 < -2A_1$, the rotation Chern number ${\cal C_R}=1$.

\subsection{Surface state and $C_2$-velocity locking}

The nontrivial value of ${\cal C_R}$ directly implies the $C_2$-protected 4D crystalline topology in $h_\text{TCI}$, despite a vanishing second Chern number. To explore the corresponding topological boundary modes for this class of 4D TCI, we consider a 4D slab with periodic boundary conditions in $x,y,z$-directions but an open boundary condition along the $w$-direction. Expanding $h_\text{TCI}$ around $k_x=k_y=k_z=0$ to ${\cal O}(k)$, we have 
\begin{equation}
h_\pm = v (k_x \gamma_1 \pm k_y \gamma_2 + k_z \gamma_3) -iv\partial_w \gamma_4 + M_0 \gamma_5,
\end{equation}  
where $M_0 = A_0 + 4 A_1 >0$. Then both $h_\pm$ admit the same zero-mode equation at $k_x=k_y=k_z=0$,
\begin{equation}
(-iv \partial_w \gamma_4 + M_0 \gamma_5) \psi(w) = 0,
\end{equation}
which can be rewritten as
\begin{equation}
( \partial_w - \frac{M_0}{v} \gamma_{45})\psi(w) = 0.
\end{equation}
We now consider a trial wavefunction
\begin{equation}
\psi(w) = f(w) \xi_s,
\end{equation}
where the spatial part $f(w) = {\cal N} e^{-\lambda w}$ and the spinor part $\gamma_{45}\xi_s = s\xi_s$ for $s=\pm$. It is easy to solve that $\lambda = -\frac{m_0 s}{v}$, which is required to be positive for a normalized wavefunction. Then for $v>0$, we have $s=-1$ and thus have two choices for $\xi_s$,
\begin{equation}
\xi_-^{(1)} = \frac{1}{\sqrt{2}}(0,0,1,1)^T, \ \ \xi_-^{(2)} = \frac{1}{\sqrt{2}}(-1,1,0,0)^T.
\end{equation}
By projecting $h_{\pm}$ onto the zero-mode basis, we obtain the effective Hamiltonians for the corresponding surface state
\begin{equation}
h_\text{surface} = v\begin{pmatrix}
k_z & -k_+ & 0 & 0 \\
-k_- & -k_z & 0 & 0 \\
0 & 0 & k_z & -k_- \\
0 & 0 & -k_+ & -k_z \\
\end{pmatrix},
\label{Eq: TCI surface}
\end{equation}
while the projected $C_2$ symmetry is 
\begin{equation}
\widetilde{C}_2 = \begin{pmatrix}
-1 & & & \\
& 1 & & \\
& & -1 &  \\
& & & 1 \\
\end{pmatrix}. 
\label{Eq: TCI surface C2}
\end{equation}
From Eq.~(\ref{Eq: TCI surface}) and Eq.~(\ref{Eq: TCI surface C2}), it is easy to see that 
\begin{itemize}
	\item the surface state consists of two 3D Weyl nodes with opposite monopole charges;
	\item the surface state features $C_2$-velocity locking effect along the $C_2$-invariant axis ($k_x=k_y=0$) in the surface Brillouin zone.
\end{itemize}
Therefore, the surface state of a 4D $C_2$-protected TCI with ${\cal C_R}=1$ exactly reproduces the key features of the dynamical Weyl pair found in a $C_2$-protected 2D AFHOTI.

\section{Topological charges of $D_n$ single groups with $(C_n)^n=1$}
\label{app:dn}

In this section, we derive the topological charges for $D_n$ single groups with $(C_n)^n=1$. As discussed in the main text, $M_x$ maps a phase band eigenstate with $j_z=J$ to another eigenstate at the same location with $j_z=-J$. When $J$ takes integer values, however, not all possible $J$ leads to a two-dimensional irrep. In particular, when $-J = J$ mod $n$, the corresponding eigenstate remains to be a one-dimensional irrep. of $C_n$. Below we list the irreps for each $D_n$ single group:
\begin{itemize}
\item $D_2$: only 1-d irreps for $J=0, 1$;

\item $D_4$: two 1-d irreps for $J=0, 2$, and one 2-d irrep $\{ J=1, J=3 \}$;

\item $D_6$: two 1-d irreps for $J=0, 3$, and two 2-d irreps $\{ J=1, J=5 \}$, $\{ J=2, J=4\}$.
\end{itemize}

First, the condition for an unstable Dirac point: $J+1 = -J$ mod $n$, which leads to $Q^{(n)}_{\frac{n-1}{2}}=0$, places no constraint for integer-valued $J$'s. Second, since $D_2$ shares the same irreps as $C_2$, mirror symmetry does not introduce further constraints in this case. So the classification for $D_2$ single group is still $Q^{(2)}_0 \in \mathbb{Z}$. For $D_4$, there is only one 2-d irrep, so there will be a three-fold degenerate Dirac point formed by the phase band states $( |J=1\rangle, |J=2\rangle, |J=3\rangle )^T$. This leads to the constraint $Q^{(4)}_0 = Q^{(4)}_1$, which reduces the classification to $Q^{(4)}_0 \in \mathbb{Z}$. For $D_6$, there are two 2-d irreps, which form a four-fold degenerate Dirac point under the basis $(|J=1\rangle, |J=2\rangle, |J=4\rangle, |J=5 \rangle )^T$. This leads to the constraint $Q^{(6)}_1 = -Q^{(6)}_4$ which is identical to that from the chiral symmetry. However, it is also possible to form a three-fold degenerate Dirac point under $(|J=0\rangle, |J=1\rangle, |J=5\rangle)^T$, which imposes $Q^{(6)}_0 = -Q^{(6)}_5 = Q^{(6)}_2$. Therefore, the classification for $D_6$ single group reduces to $\{ Q^{(6)}_0, Q^{(6)}_1\} \in \mathbb{Z}^2$. We summarize the results below:

\begin{itemize}
\item $D_2$: $Q^{(2)}_0 \in \mathbb{Z}$;

\item $D_4$: $Q^{(4)}_0 \in \mathbb{Z}$;

\item $D_6$: $\{ Q^{(6)}_0, Q^{(6)}_1 \} \in \mathbb{Z}^2$.
\end{itemize}

\section{Angular momentum basis and mirror symmetry}
\label{app:mirror_angular momentum}
We now prove the relation between $M_x(\frac{n-1}{2})$ and $M_x(-\frac{1}{2})$ in Eq.~(\ref{eq:mirror_angular momentum}) of Sec.~\ref{subsec:classification of D_n singularity}.

Recall that a single DDP compatible with $D^-_{n,-,-}$ is described by the basis $\Psi_J=(|J\rangle,|J+1\rangle,|-J\rangle,|-J-1\rangle)^T$, where $J=-\frac{1}{2}$ or $J=\frac{n-1}{2}$. In particular, an even $n$ requires $\frac{n-1}{2}$ to be a half-integer, which agrees with the double group nature of $D^-_{n,-,-}$.

For spinful fermions, an angular momentum state can be formally written as  
\begin{equation}
	|j_z\rangle = |l_z, s_z\rangle
	\label{eq:jz}
\end{equation} 
where $l_z\in\mathbb{Z}$ and $s_z=\uparrow,\downarrow$ denote the orbital and spin contributions to $j_z$. As a result, a state $|J\rangle$ can come from the following two microscopic possibilities:
\begin{equation}
	|J\rangle = |J-\frac{1}{2},\uparrow\rangle,\text{ or } |J\rangle = |J+\frac{1}{2},\downarrow\rangle,
	\label{eq:J_microscopic}
\end{equation}
according to Eq.~\ref{eq:jz}. We note $M_x$ flips the spin as a $\pi$ rotation around $y$ axis, and further flips $l_z$ with an additional factor of $(-1)^{l_z}$. Therefore, we have
\begin{equation}
	M_x \begin{pmatrix}
	|l_z,\uparrow\rangle \\
	|-l_z,\downarrow\rangle \\
	\end{pmatrix} = (-1)^{l_z}i\sigma_y \begin{pmatrix}
	|l_z,\uparrow\rangle \\
	|-l_z,\downarrow\rangle \\
	\end{pmatrix}.
\end{equation}
Consequently, we find that for $|J\rangle$, 
\begin{equation}
		M_x \begin{pmatrix}
	|J\rangle \\
	|-J\rangle \\
	\end{pmatrix} = (-1)^{J-\frac{1}{2}}i\sigma_y \begin{pmatrix}
	|J\rangle \\
	|-J\rangle \\
	\end{pmatrix}
\end{equation}
always holds, in spite of the choices in Eq.~\ref{eq:J_microscopic}. Then it is easy to show that 
\begin{eqnarray}
	M_x(-\frac{1}{2}) &=& -i\tau_z\otimes \sigma_y, \nonumber \\
	M_x(\frac{n-1}{2}) &=& -i (-1)^{\frac{n}{2}}\tau_z\otimes \sigma_y,
\end{eqnarray} 
leading to
\begin{equation}
	M_x(\frac{n-1}{2}) = (-1)^{\frac{n}{2}} M_x(-\frac{1}{2})
\end{equation}
as shown in Eq.~(\ref{eq:mirror_angular momentum}).

\section{Fragility of dynamical Weyl quadrupole in Ref. \cite{hu2020dynamical}}
\label{app:huang}

We will show in this section that the “dynamical Weyl quadrupole" introduced in Ref.~\cite{hu2020dynamical} as a characterization for the anomalous quadrupole phase is in fact unstable. The model studied in Ref.~\cite{hu2020dynamical} is similar to the $D_4$-symmetric model discussed in the main text as well as in Ref.~\cite{huang2020floquet}, but only has $D_2$ symmetry. The time dependent Hamiltonian of Ref.~\cite{hu2020dynamical} is given by:
\begin{eqnarray}
&&H(t) = h_0,  \ t \in T_1;  \quad  \quad  H(t) = h_y, \ t \in T_2; \nonumber  \\
&&H(t) = h_x,  \ t \in T_3; \quad  \quad  H(t) = h_0, \ t\in T_4;
\label{eq:D2}
\end{eqnarray}
where $T_s = [(s-1)T/4, sT/4)$. The Hamiltonian in each time duration is give by: $h_0 = t_0(\tau_1\sigma_0+ \tau_2\sigma_2)$, $h_x = t_x({\rm cos}\ k_x \tau_1\sigma_0 - {\rm sin}\ k_x \tau_2 \sigma_3)$, and $h_y = t_y ({\rm cos}\ k_y \tau_2\sigma_2 + {\rm sin} \ k_y \tau_2 \sigma_1)$, where $\tau$, $\sigma$ are Pauli matrices, and the choice of basis is the same as in the $D_4$ symmetric model in the main text. The important symmetries for our purpose are:
\begin{equation}
M_x = i \tau_1 \sigma_3,      \quad     M_y = i \tau_1\sigma_1.
\end{equation}
Despite the similarity with the $D_4$ symmetric model discussed in the main text, this model does not have $C_4$ rotation symmetry, nor does it have chiral symmetry. Ref.~\cite{hu2020dynamical} claims that the phase band Weyl nodes form a quadrupole pattern protected by $M_x$ and $M_y$, and defines the total Weyl charge within the first quadrant of the Brillouin zone mod 2 as a topological invariant for the corner mode at quasienergy $\pi/T$.
However, since there is no chiral symmetry in this model, mirror symmetry alone cannot protect dynamical singularities, according to our discussions in Sec.~\ref{subsec:mirror}.
Therefore, one can adiabatically move the quadruplet of Weyl nodes altogether such that they coincide at one point in $({\bf k}, t)$ space, and gap them out with symmetry-preserving perturbations. We now show this explicitly for this model.

\begin{figure}
	(a)
	\includegraphics[width=0.43 \textwidth]{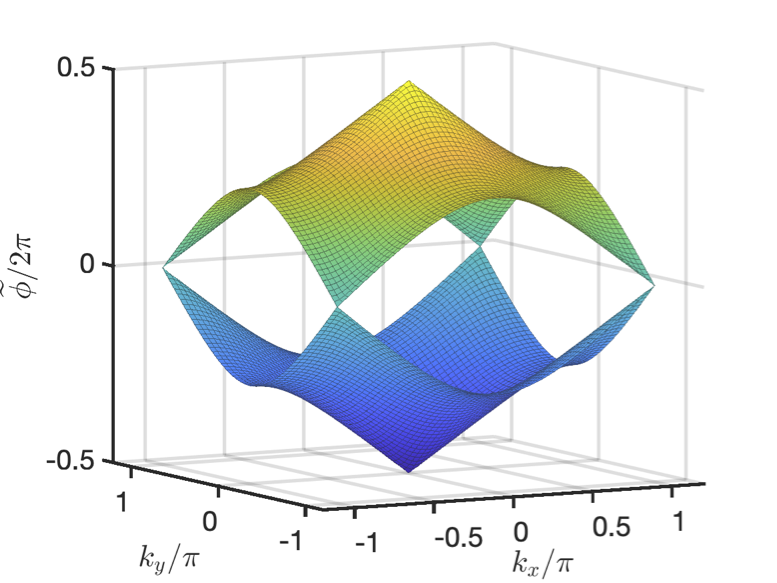}
	(b)
	\includegraphics[width=0.43 \textwidth]{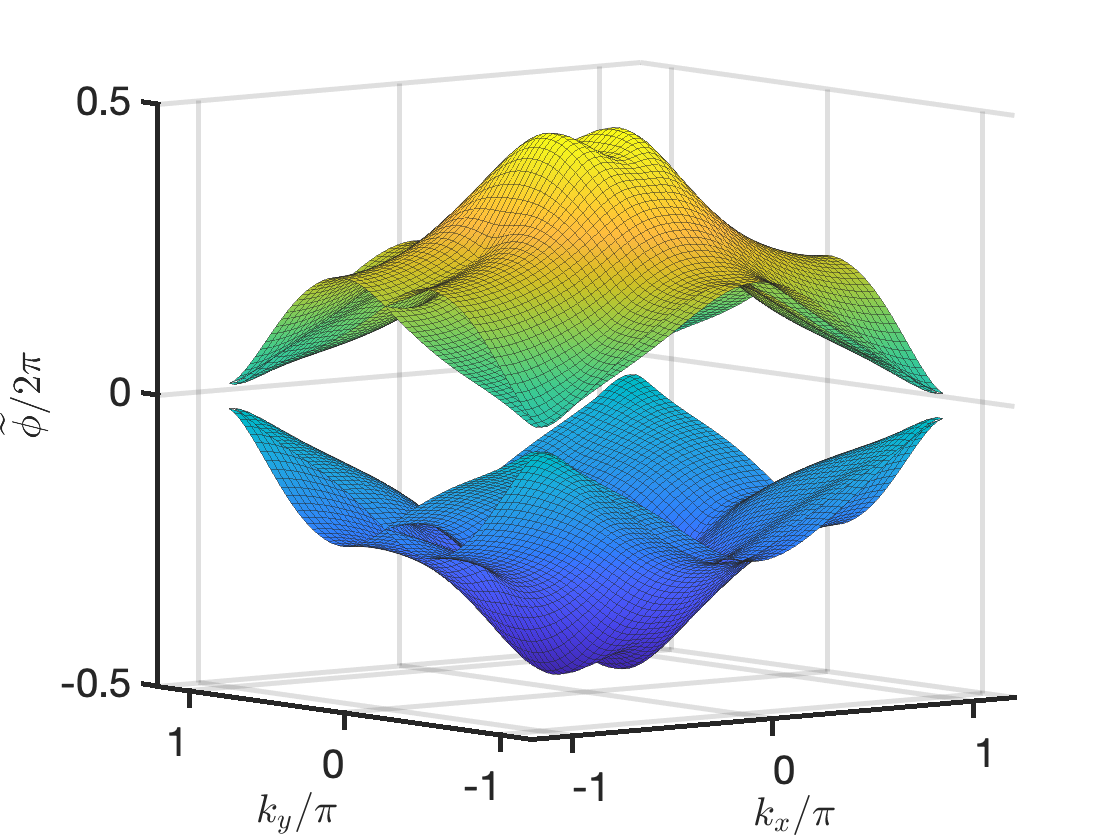}
	\caption{(a) Phase band $\widetilde{U}({\bf k}, \frac{T}{2})$ of the $D_2$ model~(\ref{eq:D2}). We choose $t_x=t_y=\pi/T$, $t_0=\sqrt{2}\pi/T$, in accordance with Ref.~\cite{hu2020dynamical}, and the model hosts corner mode at quasienergy $\pi/T$. The phase band shows a four-fold Dirac singularity at the $\Gamma$ point. (b) The singularity is gapped out upon adding a perturbation~(\ref{eq:perturb}) that respects $D_2$ symmetry. We choose $\delta=0.4\pi/T$. Although only the phase band at $t=\frac{T}{2}$ is shown here, we have also checked the phase band at arbitrary $t$ and confirmed that there is no singularity anywhere for $0<t<T$.}
	\label{fig:D2}
\end{figure}

We choose $t_x=t_y = \pi/T$ as in Ref.~\cite{hu2020dynamical}, and take $t_0 = \sqrt{2}\pi/T$ for which the model hosts corner modes at quasienergy $\pi/T$. As shown in Fig.~\ref{fig:D2}(a), the phase band exhibits a four-fold degenerate Dirac singularity at $(k_x=k_y=0, t=\frac{T}{2})$. Slightly changing $t_0$ will split this Dirac node and form the Weyl quadrupole configuration as discussed in Ref.~\cite{hu2020dynamical}. Therefore, the situation shown in Fig.~\ref{fig:D2}(a) corresponds to smoothly deforming the quadrupole pattern such that they coincide at one point. Next, we introduce the following perturbation to Hamiltonian~(\ref{eq:D2}):
\begin{equation}
h_0 \rightarrow h_0 - \delta( \tau_2 \sigma_2 + {\rm cos} k_x \tau_3 \sigma_2 + {\rm sin}k_y \tau_3 \sigma_1).
\label{eq:perturb}
\end{equation}
One can easily check that this perturbation respects $D_2$ symmetry. However, upon adding this perturbation, the singularity is gapped out, as shown in Fig.~\ref{fig:D2}(b). Therefore, we conclude that the dynamical Weyl quadrupole configuration introduced in Ref.~\cite{hu2020dynamical} is unstable under generic $D_2$-symmetric perturbations, consistent with our previous general discussions. 

\section{Phase diagram of the $C_2$-symmetric model}
\label{app:c2}

\begin{figure*}
	\includegraphics[width=0.6\textwidth]{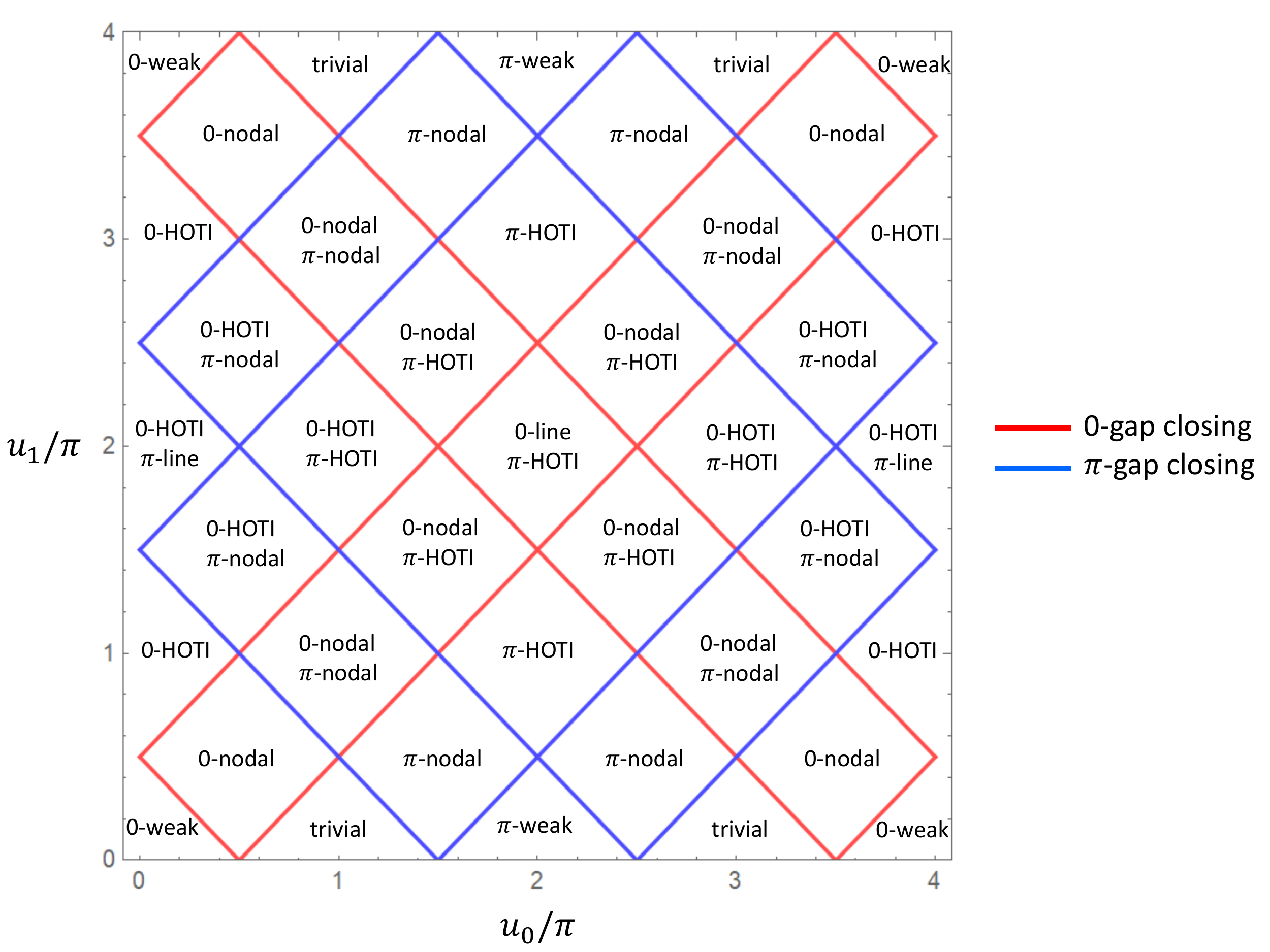}
	\caption{Phase diagram of $H^{(2)}$ for $u_2=0.5\pi$}
	\label{Fig: S1}
\end{figure*}

We first present the matrix representation of $H^{(2)}$ in momentum space,
\begin{eqnarray}
h_1 &=& u_0 \tau_0 \otimes \sigma_1,\nonumber \\
h_2 &=& \begin{pmatrix}
0 & u_1 e^{ik_x} & u_2 & 0 \\
u_1 e^{-ik_x} & 0 & 0 & u_2 e^{i k_y} \\
u_2 & 0 & 0 & u_1 e^{ik_x} \\
0 & u_2 e^{-ik_y} & u_1 e^{-ik_x} & 0 \\
\end{pmatrix}
\end{eqnarray}
The analytical expression for $U({\bf k}, t)$ at a general $({\bf k}, t)$ is difficult to obtain, which nonetheless is indeed solvable for each rotation-invariant axis. We further note that the Floquet quasienergy gap can close only at ${\bf k}=(0,0)$ or ${\bf k}=(\pi,0)$. Then the phase boundaries are found to be
\begin{eqnarray}
	u_0 \pm u_1 + u_2 = \frac{2m\pi}{T} \text{  and  }u_0 \pm u_1 - u_2 = \frac{2m\pi}{T}
\end{eqnarray}
for $m \in \mathbb{Z}$. These phase boundaries lead to a rich topological phase diagram, as shown in Fig. \ref{Fig: S1}. In particular, the blue and red phase boundaries in Fig. \ref{Fig: S1} denote the Floquet gap closing at quasieneries $0$ and $\pi$, respectively.
Here are the meaning of each phase in Fig. \ref{Fig: S1}
\begin{itemize}
	\item {\it Trivial}: A topologically trivial phase with no boundary feature.
	\item {\it 0-weak or $\pi$-weak}: A weak topological insulator which hosts 1d flat edge band at 0 or $\pi$ energy. 
	\item {\it 0-HOTI or $\pi$-HOTI}: A higher-order topological insulator with 0 or $\pi$ corner modes. 
	\item {\it 0-nodal or $\pi$-nodal}: A gapless system with point nodes at energy $0$ or $\pi$.
	\item {\it 0-line or $\pi$-line}: A gapless system with line nodes at energy $0$ or $\pi$.
\end{itemize}
For example, a $\pi$-HOTI only hosts $\pi$-corner modes instead of $0$-corner modes. On the other hand, a 0-HOTI\&$\pi$-HOTI hosts both $0$ and $\pi$ corner modes, which is exactly the AFHOTI phase in the main text. For our purpose, we focus on the AFHOTI phase in the phase diagram. A full topological characterization of the other exotic Floquet phases is left for future work.

\section{Analytical details of $D_4$-symmetric AFHOTI}
\label{app:d4}

\subsection{Analytical expression of $U({\bf k}, t)$}

We are interested in the time-evolution operator $U({\bf k}, t)$ for $t\in (\frac{T}{4},T]$, since $U({\bf k}, t<\frac{T}{4})$ cannot host any DDP. For convenience, we define $U_{2} = U({\bf k}, \frac{T}{4}<t\leq \frac{3T}{4})$ and $U_{3} = U({\bf k}, \frac{3T}{4}<t\leq T)$. In what follows, we will set $T=1$ for simplicity. We switch to a rotation-symmetric basis by finding a unitary matrix $V$, s.t.
\begin{equation}
V C_4 V^{\dagger} = \begin{pmatrix}
e^{i\frac{\pi}{4}} & 0 & 0 & 0 \\
0 & e^{-i\frac{\pi}{4}} & 0 & 0 \\
0 & 0 & e^{i\frac{3\pi}{4}} & 0 \\
0 & 0 & 0 & e^{-i\frac{3\pi}{4}} \\
\end{pmatrix} = e^{i\frac{\pi}{2} J_z}, 
\end{equation}
where $J_z=\text{diag}[\frac{1}{2},-\frac{1}{2},\frac{3}{2},-\frac{3}{2}]$.
Therefore, the Hamiltonian $\tilde{H}^{(4)}({\bf k}, t)=VH^{(4)}({\bf k}, t)V^{\dagger}$ is under the following new basis
\begin{equation}
\Psi = \left(|\frac{1}{2}\rangle,|-\frac{1}{2}\rangle,|\frac{3}{2}\rangle,|-\frac{3}{2}\rangle \right)^T.
\end{equation}
We expand $U_{2t}$ and $U_{3t}$ around $\Gamma$ and find that they have a universal form,
\begin{equation}
U_{i} ({\bf k}, t) = \begin{pmatrix}
a_i & 0 & b_i & 0 \\
0 & a_i & 0 & c_i \\
-b_i^* & 0 & a_i^* & 0 \\
0 & -c_i^* & 0 & a_i^* \\
\end{pmatrix}, \text{ for }i=2,3,
\end{equation}
where 
\begin{eqnarray}
a_{2} &=& e^{-\frac{i}{2\sqrt{2}}[w_0+w_1(4t-1)]},\nonumber  \\  
b_{2} &=& -\frac{1+i}{2} \sin \left[\sqrt{2}w_1(t-\frac{1}{4})\right] e^{\frac{iw_0}{2\sqrt{2}}}  k_+, \nonumber \\
c_{2} &=& -\frac{1-i}{2} \sin \left[\sqrt{2}w_1(t-\frac{1}{4})\right] e^{\frac{iw_0}{2\sqrt{2}}}  k_-, 
\end{eqnarray}
and
\begin{eqnarray}
a_{3} &=& e^{-\frac{i}{\sqrt{2}}[w_1+w_0(2t-1)]},\nonumber  \\  
b_{3} &=& -\frac{1+i}{2} \sin\left[\frac{w_1}{\sqrt{2}}\right] e^{i\sqrt{2}w_0(1-t)}  k_+, \nonumber  \\ 
c_{3} &=& -\frac{1-i}{2} \sin \left[\frac{w_1}{\sqrt{2}} \right] e^{i\sqrt{2}w_0(1-t)}  k_-.
\end{eqnarray}
When a DDP occurs at $({\bf k}_0, t_0)$, we expect the time-evolution operator
\begin{equation}
U({\bf k}_0, t_0) = -\mathbb{1}_4
\end{equation}
Take ${\bf k}_0=\Gamma$, the condition for a DDP to occur in $U_i$ is simply
\begin{equation}
a_i = -1.
\end{equation}

\subsection{Condition for DDP}
Let us first check $U_2$. When imposing $a_2=-1$, we have 
\begin{equation}
w_0+w_1(4t-1) =2\sqrt{2} \pi,\ \ \frac{T}{4}\leq t \leq \frac{3T}{4},
\label{Eq: U2 DDP constraint}
\end{equation}
which is only possible when
\begin{equation}
2w_1 + w_0 \geq 2\sqrt{2}\pi.
\label{Eq: U2 DTS condition}
\end{equation}
We have focused on the parameter regime where $w_0, w_1 \in [0,\sqrt{2}\pi]$, due to the periodicity of the phase diagram \cite{huang2020floquet}.

When DDP exists in $U_3$, we have $a_3=-1$, leading to
\begin{equation}
\sqrt{2}(w_1 + w_0 (2t-1)) = 2\pi,\ \ \frac{3T}{4}\leq t \leq T.
\end{equation}
By solving the above inequalities, we arrive at the DTS condition for $U_3$ as 
\begin{eqnarray}
2w_1 + w_0 &\leq& 2\sqrt{2}\pi \nonumber \\
w_1 + w_0 &\geq& \sqrt{2}\pi.
\end{eqnarray}
Together with the condition for $U_2$, we conclude that the DDP always exists when $w_0 + w_1 \geq \sqrt{2}\pi$.
As shown in Ref.~\cite{huang2020floquet}, $w_0 + w_1 = \sqrt{2}\pi$ is exactly the phase boundary separating the phases with and without $\pi$ corner modes for $H^{(4)}$. Thus, we have proved for $H^{(4)}$ that 
\begin{itemize}
	\item the DDP occurs {\it if and only if} the system hosts symmetry-protected $\pi$ corner modes.
\end{itemize}

\subsection{Effective Hamiltonian of $D_4$-protected DDP}

Now we show that the topological singularity found above by is indeed a $D_4$-protected four-fold-degenerate Dirac point. As an example, we consider the singularity at $t=t_0\in (\frac{T}{4}, \frac{3T}{4}]$ for $U_2$. Following Eq. \ref{Eq: U2 DDP constraint}, we have
\begin{equation}
	w_0+w_1(4t_0-1) =2\sqrt{2} \pi
\end{equation}
By expanding $U_2({\bf k}, t)$ around $t_0$ with $t=t_0+\Delta t$, and denote $\Delta t \equiv t$, we arrive at
\begin{widetext}
\begin{equation}
U_2({\bf k}, t) \approx -\mathbb{1}_4 + \begin{pmatrix}
i v_t t & 0 & -e^{i\frac{\pi}{4}} v_k k_+ & 0 \\
0 & i v_t t & 0 & -e^{-i\frac{\pi}{4}} v_k k_- \\
e^{-i\frac{\pi}{4}} v_k^* k_- & 0 & -i v_t t & 0 \\
0 & e^{i\frac{\pi}{4}} v_k^* k_+ & 0 & -i v_t t \\
\end{pmatrix}
\end{equation}
where 
\begin{eqnarray}
v_t = \frac{w_1}{\sqrt{2}},\ \ v_k = \frac{1}{\sqrt{2}} e^{i\frac{w_0}{2\sqrt{2}}} \sin \frac{w_0}{2\sqrt{2}}. 
\end{eqnarray}
This immediately leads to
\begin{equation}
h_D = i \log U_2({\bf k}, t) = \begin{pmatrix}
v_t t & 0 & -i e^{i\frac{\pi}{4}} v_k k_+ & 0 \\
0 & v_t t & 0 & -i e^{-i\frac{\pi}{4}} v_k k_- \\
i e^{-i\frac{\pi}{4}} v_k^* k_- & 0 & -v_t t & 0 \\
0 & i e^{i\frac{\pi}{4}} v_k^* k_+ & 0 & -v_t t \\
\end{pmatrix},
\end{equation}
which is exactly a 3d Dirac point in the $({\bf k}, t)$ parameter space.
\end{widetext}

\bibliography{singularity}

\end{document}